\documentclass[a4paper,11pt]{article}
\pdfoutput=1 

\usepackage{jcappub}
\usepackage{amsmath}
\usepackage{amssymb}
\usepackage{amsfonts}
\usepackage{epsfig}
\usepackage{epsfig,multicol,bbm}
\usepackage[printonlyused]{acronym}
\usepackage{subfig}
\usepackage{multirow}
\usepackage{bm}
\usepackage{float}
\usepackage{array}
\usepackage{graphicx}
\usepackage[T1]{fontenc}
\usepackage{cancel}
\usepackage{mathrsfs}
\usepackage{listings}
\usepackage{slashed}
\usepackage{url}
\usepackage[usenames,dvipsnames]{color}
\usepackage{hyperref}
\usepackage{xcolor}
\usepackage{hhline}
\usepackage{textalpha}
\usepackage{mathtools}
\usepackage{verbatim}
\usepackage{diagbox}

\newcommand \eq {{\rm eq}}
\newcommand{\be}{\begin{equation}}
\newcommand{\ee}{\end{equation}}

\def\dr{X}
\def\TD{T_{\rm D}}
\def\neff{N_{\rm eff}}
\def\dneff{\Delta N_{\rm eff}}

\newcommand{\ttt}[1]{\texttt{#1}}

\newcommand{\Mpl}{M\unt{Pl}}
 
\newcommand{\bath}{\mathcal{B}}

\newcommand{\gst}{g_{* \rho}}
\newcommand{\gsts}{g_{* s}}

\newcommand{\mean}[1]{\left\langle #1\right\rangle}
\newcommand{\algn}[1]{\begin{aligned} #1\end{aligned}}

\newcommand{\unt}[1]{_{\mathrm{#1}}}

\newcolumntype{?}{!{\vrule width 1pt}}
\usepackage{makecell}

\title{\boldmath Dark Radiation from the Primordial Thermal Bath in Momentum Space}
\author[a,b]{Francesco D'Eramo}
\author[c]{, Fazlollah Hajkarim}
\author[d,e]{, \\ Alessandro Lenoci}
\affiliation[a]{Dipartimento di Fisica e Astronomia, Universit\`a degli Studi di Padova, \\ Via Marzolo 8, 35131 Padova, Italy}
\affiliation[b]{Istituto Nazionale di Fisica Nucleare (INFN), Sezione di Padova, \\ Via Marzolo 8, 35131 Padova, Italy}
\affiliation[c]{Department of Physics and Astronomy, University of Oklahoma, Norman, OK 73019, USA}
\affiliation[d]{Deutsches Elektronen-Synchrotron DESY, Notkestr. 85, 22607 Hamburg, Germany }
\affiliation[e]{Racah Institute of Physics, The Hebrew University, 91904, Jerusalem, Israel }
\emailAdd{francesco.deramo@pd.infn.it}
\emailAdd{fazlollah.hajkarim@ou.edu}
\emailAdd{alessandro.lenoci@mail.huji.ac.il}
\preprint{DESY-23-177}

\abstract{Motivated by the stunning projections for future CMB surveys, we evaluate the amount of dark radiation produced in the early Universe by two-body decays or binary scatterings with thermal bath particles via a rigorous analysis in momentum space. We track the evolution of the dark radiation phase space distribution, and we use the asymptotic solution to evaluate the amount of additional relativistic energy density parameterized in terms of an effective number of additional neutrino species $\Delta N_{\rm eff}$. Our approach allows for studying light particles that never reach equilibrium across cosmic history, and to scrutinize the physics of the decoupling when they thermalize instead. We incorporate quantum statistical effects for all the particles involved in the production processes, and we account for the energy exchanged between the visible and invisible sectors. Non-instantaneous decoupling is responsible for spectral distortions in the final distributions, and we quantify how they translate into the corresponding value for $\Delta N_{\rm eff}$. Finally, we undertake a comprehensive comparison between our exact results and approximated methods commonly employed in the existing literature. Remarkably, we find that the difference can be larger than the experimental sensitivity of future observations, justifying the need for a rigorous analysis in momentum space.}

\begin{document} 

\maketitle

\flushbottom
\section{Introduction}
\label{sec:intro}

Motivated extensions of the standard model (SM) often incorporate new light and feebly interacting degrees of freedom~\cite{Alexander:2016aln,Battaglieri:2017aum,Beacham:2019nyx}. Sometimes, these elusive particles are even stable. Whether their stability follows from the symmetry structure of the underlying theory (e.g., new global symmetries) or their lifetime is just much longer than the age of the Universe, it is important to quantify accurately and precisely their relic abundance as well as the associated detectable signals. Explicit realizations include frameworks where these new particles serve as dark matter candidates~\cite{Hall:2009bx,Bernal:2017kxu,Boyarsky:2018tvu} with mass bound to be above the keV range if they are produced via processes involving particles from the primordial bath~\cite{Heeck:2017xbu,Boulebnane:2017fxw,Ballesteros:2020adh,DEramo:2020gpr,Decant:2021mhj}.

In this work, we focus on new physics scenarios with weakly-coupled particles that never contribute to the energy density of non-relativistic matter. They are instead always in the relativistic regime and therefore provide an additional radiation component in the early Universe. We probe the energy density stored in relativistic degrees of freedom at two key moments in cosmic history. The first one is Big Bang Nucleosynthesis (BBN), which is not a sudden process, and it is sensitive to the radiation energy density when the Universe was one second old and the temperature $T_{\rm BBN} \sim 1 \, {\rm MeV}$. Another crucial probe is the last scattering surface, namely the moment when protons and electrons form neutral hydrogen, and photons could travel undisturbed afterward. This happens much later than BBN although the Universe is still relatively young: its age is approximately 380,000 years, and the photon temperature $T_{\rm CMB} \sim 0.3 \, {\rm eV}$. The photons from the last scattering surface that we detect today are dubbed the Cosmic Microwave Background (CMB) radiation. These two cosmic probes provide two independent measurements of the radiation energy density at early times. Incidentally, the measured number of effective relativistic degrees of freedom does not have to be the same since things may change as the Universe evolves between $T_{\rm BBN}$ and $T_{\rm CMB}$.

The only SM relativistic degrees of freedom at both BBN and CMB are photons and neutrinos. However, theories motivated from the top down often feature the presence of one (or more) light particle $\dr$ that can be present in the relativistic regime and with a significant cosmic abundance. Historically, the presence of this \textit{dark radiation} is quantified in terms of an effective number of additional neutrino species. More explicitly, the energy density of relativistic particles at the time of CMB formation can be written as
\be
\rho_R(T_{\rm CMB})  = \rho_\gamma + \rho_\nu + \rho_\dr = 
\left[ 1 + \frac{7}{8} N_{\rm eff} \left( \frac{4}{11} \right)^{4/3}\right]  \rho_\gamma  \ .
\label{eq:rhoRCMB}
\ee
The first term in the bracket accounts for both photon polarizations, whereas the second one is proportional to the statistical Fermi-Dirac factor of $7/8$ and to the fourth power of the neutrino to photon temperature ratio $T_\nu / T_\gamma = (4/11)^{1/3}$. This relation defines the dimensionless factor $\neff$ that quantifies the number of effective neutrino species. If we stick to the SM alone, one may naively conclude that this quantity is just the number of fermion generations (in this case three). In fact, a careful study of neutrino decoupling leads to a slightly larger number~\cite{Bennett:2019ewm,EscuderoAbenza:2020cmq,Akita:2020szl,Froustey:2020mcq,Bennett:2020zkv}. The recent analysis in Ref.~\cite{Cielo:2023bqp} reports the value $N^{\rm SM}_{\rm eff} = 3.043$.

How well do we measure the radiation energy density in the early Universe? Or, equivalently, how well do we measure $\neff$? BBN and CMB have similar sensitivities at the moment. The analysis of light element abundances, where the effective number of neutrino species is usually denoted by $N_\nu$, puts the constraint $N^{\rm BBN}_\nu = 2.889 \pm 0.229$~\cite{Pisanti:2020efz,Yeh:2022heq}. From the CMB side, the most constraining result comes from the Planck satellite, $\neff^{\rm Planck} = 2.99 \pm 0.17$~\cite{Planck:2018vyg}. Other ground-based experiments, such as ACT~\cite{ACT:2020gnv} and SPT-3G~\cite{SPT-3G:2021wgf}, probe $\neff$ with a lower sensitivity and are consistent with Planck. The measured values at BBN and CMB are consistent with each other, and they are both consistent with the SM prediction. This does not have to be the case necessarily because extra dark radiation components can alter the SM prediction, and things may change between BBN and CMB formation. As a result, the presence of dark radiation in the early Universe is severely limited. 

Thus $\neff$ is a powerful probe of SM extensions containing light degrees of freedom~\cite{Brust:2013ova}, and this is the first motivation for our work. The second one is supported by the expected improvements for the future. At the moment, the best sensitivity available is the one due to Planck and it results in $\sigma(N_{\rm eff}) \simeq 0.17$. In the near future, the first significant advancement will be due to the Simons Observatory~\cite{SimonsObservatory:2018koc} with $\sigma(N_{\rm eff}) \simeq 0.05$. Later on, CMB-S4~\cite{CMB-S4:2016ple,Abazajian:2019eic,CMB-S4:2022ght} will reach $\sigma(N_{\rm eff}) \simeq 0.03$. There are also futuristic proposals~\cite{Sehgal:2020yja} with the target $\sigma(N_{\rm eff}) \simeq 0.014$. This astonishing program will make $\neff$ an even more robust new physics probe. Even in the absence of new physics, the increased sensitivity will make the difference between the naive expectation for $N^{\rm SM}_{\rm eff}$ and its actual value detectable. This known SM case is an unambiguous illustration of how important it is to provide not only accurate but also precise predictions. 

Our main goal is to develop the framework needed to provide robust predictions for the amount of dark radiation within SM extensions. We focus on a specific injection source: decays and/or scatterings involving degrees of freedom belonging to the primordial thermal bath. Our study is based upon two rather mild assumptions: there was a thermal bath before BBN, and the dark radiation candidate $\dr$ has feeble interactions with the thermal bath itself (e.g., SM particles). The first one, which is not supported by observations, is a reasonable extrapolation of what we probe at BBN~\cite{Kawasaki:2004qu}. Although our analysis assumes that the thermal bath dominates the energy budget, it is straightforward to extend our results to cases where the thermal bath is a sub-dominant component but it is still responsible for the production of dark sector particles~\cite{Co:2015pka,DEramo:2017gpl,Redmond:2017tja,Hamdan:2017psw,DEramo:2017ecx,Hardy:2018bph,Bernal:2018kcw,Calibbi:2021fld,Bernal:2022wck}. The second one is rather common in UV complete models. 

This \textit{thermal production} is in some sense unavoidable, and the way it works is rather simple from a qualitative point of view. Processes from the bath dump $\dr$ particles into the early Universe, and if they happen often enough they may even lead to the thermalization of dark radiation. Whether this happens or not, at some point the Universe is so cold and diluted that nothing can happen to $\dr$ particles anymore, they just free-stream uninterrupted. A rapid way to estimate the amount of dark radiation is via the instantaneous decoupling approximation. This quick method leads to answers that are definitely wrong by an amount larger than future experimental sensitivities, but also larger than the current ones if one considers production around the time of the QCD crossover where the number of SM effective degrees of freedom changes rapidly. Alternative methods based on ordinary differential equations that allow to tracking of the $\dr$'s number or energy densities improve the accuracy of the predictions. However, as we discuss in this paper, they are still based upon some assumptions that are not always justified. And the well-known example of SM neutrinos illustrates how important is to investigate the decoupling epoch carefully. 

We present here a general method to investigate the thermal production of dark radiation in momentum space based on an integro-differential Boltzmann equation that allows us to track the $\dr$'s distribution in the phase space. A study based entirely on a phase space analysis has multiple benefits. We list and explain below the main novelties of our approach.
\begin{itemize}
\item \textbf{Non-Thermalized Relics.} We always consider initial conditions at early times with no additional dark radiation. If the interaction strength is not enough, these particles may never thermalize in the early Universe but still be around today with a detectable abundance. Our methodology accounts carefully for parameter space regions where $\dr$ never reach thermalization. This has to be contrasted with other methods conventionally adopted in the literature. The instantaneous decoupling approximation cannot work in this case for obvious reasons. Even if we adopt improved procedures based on integrating the Boltzmann equation in momentum space to track a given moment of the distribution function (e.g., the number density), there are always some assumptions about reaching a thermal profile. \vspace{-0.1cm}
\item \textbf{Decoupling epoch.} Even if thermalization is achieved, the decoupling is neither instantaneous nor momentum-independent. Methods based on ordinary differential equations go beyond the instantaneous decoupling, but they still need to assume that the phase space profile of the distribution function remains thermal after decoupling. As we will show, different momenta decouple at different times and residual spectral distortions are somewhat unavoidable. These effects are not dramatic and do not alter the predicted amount of dark radiation by orders of magnitude, but they still lead to effects larger than the CMB-S4 sensitivity and therefore need to be accounted for.  \vspace{-0.1cm}
\item \textbf{Quantum statistical effects.} There is no reason that forces us to employ classical statistical distributions in our framework for bath particles or dark radiation. We discuss throughout this work what difference it makes to treat classically or quantum-mechanically the different actors. Here, we emphasize how the assumption of classical statistics is never needed in our case. As we explain in the paper, this assumption is not needed if one employs the instantaneous decoupling approximation either. However, if we adopt the procedure where ordinary differential equations describe the evolution of number or energy densities, one is forced to employ the Maxwell-Boltzmann statistics for the dark radiation to reduce the original integro-differential equation to ordinary ones. Bath particles can always be treated quantum-mechanically instead.
\vspace{-0.1cm}
\item \textbf{Feedback on the thermal bath.} The Boltzmann equation in momentum space describes accurately how dark radiation particles populate the associated phase space, but it is only part of the full story. Production processes also subtract energy from the thermal bath and it is not possible to study the two systems independently. We complete our framework with an additional Boltzmann equation accounting for the evolution of the energy density of the thermal bath. Besides its evolution following from the geometry of the Hubble expansion, the right-hand side of the Boltzmann equation contains a collision term that quantifies the energy exchanged with the dark radiation sector. It is enough for our purposes to track the energy density of the primordial bath without going to the phase space in this case; SM interactions are quite efficient at early times and ensure thermalization. 
\end{itemize}

We structure the presentation of our results as follows. Sec.~\ref{sec:formalism} contains the general formalism to track the phase space evolution of any thermally produced dark radiation particle. This work focuses on a broad class of production channels, the ones where only one dark radiation particle is produced in the final state of a generic process involving bath particles. We dub this scenario \textit{single production}, and the Boltzmann equation analysis simplifies substantially as explained in Sec.~\ref{sec:single}. Furthermore, we develop the formalism for two explicit production channels: two-body decays and binary collisions. Technical details about phase space integrals appearing in the collision operators of the Boltzmann equation are relegated to App.~\ref{app:PSint}. Our results are collected in Sec.~\ref{sec:results} where we solve for the phase space distribution and integrate over the particle momenta to compute the effective number of neutrino species as a function of dark sector masses and couplings. We perform the analysis for both classical and quantum statistics, and we highlight the differences when there are any. An undeniable advantage of a phase space analysis is the opportunity to track the evolution of each momentum bin, and in particular, analyze how different momenta decouple at different epochs in the cosmic history. This is the origin of spectral distortions we find in our solutions, and ultimately the reason why our predictions differ slightly from the ones obtained via standard methods. Thermalization and decoupling in each momentum bin are scrutinized in the App.~\ref{app:thermalization}. We put our study into the context of the current literature in Sec.~\ref{sec:comparison} where we compare our results with others derived via approximate methods. These standard procedures, which we review in App.~\ref{app:approx}, are the instantaneous decoupling approximation, and Boltzmann equations for the number density and the energy density of the dark radiation candidates. We explain in the paper all the assumptions needed to recover these approximate methods from the complete phase space framework. Remarkably, we find parameter space regions where our results differ from conventional methods by more than the sensitivity of future experiments. Finally, we summarize our findings and discuss future applications of our general framework in Sec.~\ref{sec:conclusion}.
 
\section{General framework for a phase space analysis}
\label{sec:formalism}

\begin{figure}
\centering
\includegraphics[width=0.6\textwidth]{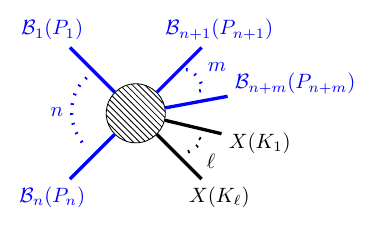}
\caption{The most general process for thermal production of dark radiation considered in this work: $n$ bath degrees of freedom ($\mathcal{B}_i$) collide and lead to a final state with $m$ bath and $\ell$ dark radiation ($\dr$) particles, respectively. We also include production via decays ($n = 1$).}
\label{fig:general}
\end{figure}

We develop in this Section the general formalism to investigate the thermal production of dark radiation in momentum space. The primordial bath is assumed to be a thermalized gas of weakly-coupled particles $\bath_i$ with temperature $T$ filling the early Universe. Our framework is applicable if such gas is made only of SM particles as well as for extensions with additional particles in thermal equilibrium. We consider the most general process producing an arbitrary number of dark radiation particles $\dr$ from initial and final states with $n$ and $m$ bath degrees of freedom, respectively. The process is illustrated in Fig.~\ref{fig:general} and it explicitly reads 
\be
\underbrace{\bath_1(P_1)+\dots+\bath_n(P_n)}_{n} \longleftrightarrow \underbrace{\bath_{n+1}(P_{n+1})+\dots+\bath_{n+m}(P_{n+m})}_m +
\underbrace{\dr(K_1) + \ldots \dr(K_\ell)}_{\ell} \ .
\label{eq:genprocess}
\ee
We put in parenthesis the associated four-momenta and we use the following parameterization: bath and dark radiation particles have four-momenta $P^\mu_i = (E_i, \vec{p}_i)$ and $K^\mu_r = (\omega_r, \vec{k}_r)$, respectively. Furthermore, we denote the magnitude of the spatial momentum with the same letter without the vector symbol (i.e., $\left| \vec{p}_i \right| = p_i$ and $| \vec{k}_r | = k_r$).

Our goal is to track the evolution of the phase space distribution (PSD) function $f_\dr$, and there are two distinct reasons why it changes. On the one hand, there is the dilution effect due to the Hubble expansion, as described by the Friedmann-Lemaître-Robertson-Walker (FLRW) geometry. On the other hand, processes such as the one  in Fig.~\ref{fig:general} create and destroy $\dr$ particles and this is an effect controlled by the underlying dynamics. 

In the remaining of this Section, we write down the collision operator for the process in Eq.~\eqref{eq:genprocess} and the Boltzmann equation describing the PSD evolution with time. However, the evolution of $f_\dr$ cannot be determined independently on the one of the thermal bath because of the energy exchanged between the two sectors. For this reason, we derive another Boltzmann equation governing the evolution of the thermal bath that accounts for the feedback of dark radiation particles, and we provide a system of Boltzmann equations that can be solved once we couple them with the Friedmann equation quantifying the Hubble rate. Finally, we provide a recipe to compute the additional number of effective neutrino species from the numerical solution of the Boltzmann system.

\subsection{Dark radiation evolution in momentum space}
\label{sec:PS}

Without any loss of generality, we focus on the $\dr$ particle with four-momentum $K_1$ in Fig.~\ref{fig:general}. Homogeneity and isotropy of the FLRW geometry ensure that the PSD can only depend on the magnitude of the spatial momentum $k_1$ and the cosmic time $t$. Thus the Boltzmann equation governing the evolution of $f_\dr(k_1, t)$ takes the schematic form $L[f_\dr(k_1, t)] = C[f_\dr(k_1, t)]$. The Liouville operator $L$ acting on the PSD takes care of the geometry of the expanding universe whereas the collision operator $C$ encodes the dynamics mediating the process in Fig.~\ref{fig:general}. For an FLRW background, the equation reads
\be
\omega_1 \frac{df_\dr(k_1, t)}{dt}  = C[f_\dr(k_1, t)] \ .
\label{eq:BEforF}
\ee
The collision operator accounting for Eq.~\eqref{eq:genprocess} explicitly reads (see, e.g., App.~B of Ref.~\cite{DEramo:2020gpr})
\be
\begin{split}
C[f_\dr(k_1, t)]  = & \frac{\ell}{2} \int \prod_{i = 1}^{n} d\Pi_i  \prod_{j = n+1}^{n+m}  d\Pi_j \prod_{r = 2}^{\ell}  d\mathcal{K}_r \; (2\pi)^4\delta^4 \left(P_{\rm in} - P_{\rm fin}  \right)  \\ &  \times 
\left[|\mathcal{M}_\rightarrow|^2  \prod_{i = 1}^n f_{\bath_{i}}\prod_{j=n+1}^{n+m} (1\pm f_{\bath_{j}})  \prod_{r = 1}^{\ell} \left(1 \pm f_{\dr}(k_r) \right) \, + \right. \\ & \left. \quad\quad 
- |\mathcal{M}_\leftarrow|^2 \prod_{j=n+1}^{n+m} f_{\bath_{j}} \prod_{r = 1}^{\ell}  f_{\dr}(k_r) \prod_{i = 1}^n \left(1 \pm f_{\bath_{i}} \right)   \right] \ .
\label{eq:Cfgeneral}
\end{split}
\ee
with $P_{\rm in} = \sum_{i=1}^n P_i$ and $P_{\rm fin} = \sum_{j=n+1}^{n+m} P_j + \sum_{r=1}^{\ell} K_r$. The Lorentz invariant relativistic phase space factors are defined as $d\Pi_i = g_i d^3 p_i / [(2\pi)^3 2 E_i]$ and $d\mathcal{K}_r = g_\dr d^3 k_r / [(2\pi)^3 2 \omega_r]$ with $g_i$ and $g_X$ the number of internal degrees of freedom for $\bath_i$ and  $\dr$, respectively. The transition amplitudes for the production ($\mathcal{M}_\rightarrow$) and destruction ($\mathcal{M}_\leftarrow$) are squared and then averaged over initial and final states with the appropriate symmetry factors for identical particles. So far, we have kept the matrix element for the direct and inverse processes distinct. However, if the interactions mediating the process in Eq.~\eqref{eq:genprocess} conserve CP or equivalently T, we have the identity $|\mathcal{M}_\rightarrow|^2 = |\mathcal{M}_\leftarrow|^2 = |\mathcal{M}|^2$. From now on, we assume CP conservation.

Bath particles follow equilibrium PSDs that can be either Bose-Einstein (BE) or Fermi-Dirac (FD). We will discuss throughout this paper when quantum statistical effects can be neglected and the Maxwell-Boltzmann (MB) statistics can serve as a valid approximation. Chemical potentials are negligible at high temperatures, and the explicit expressions for the distribution functions of bath particles in thermal equilibrium read
\be
f_{\bath_i} = 
\left\{ \begin{array}{ccl} \left( \exp[E_i / T] - 1 \right)^{-1} & $\qquad \qquad$  & \text{BE} \\ 
\left( \exp[E_i / T] + 1 \right)^{-1} & $\qquad$ & \text{FD} \\ 
\exp[- E_i / T] & $\qquad$ & \text{MB} \end{array}\right. \ .
\label{eq:feq}
\ee
It is possible to simplify the collision operator in Eq.~\eqref{eq:Cfgeneral} by plugging the explicit expressions given in Eq.~\eqref{eq:feq} for the PSDs of bath particle. Upon using the identity
\be
\prod_{i = 1}^n \frac{1 \pm f_{\bath_{i}}}{f_{\bath_{i}}} \prod_{j=n+1}^{n+m} \frac{f_{\bath_{j}}}{1\pm f_{\bath_{j}}} = 
\prod_{i = 1}^n e^{E_i/T} \prod_{j=n+1}^{n+m} e^{- E_j/T} = \prod_{r = 1}^l  e^{\omega_j/T} \ ,
\ee
where in the last step we impose the conservation of energy, we find~\footnote{One can derive the same result by invoking the detailed balance principle that for the CP conserving case reads $\prod_{i = 1}^n f_{\bath_{i}}\prod_{j=n+1}^{n+m} (1\pm f_{\bath_{j}}) \prod_{r = 1}^{\ell} \left(1 \pm f_{\dr}^\eq(k_r) \right)  =
\prod_{j=n+1}^{n+m} f_{\bath_{j}} \prod_{r = 1}^{\ell}  f_{\dr}^\eq(k_r) \prod_{i = 1}^n \left(1 \pm f_{\bath_{i}} \right) $.}
\be
\begin{split}
C[f_\dr(k_1, t)] = & \frac{\ell}{2} \int \prod_{i = 1}^{n} d\Pi_i  \prod_{j = n+1}^{n+m}  d\Pi_j \prod_{r = 2}^{\ell}  d\mathcal{K}_r \; (2\pi)^4\delta^4 
\left(P_{\rm in} - P_{\rm fin}  \right) |\mathcal{M}|^2  \\ &  \times 
 \prod_{i = 1}^n f_{\bath_{i}}\prod_{j=n+1}^{n+m} (1\pm f_{\bath_{j}})   \prod_{r = 1}^{\ell} \left(1 \pm f_{\dr}(k_r) \right) 
\left[ 1 - \prod_{r = 1}^{\ell} e^{\omega_r / T} \frac{f_{\dr}(k_r)}{1 \pm f_{\dr}(k_r)}  \right] \ .
\label{eq:Cfgeneralfinal}
\end{split}
\ee
This is the most general form for the collision operator once we account for CP-conserving processes producing an arbitrary number of dark radiation particles. 

\subsection{Feedback on the thermal bath}

Even without dark radiation, the bath energy density $\rho_\bath$ is not constant in time but it gets diluted by the expansion. Processes such as the one in Fig.~\ref{fig:general} also impact how $\rho_\bath$ evolves, and we quantify the energy exchanged between visible and dark sectors. Strictly speaking, we would need to couple Eq.~\eqref{eq:BEforF} with the analogous ones for each $ \bath_i$ to account for the dark radiation feedback on the thermal bath evolution. In practice, bath particles are always in equilibrium and it is sufficient to focus on their energy densities that are collectively summed to give the total bath energy density. The starting point to determine the $\rho_\bath$ evolution is the integration of both sides of Eq.~\eqref{eq:BEforF} over the phase space
\be
\frac{d\rho_\dr}{dt} + 4H\rho_{\dr} = g_\dr \int \frac{d^3 k_1}{(2\pi)^3} \omega_1 \frac{df_\dr}{dt} = g_\dr  \int \frac{d^3 k_1}{(2\pi)^3}  C[f_\dr(k_1, t)] \ .
\label{eq:Cfsingle_int}
\ee
The first equality follows from the identification of the dark radiation energy density $\rho_\dr$ as a function of time by integrating the PSD times the energy over the phase space
\be 
\rho_\dr(t) = g_\dr \int \frac{d^3 k_1}{(2\pi)^3} \omega_1  f_\dr(k_1, t)   \ .
\label{eq:rhodrmain}
\ee
Energy conservation imposes that the quantity appearing on the left-hand side of Eq.~\eqref{eq:Cfsingle_int} has to be, with the opposite sign, equal to the right-hand side of the equation accounting for the red-shift of the bath energy density. Thus the bath evolves according to
 \be
\frac{d\rho_{\cal B}}{dt} +3 H (\rho_{\cal B} + p_{\cal B}) = - \left( \frac{d\rho_\dr}{dt} + 4H\rho_{\dr} \right) = -  g_\dr  \int \frac{d^3 k_1}{(2\pi)^3}  C[f_\dr(k_1, t)]  \ . 
\label{eq:BEforSM}
\ee 
Energy density $\rho_\bath$ and pressure $p_\bath$ are related via the equation of state $p_\bath = w_\bath \rho_\bath$.

\subsection{Final system of Boltzmann equations}

The evolution of the dark radiation PSD $f_\dr$ and the bath energy density $\rho_\bath$ is controlled by the Boltzmann equations in Eqs.~\eqref{eq:BEforF} and \eqref{eq:BEforSM}, respectively. A generic collision operator $C$ appears on the right-hand side of both equations and for the process in Eq.~\eqref{eq:genprocess} and illustrated in Fig.~\ref{fig:general} we can use the simplified form provided in Eq.~\eqref{eq:Cfgeneralfinal} if CP is conserved. Both equations are differential in the cosmic time $t$ and they feature PSDs that depend on the bath temperature $T$. Thus we need a relation connecting time and temperature, and this is provided by the Friedmann equation accounting for the Hubble rate $H \equiv \dot{a} / a$, where $a(t)$ is the FLRW scale factor. To summarize, the full evolution is described by the system 
\be
\left\{
\begin{split}
& \, \omega_1 \frac{df_\dr(k_1, t)}{dt}  =  C[f_\dr(k_1, t)]  \\
& \, \frac{d\rho_{\cal B}}{dt} + 3 H (1 + w_{\cal B})  \rho_{\cal B} =  -  g_\dr  \int \frac{d^3 k_1}{(2\pi)^3}  C[f_\dr(k_1, t)] \\
& \, H  =\frac{\sqrt{\rho_{\cal B} + \rho_\dr}}{\sqrt{3} M_{\rm Pl}} 
\end{split} \right. \qquad  \ .
\label{eq:BoltzSystemGeneral}
\ee
The last equation of the system assumes that there is no other contribution to the energy density besides the ones from the thermal bath and the dark radiation. In other words, we are assuming that the thermal bath itself is dominating the energy budget at early times as is the case for a standard cosmological history that extrapolates the BBN snapshot of the Universe. Thus dark radiation production happens during a radiation-dominated epoch. 

As long as collisions happen at a significant rate, the evolution is nontrivial and the system above has to be solved numerically. However, once number-changing processes stop being effective the phase space evolution is just due to free-streaming and the dark radiation energy density defined in Eq.~\eqref{eq:rhodrmain} scales as $\rho_\dr \propto a^{-4}$. This is also the scaling of the photon energy density below the $e^{\pm}$ threshold (at temperatures below the electron mass). Thus once we are well below the MeV scale, the ratio $\rho_\dr / \rho_\gamma$ reaches a constant value, and this ratio is crucial to quantify the amount of additional radiation at the time of CMB formation.  As it is conventionally done in the literature, we isolate the SM contribution to $\neff$ and we split the total amount by defining $N_{\rm eff} = N^{\rm SM}_{\rm eff}  + \dneff$. We can identify the contribution to $\dneff$ by direct comparison between the two terms on the opposite sides of the last equality of Eq.~\eqref{eq:rhoRCMB}, and it explicitly reads
\be
\Delta N_{\rm eff} = \frac{8}{7} \left( \frac{11}{4} \right)^{4/3} \frac{\rho_{\dr}(t_{\rm CMB})}{\rho_\gamma(t_{\rm CMB})} \ . 
\label{eq:dneff}
\ee

\section{A concrete scenario: single production}
\label{sec:single}

We study quantitatively a scenario that we call \textit{single production}: dark radiation is produced via processes like the one in Fig.~\ref{fig:general} with only one $\dr$ particle in the final state (i.e., $\ell = 1$). This is often the leading production channel in concrete models due to a severe price to pay in terms of a small coupling for each final state $\dr$.

A prominent class of this kind of candidates are axion-like-particles (ALPs) with dimension 5 interactions to standard model fields suppressed by a high scale $\Lambda$. Interactions with gauge fields and fermions are both allowed whereas couplings to the Higgs currents can be redefined away. More concretely, for an ALP field $\phi$ we have the interactions
\be
\mathcal{L}_\phi = \frac{\phi}{8 \pi \Lambda} \sum_i \alpha_i F_i^{\mu\nu} \widetilde{F}_{i \, \mu\nu} + 
\frac{\partial_\mu \phi}{2 \Lambda} \left( \sum_{i j} \overline{\psi}_i \gamma^\mu \left( c_{ij}^V + c_{ij}^A \gamma^5 \right) \psi_j + {\rm h.c.} \right) \ .
\ee
The first class of operators contains the ALP coupled to standard model gauge fields $F_i^{\mu\nu}$ with the sum running over the three gauge groups. We follow the standard conventions adopted in the literature and we normalize these couplings proportionally to the fine structure constant $\alpha_i$ of the associated gauge group. The second contribution contains derivative couplings for the ALPs to vector and axial-vector currents of standard model fermions. If $\Lambda$ is much higher than the temperature under consideration, single production is clearly the leading production channel. Flavor-diagonal operators (i.e., $i = j$) mediate only production via binary scatterings. If flavor-violating interactions are allowed, two-body decays are also possible. The QCD axion~\cite{Peccei:1977hh,Peccei:1977ur,Wilczek:1977pj,Weinberg:1977ma} is a case motivated from the top-down by the strong CP problem.
 
We suppress the index labeling $\dr$ particles and we have the four-momentum $K^\mu = (\omega, \vec{k})$ with the relativistic dispersion relation $k = \omega$. The Boltzmann equation for $f_\dr$ simplifies in this case since the PSD appears in the collision operator only via the combination
\be
\left(1 \pm f_{\dr}(k) \right) 
\left[ 1 - e^{\omega / T} \frac{f_{\dr}(k)}{1 \pm f_{\dr}(k)}  \right] = 1 - f_{\dr}(k) \left( e^{\omega / T} \mp 1 \right) =  1 - \frac{f_\dr(k)}{f_\dr^{\rm eq}(k)} \ ,
\ee
where we identify the equilibrium PSD $f_\dr^{\rm eq}(k)$ for $\dr$ without the chemical potential. This identity holds for both Bose-Einstein and Fermi-Dirac statistics, and it is valid also when we neglect completely quantum effects and adopt Maxwell-Boltzmann statistics for $\dr$. The Boltzmann equation describing the PSD's evolution takes the simple form
\be
\frac{df_\dr(k, t)}{dt} = \mathcal{C}(k, t) \left( 1 - \frac{f_\dr(k, t)}{f_\dr^{\rm eq}(k, t)}  \right)  \ ,
\label{eq:Cfsingle1}
\ee
where we introduce the collision term
\be
\mathcal{C}(k, t) =   \, \frac{1}{2\omega}\int \prod_{i = 1}^{n} d\Pi_i  \prod_{j = n+1}^{n+m}  d\Pi_j  (2\pi)^4\delta^4 \left( P_{\rm in} -P_{\rm fin}\right) |\mathcal{M}|^2 \prod_{i = 1}^n f_{\bath_{i}}\prod_{j=n+1}^{n+m} (1\pm f_{\bath_{j}}) \ .
\label{eq:colltermsingle}
\ee
We stress the difference between the collision operator $C[f_\dr(k_1, t)]$ and the collision term $\mathcal{C}(k, t)$ defined in Eqs.~\eqref{eq:Cfgeneralfinal} and \eqref{eq:colltermsingle}, respectively; the former is a functional of the PSD $f_\dr$, the latter is a regular function of the physical momentum $k$ and the cosmic time $t$. The result in Eq.~\eqref{eq:Cfsingle1} is the most general Boltzmann equation describing the single production of dark radiation, and there is no assumption whatsoever about the functional form of the unknown PSD $f_\dr$ (e.g., we do not have to assume kinetic equilibrium). The quantity $\mathcal{C}(k, t)$ has the units of energy and it can be interpreted as the collision rate for a given momentum bin. We focus now on two specific production channels.

\begin{table}
	\centering
	\begin{tabular}{|c|c|c|}\hline
{$\bath_1$}&{$\bath_2 $}&$D(k,T)$\\\hline  \hline
MB & MB & $e^{-(k+E_2^-)/T}$\\\hline\hline
BE & BE &
$ -\frac{1}{e^{k/T}-1}\left[- \log \left({e^{-(E_2^-+k)/T}-1} \right)+\log\left(1-e^{{-E_2^-}/{T}} \right) \right]$\\\hline
BE & FD &
$ \frac{1}{e^{k/T}+1}\left[- \log \left({e^{-(E_2^-+k)/T}-1} \right)+\log\left(1+e^{{-E_2^-}/{T}} \right) \right]$\\\hline
FD & BE &
$ \frac{1}{e^{k/T}+1}\left[\log\left(1+{e^{-(E_2^-+k)/T}} \right)-\log\left(1-e^{{-E_2^-}/{T}} \right) \right]$\\\hline
FD & FD &
$ \frac{1}{e^{k/T}-1}\left[-\log\left(1+{e^{-(E_2^-+k)/T}} \right)+\log\left(1+e^{{-E_2^-}/{T}} \right) \right]$\\\hline
	\end{tabular}
	\caption{Explicit expressions for the function $D(k,T)$ appearing in in Eq.~\eqref{eq:C12final} for different bath particles statistics with $E_2^- = (m_1^2-m_2^2) / (4 k) + m_2^2 k / (m_1^2-m_2^2)$.}
	\label{tab:stats}
\end{table}

\subsection{Two-body decays}
\label{sec:2bodydec}

The first channel we investigate is $\dr$ production via two-body decays $\bath_1 \to \bath_2+ \dr$, and the general collision term in Eq.~\eqref{eq:colltermsingle} takes the following form 
\be
\mathcal{C}_{1 \rightarrow 2}(k, t)  =   \, \frac{1}{2\omega}\int d\Pi_1  d\Pi_2  (2\pi)^4\delta^4 \left( P_1-P_2-K\right) |\mathcal{M}_{\bath_1\to \bath_2 X}|^2  f_{\bath_{1}} (1\pm f_{\bath_{2}}) \ .
\label{eq:C_decays}
\ee
The partial decay width for this process results in
\be
\Gamma_{1} = \frac{g_2 \, g_\dr}{16 \pi m_1} |\mathcal{M}_{\bath_1\to \bath_2 X}|^2 \, \mathcal{Y}_2  \,  ,
\ee
The bath particles $\bath_1$ and $\bath_2$ have masses $m_1$ and $m_2$, respectively, and the degeneracy factors $g_2$ and $g_\dr$ are due to our conventions with the squared matrix element averaged also over final states. Finite $\bath_2$ mass effects are captured by the phase space factor $\mathcal{Y}_2 \equiv 1 - (m_2/m_1)^2$. The squared matrix element is constant (monochromatic final states), and we can take it outside of the integral and trade with the partial decay width. We spell out the remaining integrations over the phase space in App.~\ref{app:PSintA}, and we write the collision term as follows
\begin{equation} 
\mathcal{C}_{1 \rightarrow 2}(k,T) = \frac{g_1}{g_\dr} \frac{1}{{\cal Y}_2} \frac{\Gamma_1}{m_1}  \frac{m_1^2 T}{k^2} D(k,T) \  .
\label{eq:C12final}
\end{equation}
Here, we employ the bath temperature $T$ as a time variable. The dimensionless function $D(k,T)$ is defined by Eq.~\eqref{eq:Ddef}, and the explicit expressions for different statistics are shown in Tab.~\ref{tab:stats}. 

\begin{figure}
	\centering
	\includegraphics[width=.47\textwidth]{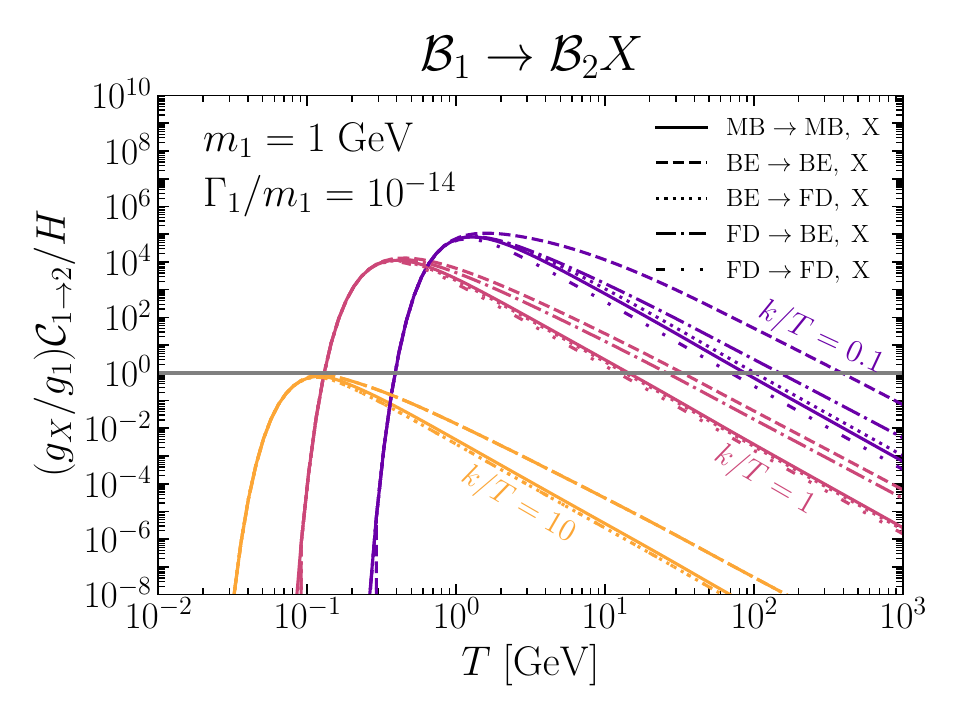} $\quad$
	\includegraphics[width=.47\textwidth]{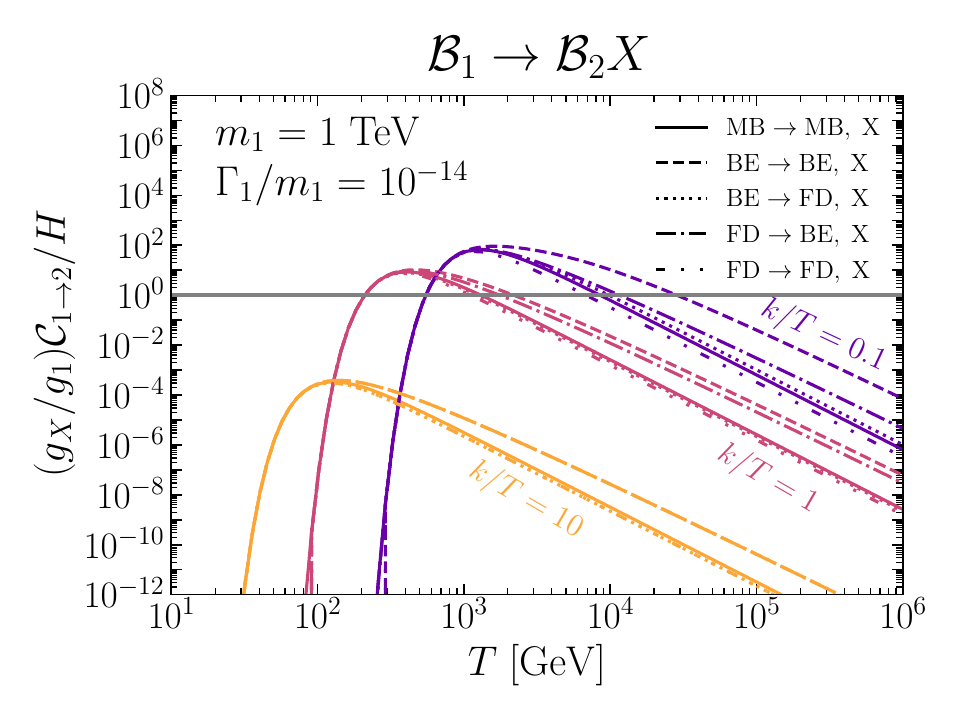}
	\caption{Collision terms for two-body decays divided by the Hubble parameter as a function of the bath temperature $T$. We fix the $\bath_1$ mass to 1 GeV (left) and 1 TeV (right), and the final state products to be massless. The decay width is set to $\Gamma_1 / m_1 = 10^{-14}$. We consider three different physical momenta $k$: $0.1 \, T$ (purple); $T$ (magenta); $10 \, T$ (yellow). The different line styles correspond to the choices for the particle statistics explained in the legend.}
	\label{fig:C_over_H}
\end{figure}

We compare the collision rate normalized as $(g_X/g_1){\cal C}_{1 \rightarrow 2}$ (to remove the  $g_i$ factors) with the expansion rate. The latter is provided by the Friedmann equation, and we evaluate it for a thermal bath of SM particles with the equation of state parameter $w_\bath = w_{\rm SM}$ as provided by Ref.~\cite{Laine:2015kra} (see also Refs.~\cite{Drees:2015exa,Saikawa:2018rcs}). Fig.~\ref{fig:C_over_H} shows the temperature dependence of the ratio $(g_X/g_1){\cal C}_{1 \rightarrow 2}(k,T) / H(T)$ for $m_1 = 1 \, {\rm GeV}$ (left panel) and $m_1 = 1 \, {\rm TeV}$ (right panel). For both cases, we fix the dimensionless ratio $\Gamma_1 / m_1 = 10^{-14}$ and $ m_2 = 0$. The different colors correspond to different choices of physical momenta, $k / T = \{0.1, 1, 10\}$, and for each color (i.e., for each chosen physical momentum) we show five different lines corresponding to the cases listed in Tab.~\ref{tab:stats}. We fix the values of the physical momenta in units of the temperature $T$ because after decoupling they change just due to the cosmological red-shift, $k \propto a^{-1}$. Likewise, the bath temperature cools down in agreement with entropy conservation and, up to corrections proportional to the variation of the bath degrees of freedom, it scales as $T \propto a^{-1}$. Thus the ratio $k / T$ is approximately constant after decoupling and it can be thought of as a given comoving momentum. We notice from Fig.~\ref{fig:C_over_H} how different comoving momenta decouple (i.e., ${\cal C}_{1 \rightarrow 2}  \lesssim H$) at different temperatures. We also observe how statistical effects may alter the temperature at which the dark radiation thermalizes with the bath, while the decoupling temperature is basically independent of the statistics.

\subsection{Binary scatterings}

\begin{figure}
	\centering
	\includegraphics[width=.49\textwidth]{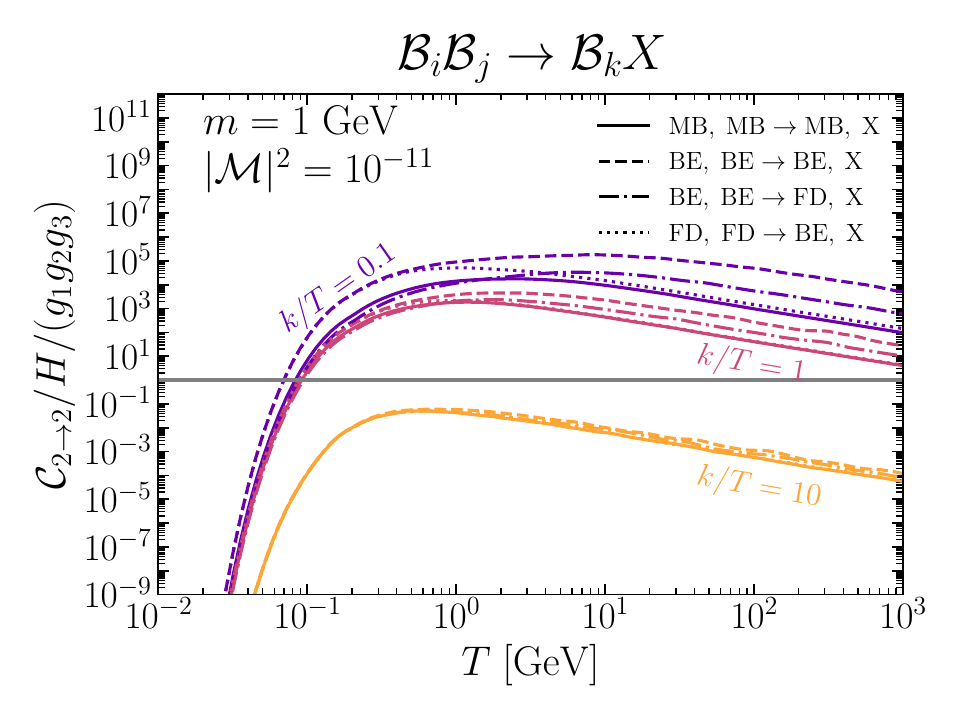}
	\includegraphics[width=.49\textwidth]{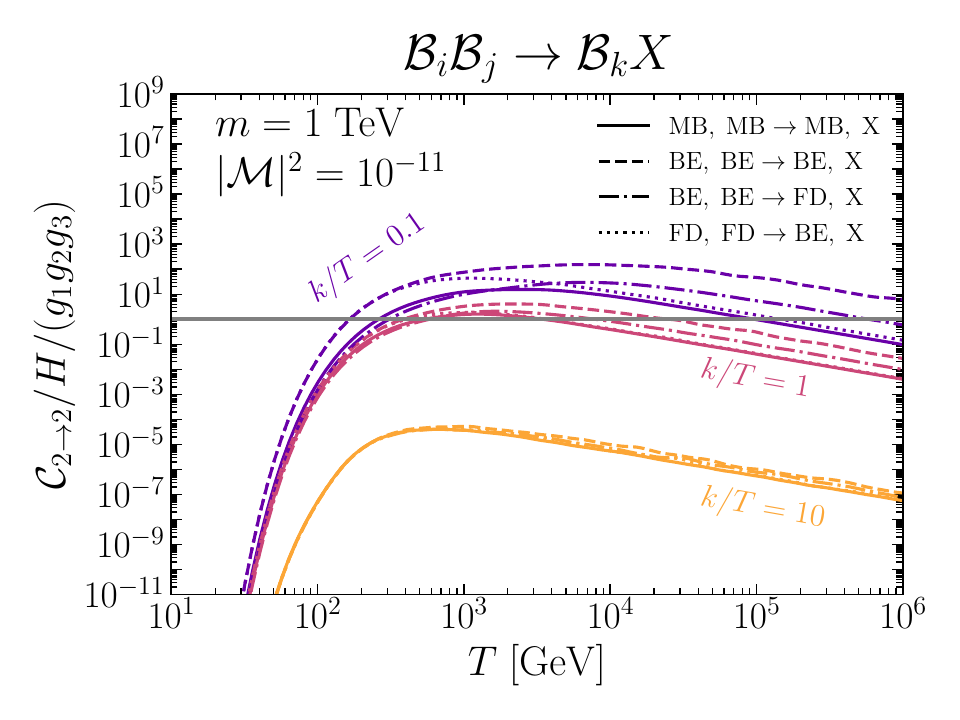}
	\caption{Collision terms for binary scatterings divided by the Hubble parameter as a function of the bath temperature $T$. The particle $\bath_3$ is massless, whereas $\bath_1$ and $\bath_2$ have the same mass $m$ equal to 1 GeV (left) and 1 TeV (right). We fix the scattering matrix element to $|{\cal M}|^2 = 10^{-11}$. The notation is the same as the one adopted in Fig.~\ref{fig:C_over_H}.}
	\label{fig:C_over_H_sca}
\end{figure}

The second example we consider is single production of dark radiation from binary scatterings. Once we identify a specific collision (e.g., $\bath_1 \bath_2 \rightarrow \bath_3 \dr$), we have to include all processes that are allowed by crossing symmetry and sum over them to get the collision rate
\be
\mathcal{C}_{2 \rightarrow 2}(k, t) = {\cal C}_{\bath_1\bath_2\to \bath_3 \dr}(k, t) + {\cal C}_{\bath_1\bath_3\to \bath_2 \dr}(k, t) +  {\cal C}_{\bath_2\bath_3\to \bath_1 \dr}(k, t) \ . 
\label{eq:collratescattering}
\ee
Each contribution can be obtained from the general result in Eq.~\eqref{eq:Cfsingle1}, and we find
\begin{equation}\algn{
	{\cal C}_{\bath_i\bath_j\to\bath_\ell\dr}(k,t) = &{}&{}\frac{1}{2 \omega}  \int d\Pi_id\Pi_j d\Pi_\ell\ (2\pi)^4 \delta^{(4)}(P_i + P_j  -P_\ell -K) \\
&	&\times |{{\cal M}_{\bath_i\bath_j\to\bath_\ell\dr}}|^2 f_{\bath _i}f_{\bath_j}(1\pm f_{\bath_\ell}).}\ 
	\label{eq:colltermscat}
\end{equation} 
The phase space integrals can be computed with the techniques presented in App.~\ref{app:PSintB} where we develop a formalism valid for arbitrary interactions, spectra, and statistics. For the sake of illustration, we focus here on a simplified framework where we set the (dimensionless) scattering matrix elements to a constant value independent on the momenta, and this quantity is the same for each permutation: $|{\cal M}_{\bath_1\bath_2\to \bath_3\dr}|^2 = |{\cal M}_{\bath_1\bath_3\to \bath_2\dr}|^2 = |{\cal M}_{\bath_2\bath_3\to \bath_1\dr}|^2 =|{\cal M}|^2$. We choose a mass spectrum with only one overall mass scale: $\bath_1$ and $ \bath_2 $ have the same mass $m_1= m_2= m$, $\bath_3$ is massless. If we adopt MB statistics for all bath particles, the collision term can be computed analytically up to an integral over the Mandelstam variable $s$ (see Ref.~\cite{DEramo:2020gpr}), and it reads
\be
\begin{split}
\mathcal{C}_{2 \rightarrow 2}(k, T)  = \frac{g_1 g_2 g_3}{256 \pi^3} |{\cal M }|^2 \,  \frac{ T }{k^2} \, e^{-\omega/T}  & \left[  \int_{4m^2}^\infty ds \sqrt{ 1-\frac{4m^2}{s}} e^{-\frac{s}{4k T}} + \right. \\ & \left.  \quad 2\times \int_{m^2}^\infty ds \left( 1-\frac{m^2}{s}\right) e^{-\frac{s}{4k T}+ \frac{m^2}{4kT}\left(1-\frac{4k^2}{s-m^2}\right)}  \right]    \ .
\end{split}
\label{eq:CscFinal}
\ee
The factor of 2 in the second term is due to the fact that $ {\cal C}_{\bath_1\bath_3\to\bath_2\dr} = {\cal C}_{\bath_2\bath_3\to\bath_1\dr}$.

Fig.~\ref{fig:C_over_H_sca} shows the comparison between the normalized collision rate ${\cal C}_{2 \to 2}/(g_1g_2g_3)$ and the Hubble expansion rate $H$. We choose two different values for the overall mass scale, $m= 1 \, {\rm GeV}$ (left panel) and $m = 1 \, {\rm TeV}$ (right panel), and we set the squared matrix element to the constant value $ |{\cal M}|^2 = 10^{-11} $. Lines in the plot colored differently correspond to different physical momenta $k$ in units of the temperature $T$: $k / T = \{0.1, 1, 10\}$. The reason for this normalization of the momenta is the same as the one for the decays. For each color, we show four different cases corresponding to different choices for the statistics of the bath particles. We notice how different comoving momenta decouple (i.e., ${\cal C}_{2\to 2} \lesssim H$) at different temperatures and how statistical effects may significantly alter the decoupling temperature, unlike in the decay case. Compared to the decay case, the collision term for scatterings is a shallower function of temperature. Thus if some momenta reach equilibrium, they remain coupled for a longer time than in the case of decays, making it easier to retain kinetic equilibrium.

\section{Results}
\label{sec:results}

In the previous section, we have introduced the single production scenario and shown how the Boltzmann equation for the dark radiation PSD reduces to the simple form in Eq.~\eqref{eq:Cfsingle1}. In particular, the collision term $\mathcal{C}(k, t)$ as defined in Eq.~\eqref{eq:colltermsingle} is a regular function of the physical momentum $k$ and the cosmic time $t$. We have derived the explicit collision term expression for two specific production processes, two-body decays and binary scatterings, and shown their temperature dependence in Figs.~\ref{fig:C_over_H} and \ref{fig:C_over_H_sca}. We are now ready to feed the Boltzmann equation with these collision terms, solve it, and use the output to quantify $\dneff$.

First, we rewrite the Boltzmann system in Eq.~\eqref{eq:BoltzSystemGeneral} for dark radiation single production 
\be
\left\{
\begin{split}
& \, \frac{df_\dr(k, t)}{dt} = \mathcal{C}(k, t)  \left( 1 - \frac{f_\dr(k, t)}{f_\dr^{\rm eq}(k, t)}  \right) \\
& \, \frac{d\rho_{\cal B}}{dt} + 3 H (1 + w_{\cal B})  \rho_{\cal B} =  -  g_\dr  \int \frac{d^3 k}{(2\pi)^3} \, k \, \mathcal{C}(k, t) \left( 1 - \frac{f_\dr(k, t)}{f_\dr^{\rm eq}(k, t)}  \right) \\
& \, H  =\frac{\sqrt{\rho_{\cal B} + \rho_\dr}}{\sqrt{3} M_{\rm Pl}} 
\end{split} \right. \qquad \ .
\ee
We find it convenient to employ the FLRW scale factor $a(t)$ as the ``time variable'', and therefore we trade the cosmic time $t$ with $a(t)$. As a reference, we define $a_I$ to be the scale factor at some high-temperature scale $T_I$ well before production processes become efficient. For the scenarios considered in this work, physical results do not depend on this choice if $T_I$ is sufficiently high: production via decays is always IR dominated, and the same is true for production via collisions if the underlying interactions are renormalizable (see, e.g., Ref.~\cite{Hall:2009bx}).

We employ as the evolution variable the dimensionless ratio $A \equiv a/a_I$. After the processes accounted for by the collision term $\mathcal{C}(k, t)$ stop being efficient, $\dr$ particles red-shift with the expansion and their physical momenta decrease as $k \propto a^{-1}$. Thus the use of comoving momenta is particularly advantageous since they scale out the effect of the Hubble expansion, and for this reason, we introduce 
\be
q \equiv \frac{k a}{(a_I T_I)} = \frac{k A}{T_I}  \ .
\label{eq:comoving_momentum}
\ee
Likewise, we define the comoving energy densities of the thermal bath $R_{\cal B}$ and of the dark radiation $R_\dr$. We find it convenient to identify the following dimensionless quantities
\be
R_{\cal B} \equiv \frac{\rho_{\cal B} a^4}{T_I^4 a_I^4} = \frac{\rho_{\cal B}}{T_I^4} A^4 \ , \qquad \qquad \qquad
R_X \equiv \frac{\rho_X a^4}{T_I^4 a_I^4} = \frac{\rho_X}{T_I^4} A^4 \ .
\label{eq:Rvar}
\ee
Strictly speaking, the former is not always constant as the Universe expands since we have mass threshold effects that alter the SM equation of state. The latter can be computed at each value of $A$ (i.e., at each moment in time) once $f_\dr$ is known since we just need to integrate Eq.~\eqref{eq:rhodrmain} over the momenta and scale out the effect of the expansion
\be
R_\dr(A) = \frac{g_\dr}{2\pi^2} \int dq \ q^3 f_\dr(q,A)\ ,
\label{eq:RX}
\ee
where we take advantage of the isotropy of the Universe to perform the integral over the solid angle in polar coordinates. 

The Boltzmann system in terms of the new variables reads
\be
\left\{
\begin{split}
& \,  \frac{df_\dr(q,A)}{d\log A} = \frac{\mathcal{C}(q,A)}{H(A)} \left( 1 - \frac{f_\dr(q,A)}{f_\dr^{\rm eq}(q,A)} \right)   \\
& \, \frac{dR_{\cal B}}{d\log A} + (3 w_{\cal B} - 1) R_{\cal B}  =  
-  \frac{g_\dr}{2\pi^2} \int dq \, q^3 \left( 1 - \frac{f_\dr(q,A)}{f_\dr^{\rm eq}(q,A)} \right) \frac{\mathcal{C}(q,A)}{H(A)} \\
& \, H(A)  = \frac{\sqrt{R_{\cal B}(A) + R_\dr(A)}}{\sqrt{3} M_{\rm Pl}} \frac{T_I^2}{A^2} 
\end{split} \right. \qquad \ .
\label{eq:BoltzSystemSingle}
\ee
If one looks at the first equation of the system, the one for the evolution of $f_\dr$, it seems that different Fourier modes are not coupled to each other and the evolution of each momentum bin can be treated independently. In other words, one may erroneously conclude that this is not an integro-differential equation because the PSD for $\dr$ does not appear in the collision term. In fact, the different Fourier modes for $\dr$ particles are still coupled. The equation for the evolution of $f_\dr$ contains on the right-hand side the Hubble parameter that is set by the energy content of the universe; this is quantified by the Friedmann equation, the third one in the Boltzmann system. If one looks at Eq.~\eqref{eq:RX}, it is manifest how the $\dr$'s contribution to the energy density results from the integral of the PSD over all the Fourier modes. Furthermore, the right-hand side of the second equation of the system also contains an integral over all the modes. The Boltzmann equation is inevitably still integro-differential. This has to be contrasted with the case of dark matter freeze-in where dark matter particles never get close to thermalization, their contribution to the energy budget is negligible and therefore one can take the standard cosmological history result for the Hubble parameter~\cite{DEramo:2020gpr}.

In what follows, we solve the Boltzmann system numerically\footnote{We use the \ttt{lsoda} routine implemented in the \ttt{scipy odeint} wrapper of the \ttt{FORTRAN odepack} library. The routine allows for automatic switch between non-stiff (\ttt{adams}) and stiff (\ttt{BDF}) solvers and change of their orders to ensure an accurate solution for both stiff and not stiff cases. The algorithm is adaptive, that is it reduces the step size to keep the estimated local truncation error smaller than \ttt{atol + abs(f\_X,R\_B)*rtol}. We choose the maximum stepsize (in $\log A$) as \ttt{hmax=0.01}, \ttt{atol=rtol=1e-6}. Integrals are performed numerically with the \ttt{scipy romberg} integrator and tolerances \ttt{atol=rtol=1e-8}. We checked the robustness of our solutions with different algorithms, tolerances and codes (e.g. with both \ttt{scipy} and \ttt{Mathematica}). With a precomputed collision term, a solution to the system is found in <60 seconds on a quad-core laptop. The conservation of the total energy density, i.e. $1+(3w_{\bath}-1)R_\bath d\log A/( dR_\bath +dR_X) \ll 1 $ is assured below the percent level.}. We bin the comoving momentum in the range  $q\in [0.005,20]$ with $N_q=64$ logarithmically-spaced bins and we evaluate the $N_q$ functions of the temperature $\mathcal{C}(q,T)$ for two-body decays and binary scatterings. Thus we deal with $N_q+1$ Boltzmann equations, $N_q$ for the unknown function $f_\dr(q,A)$ of the associated momentum momentum bin, and a last one for $R_{\cal B}$. At each ``time step'' the PSD is interpolated in momenta and used to compute integrated quantities such as $ R_\dr $, and the right-hand side of the second equation in the system \eqref{eq:BoltzSystemSingle}. The temperature of the thermal bath $T$ at each value of $A$ can be found by solving the equation
\be
R_{\cal B}(A) = \frac{\pi^2}{30} \gst(T) \left(\frac{T}{T_I}\right)^4 A^4 \ .
\label{eq:RBvsA}
\ee
We consider a SM thermal bath with thermodynamic variables as provided by Ref.~\cite{Laine:2015kra}.

Initial conditions describe the system when $A = 1$ (i.e., $a = a_I$). In the beginning, we assume that there is no dark radiation, $f_\dr (q, A = 1) =0$. The generalization to the case where there is some initial $\dr$ population (e.g., from inflaton decays) is straightforward. On the contrary, the initial bath energy density is non-vanishing and follows from the relation in Eq.~\eqref{eq:RBvsA}, $R_{\cal B}(A=1) = \frac{\pi^2}{30} \gst(T_I)$. We integrate the Boltzmann system up to a value of the scale factor $A_F$ corresponding to a temperature $T_F$ well below the electron mass and above the recombination temperature. This is necessary because the effective number of relativistic degrees of freedom does not change anymore afterward, $\gst(T \ll T_F) = {\rm const}$, and the ratio between the $\dr$'s and bath's energy densities is constant. Thus we reach our final goal of evaluating the contribution to $ \dneff $ that results in
\begin{equation}
\dneff = \frac{8}{7} \left( \frac{11}{4} \right)^{4/3} \frac{R_\dr(A_F)}{2 R_\bath(A_F) / \gst(T_F)} \ .
\label{eq:DNeffX}
\end{equation}
In the denominator of the last fraction, we identify the energy density of the two photon polarizations since this is the quantity that enters in Eq.~\eqref{eq:dneff}.

We find it convenient to identify a dark radiation temperature. One possible definition for this quantity would be to solve the differential equation for the PSD, derive the energy density $\rho_\dr$ as defined in Eq.~\eqref{eq:rhodrmain}, and under the assumption that $ f_\dr \simeq f_\dr^\eq $ define a dark radiation temperature for the dark radiation inverting the relation in Eq.~\eqref{eq:rhodrapp}. This clearly works if the interaction strength is large enough to bring $\dr$ particles to equilibrium, but it does not work in the most general case. A more general definition, which we choose in our work, does not rely upon the assumption of thermalization. We define the dark radiation temperature from the width of the PSD computing its second moment as follows
\be
T_\dr(A) =  \frac{T_I}{A} \left[\frac{\int dq\ q^4 f_\dr(q,a)}{\int dq\ q^2 f_\dr(q,a)} \right]^{\frac12}\times 
\left\{ \begin{array}{ccl}   (12\zeta(5)/\zeta(3))^{-1/2} & $\qquad \qquad$  & \text{BE} \\ 
(15\zeta(5)/\zeta(3))^{-1/2} & $\qquad \qquad$ & \text{FD} \\ 
(2\sqrt{3})^{-1} & $\qquad \qquad$ & \text{MB} \end{array}\right. \ .
\label{eq:TX}
\ee 
The overall normalization coefficients are understood by considering the limit where interactions are strong enough to ensure thermalization. In this regime, $\dr$ particles are in equilibrium with the bath and the PSD reaches the expressions
\be
f_{\dr}(q, A) = 
\left\{ \begin{array}{ccl} \left( \exp[q \, T_I / (A \, T)] - 1 \right)^{-1} & $\qquad \qquad$  & \text{BE} \\ 
	\left( \exp[q \, T_I / (A \, T)] + 1 \right)^{-1} & $\qquad \qquad$ & \text{FD} \\ 
	\exp[-q \, T_I / (A \, T)] & $\qquad \qquad$ & \text{MB} \end{array}\right. \ .
\ee
If we apply the definition in Eq.~\eqref{eq:TX}, we find that the dark radiation temperature is equal to the one of the thermal bath, $T_X = T$. Thus our general definition reproduces correctly the case when thermalization is achieved. 

The next two sub-sections contain numerical solutions of the system in Eq.~\eqref{eq:BoltzSystemSingle} for the two production mechanisms considered in this work: two-body decays and binary scatterings. First, we neglect quantum statistical effects for all the particles involved in the production and treat them all with MB statistics. This simplification allows us to rely on analytical expressions for the collision terms, and we provide predictions for $\dneff$ obtained from a precise and quasi-analytical phase space analysis. In the following sub-section, we include quantum statistical distributions for all the degrees of freedom participating in the production processes. We consider both bosonic and fermionic dark radiation, and we compare the outcome of this rigorous quantum statistical analysis with approximated results where we neglect either the statistics of all the particles or just the statistics of the bath particles.

\subsection{Classical Statistics}

We consider the decay $\bath_1 \to \bath_2 + \dr$ with all particles obeying the MB statistics, and we focus on the simple case $ g_{1} = g_{2} = g_\dr = 1$. The mass of the final state particle $\bath_2$ is assumed to be negligible, and the only mass scale in the game is $m_1$. Another key quantity is the decay width $\Gamma_1$ for this production process, and we trade it for the dimensionless ratio $\Gamma_1 / m_1$. 

\begin{figure}
	\centering
	\includegraphics[width=.49\textwidth]{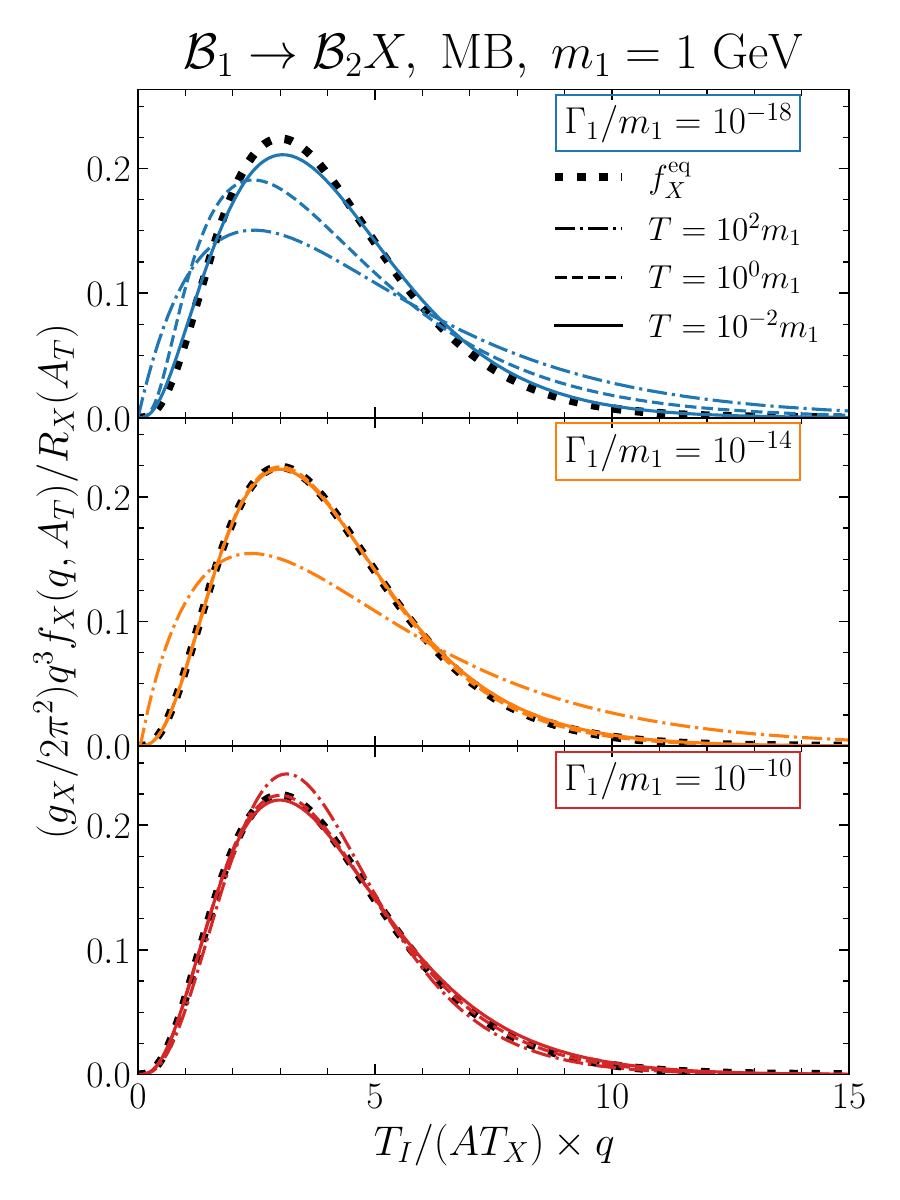}
	\includegraphics[width=.49\textwidth]{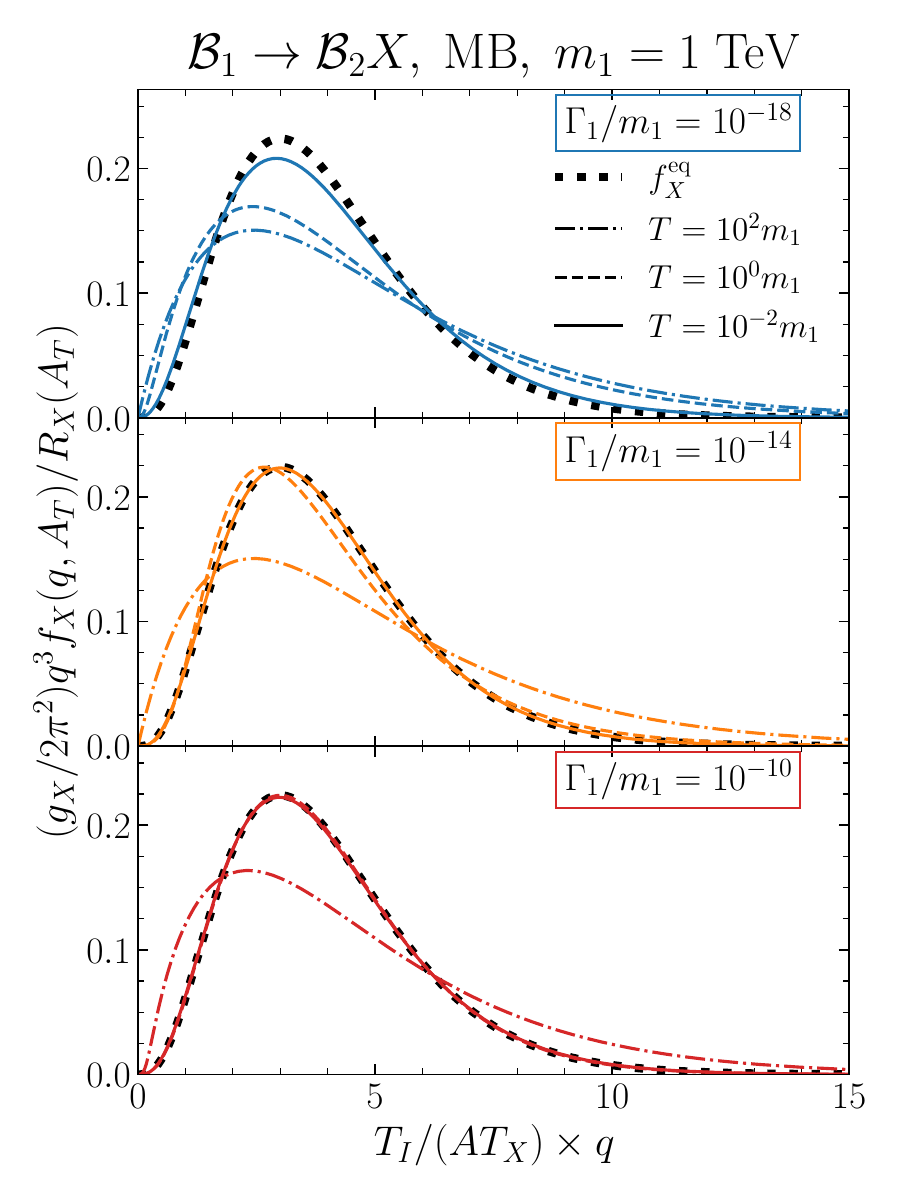}
	\caption{PSDs as a function of the comoving momentum for production via two-body decays for $ m_1 = 1 \, {\rm GeV}$ (left) and $ m_1 = 1 \, {\rm TeV}$ (right). Different rows are for $ \Gamma_1/m_1 = $ $ 10^{-18} $ (top, blue), $ 10^{-14} $ (middle, orange) and $ 10^{-10} $ (bottom, red). We show the PSD at different scale factor values $A_T$ when the bath temperature was $ T={(100, 1,0.01) \times m_1} $ (dot-dashed, dashed, solid lines). We compare them with a thermal distribution with temperature $T_X$ defined in Eq.~\eqref{eq:TX} (dotted black lines). The choices for the normalizations of both quantities on the Cartesian axes are explained in the text.}
	 \label{fig:f_sol_MB}
\end{figure}

Fig.~\ref{fig:f_sol_MB} shows the numerical solutions for the PSD. The two different panels correspond to different values for the mass of the decaying particle, $ m_1 = 1 \, {\rm GeV}$ (left) and $ m_1 = 1 \, {\rm TeV}$ (right). These two values exemplify a generic case since they allow us to study the production of dark radiation around the QCD crossover where the number of relativistic degrees of freedom changes abruptly, and at very high temperatures where this quantity is constant, respectively. For each panel, the three different rows have different values for the dimensionless ratio $\Gamma_1 / m_1$ as given in the legenda. These three choices are meant to represent three distinct physical scenarios: freeze-in of dark radiation (blue lines), thermal equilibrium barely reached (orange lines), and dark radiation strongly coupled to the thermal bath (red lines). 

The lines in our plots always show the PSD $f_X(q, A_T)$ as a function of the comoving momentum $q$ and at fixed values of the scale factor $A_T$ that is understood as the scale factor value when the bath temperature was $T$. The PSDs appearing on the vertical axis are normalized in a way that we find convenient. The multiplicative factor $(g_\dr/2\pi^2) q^3$ is what is needed to identify the integrand whose integral leads to the dimensionless comoving energy density $R_\dr$ (see Eq.~\eqref{eq:RX}). The quantity $R_\dr$ appearing in the denominator is helpful if one wants to compare the PSD at different moments; initially, before the production processes do their job, the value of $f_\dr$ is rather small, and it grows later on. Thus we are comparing the shape of the PSD at different times and not the overall normalization. The choice of the variable on the horizontal axis, where the dark radiation temperature $T_\dr(A)$ given in Eq.~\eqref{eq:TX}, is also convenient because we want to investigate when we achieve thermalization. We do it by comparing our results with the reference thermal distribution $f_X^{\rm eq}(q, A) = \exp[- T_I q /(A \, T_\dr(A))]$, and this definition corresponds to $f_X^{\rm eq} = \exp[- k / T_X]$ with $k$ the physical momentum related to the comoving momentum by Eq.~\eqref{eq:comoving_momentum}. Within these conventions, the equilibrium distribution is fixed in time and can be effectively used as a reference to quantify deviation from thermalization. 

Each plot in Fig.~\ref{fig:f_sol_MB} shows the reference thermal distribution as a dotted black line, and different `snapshots'' of the numerical solution for the PSD at moments of the expansion history when the bath temperature was $T = \{100,1,0.01\} \times m_1$. These three values are chosen to represent moments when the production of dark radiation has not started yet, is most effective, and has terminated, respectively. The early-time PSD (dot-dashed line), at $T = 100 \, m_1$, appears to be much wider than the equilibrium one but with a peak at lower comoving momenta. The PSD when $ T=m_1 $ (dashed lines) is usually the closest possible that it can get to the equilibrium one throughout the cosmic evolution as the interaction rate is maximal and, in the strongly coupled cases, it matches exactly the equilibrium expression. Finally, the late-time PSD (solid line) at low temperatures, $T = 0.01 \, m_1$, is the free-streaming one: $\dr$ particle production has ceased at this point and thus afterward the (normalized) PSD is frozen in time. Already from this analysis, we can see noticeable differences between the final PSD and the equilibrium one, flagging the presence of \textit{spectral distortions}. These distortions are due to the fact that the interaction turns off at different times (temperatures), for the different momentum bins, as already pointed out in Fig.~\ref{fig:C_over_H}. We discuss in further detail this effect and its relation with the kinetic equilibrium assumption in App.~\ref{app:thermalization}.

\begin{figure}
	\centering
	\includegraphics[width=.49\textwidth]{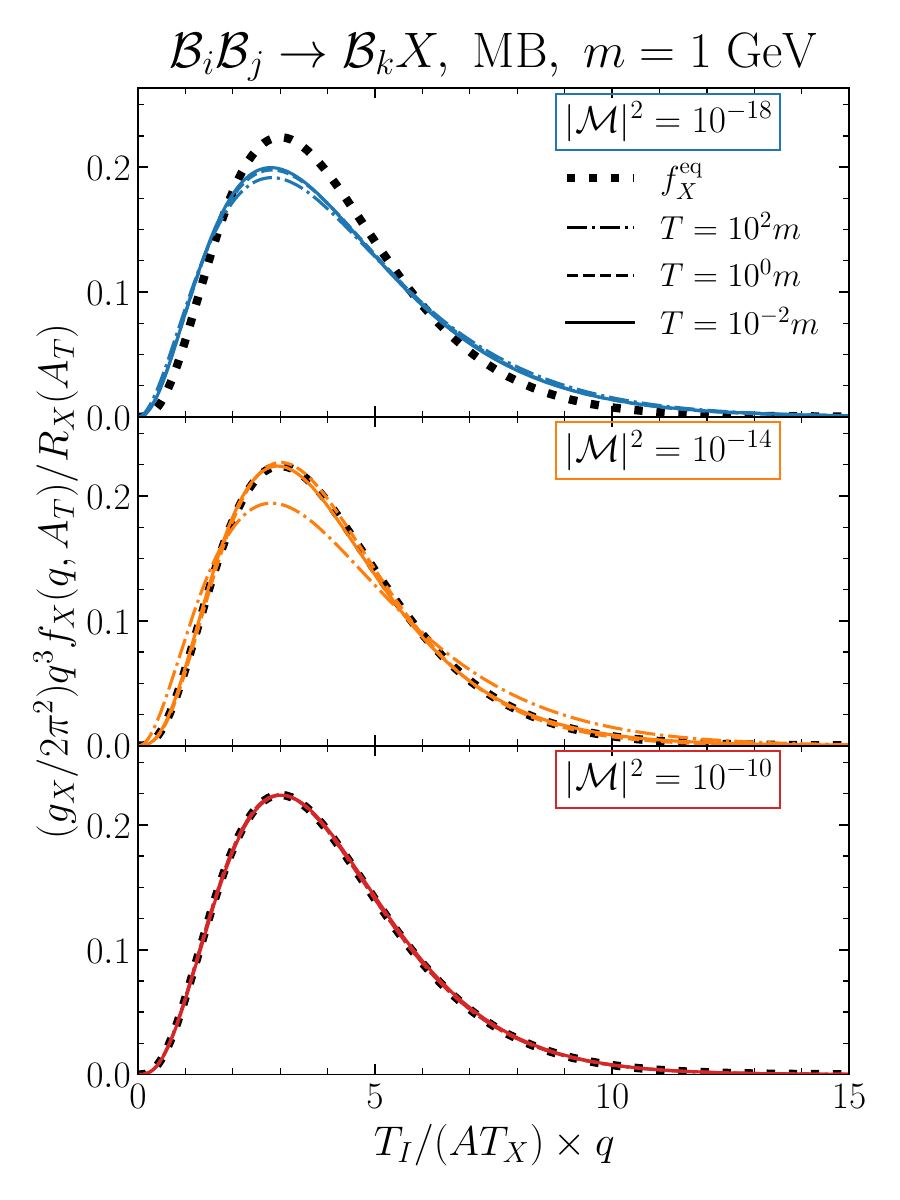}
	\includegraphics[width=.49\textwidth]{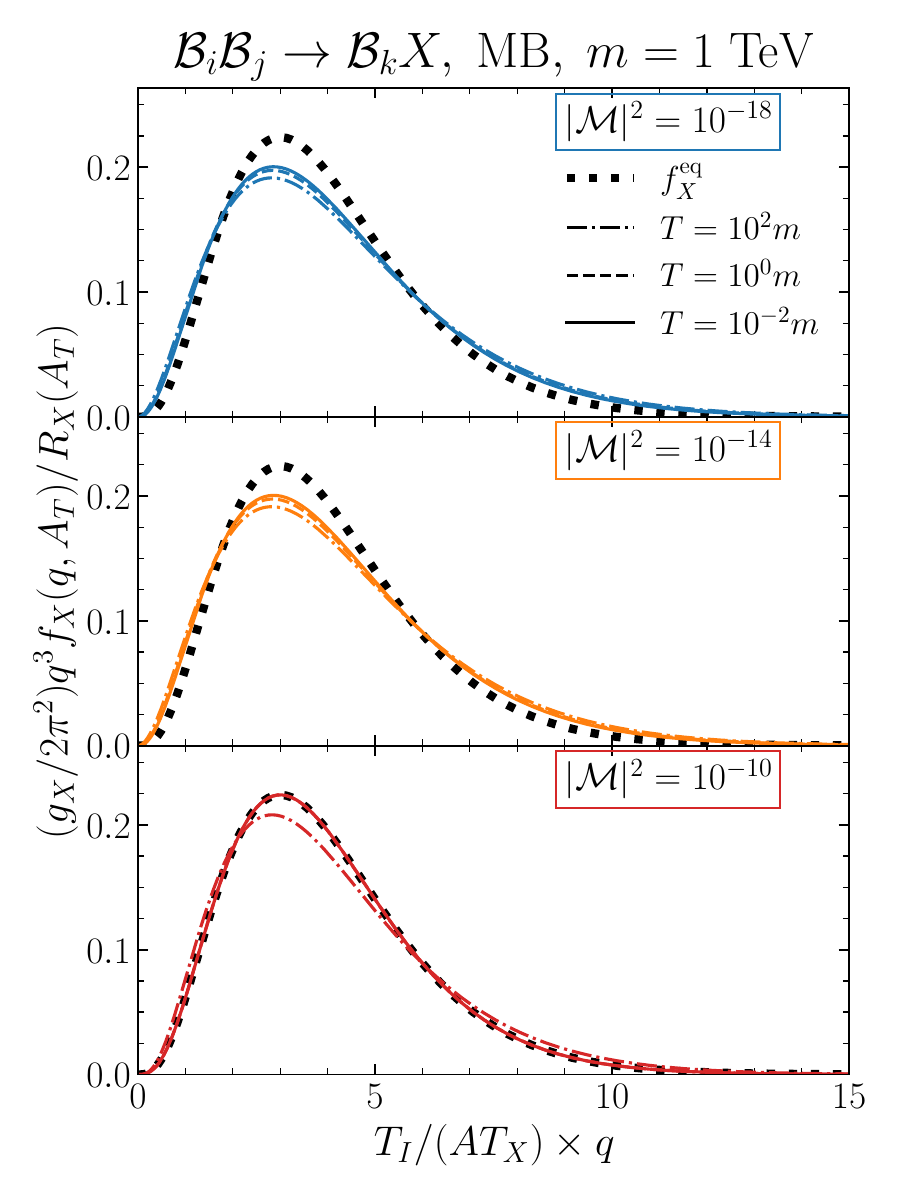}
	\caption{PSDs as a function of the comoving momentum for production via scatterings for $ m_1 = 1 \, {\rm GeV}$ (left) and $ m_1 = 1 \, {\rm TeV}$ (right). Notation as in Fig.~\ref{fig:f_sol_MB} but this time $ |{\cal M}|^2 = 10^{-18} $ (top, blue), $ 10^{-14} $ (middle, orange) and $ 10^{-10} $ (bottom, red).}
	\label{fig:f_sol_MB_sca}
\end{figure}

The discussion for production via the scattering $\bath_1 + \bath_2 \to \bath_3 +X $ (and permutations) is very similar to the decay case so we outline here the main peculiarities. As before, we employ the MB statistics for all particles and we take $ g_{1} = g_{2} = g_3 =  g_X = 1$. We take the same mass spectrum chosen in the previous section, with $\bath_1$ and $\bath_2$ having the same mass $m$ and $\bath_3$ massless. In Fig.~\ref{fig:f_sol_MB_sca}, we show the normalized PSD obtained numerically for $m = 1 \, {\rm GeV}$ (left) and $ m = 1 \, {\rm TeV}$ (right). As already done in the previous section, we assume that the squared matrix element $ |{\cal M}|^2$ is constant and we present results for the three different values $10^{-18}$ (upper row, blue lines), $10^{-14}$ (middle row, orange lines), and $10^{-10}$ (lower row, red lines). These three choices are again representative of three different scenarios where thermalization is never, barely, and completely achieved, respectively. The different snapshots are chosen at the same reference temperatures $T = \{100,1,0.01\} \times m$, and the thermal distribution is again presented as a black dotted line. This figure shows in which regimes and at what temperatures thermal equilibrium is achieved. The behavior is qualitatively very similar to the decays, but overall the distributions are much closer to the reference thermal one, signaling how in the scattering case, where three processes are involved in producing dark radiation, spectral distortions are more difficult to produce. However, we emphasize how these results are derived under the assumption that the squared matrix element is constant and it is not obvious that it still holds if the underlying dynamics leads to transition amplitudes with a non-trivial momentum dependence. A well-known example is the QCD axion because its Nambu-Goldstone nature implies coupling proportional to the derivative of the axion field itself, and the various momentum bins feel interaction strengths that are rather different. 

\begin{figure}
	\centering
	\includegraphics[width=.49\textwidth]{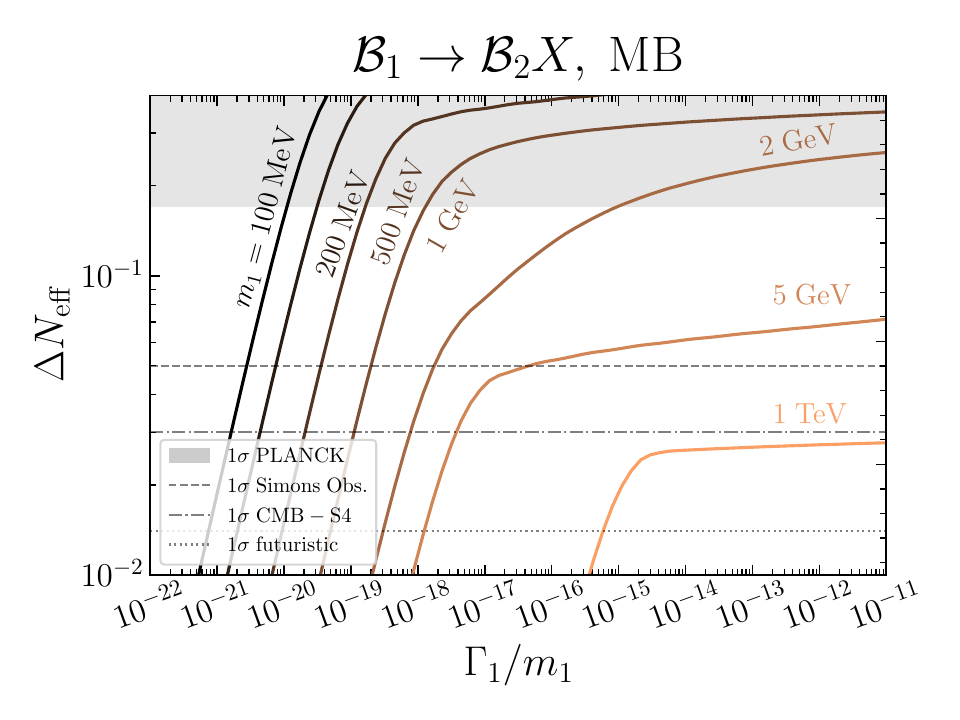}
	\includegraphics[width=.49\textwidth]{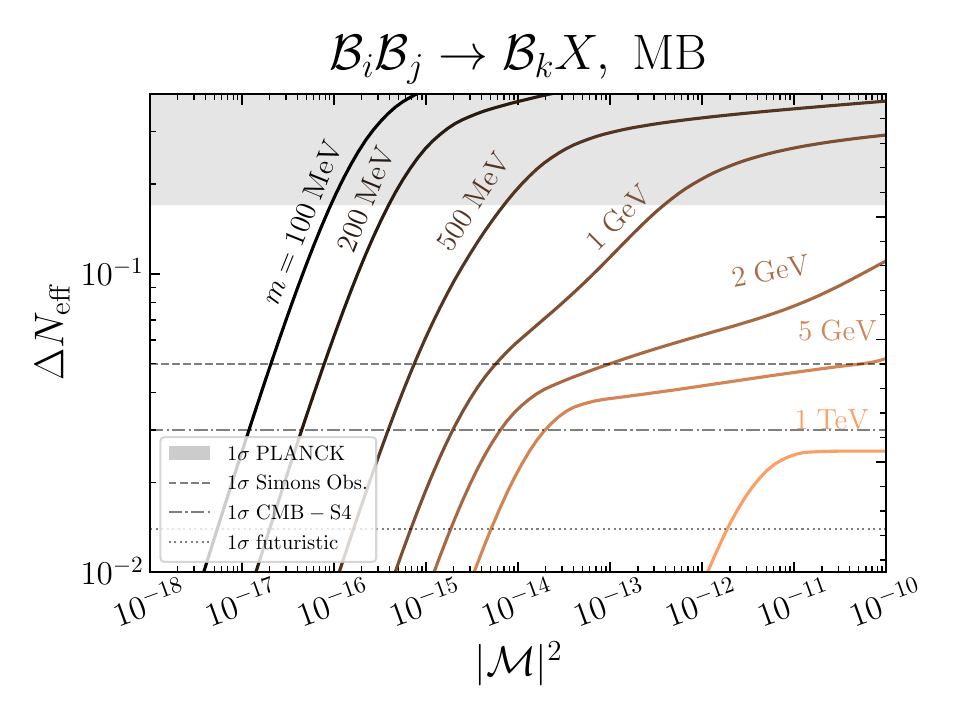}
	\caption{Predicted values of $ \dneff $ for $\dr$ production via decays (left) and scatterings (right) obtained with the Maxwell-Boltzmann (MB) statistics for all particles involved in the production process. For the decay case, we show how $ \dneff $ depends on the dimensionless ratio $ \Gamma_1/m_1 $ for different choices of the decaying particle mass. For the scattering case we show the dependence is on the constant squared matrix element $ |{\cal M}|^2 $ for the collision.}
	\label{fig:dneff_f}
\end{figure}

Regardless of the production mechanism, once the PSD is obtained we can integrate it over the momenta to find the energy density and the resulting $ \dneff $. In the left panel of Fig.~\ref{fig:dneff_f}, we show the result of $ \dneff $ as a function of the parameter $ \Gamma_1/m_1 $ for different choices of the decaying particle mass $ m_1 $. Notice how the rapid changes of relativistic degrees of freedom in the plasma is responsible for a huge variation in the value of $ \dneff $ from $m_1 = 100 \, {\rm MeV}$ to $m_1 = 5 \, {\rm GeV}$. Likewise, the right panel of Fig.~\ref{fig:dneff_f} shows the prediction for $\dneff$ as a function of the constant scattering amplitude for different values of the mass scale $m$ (chosen to be the mass of $\bath_1$ and $\bath_2$ whereas $\bath_3$ is assumed to be massless).

\subsection{Quantum statistics}
\label{sec:res2}

\begin{figure}
	\centering
	\includegraphics[width=.49\textwidth]{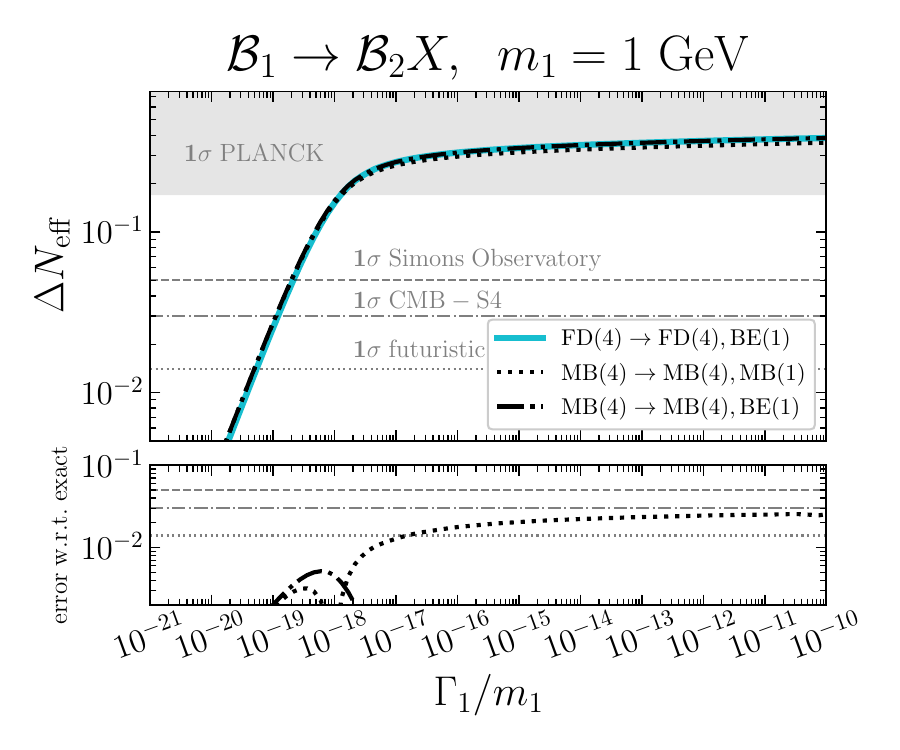}
		\includegraphics[width=.49\textwidth]{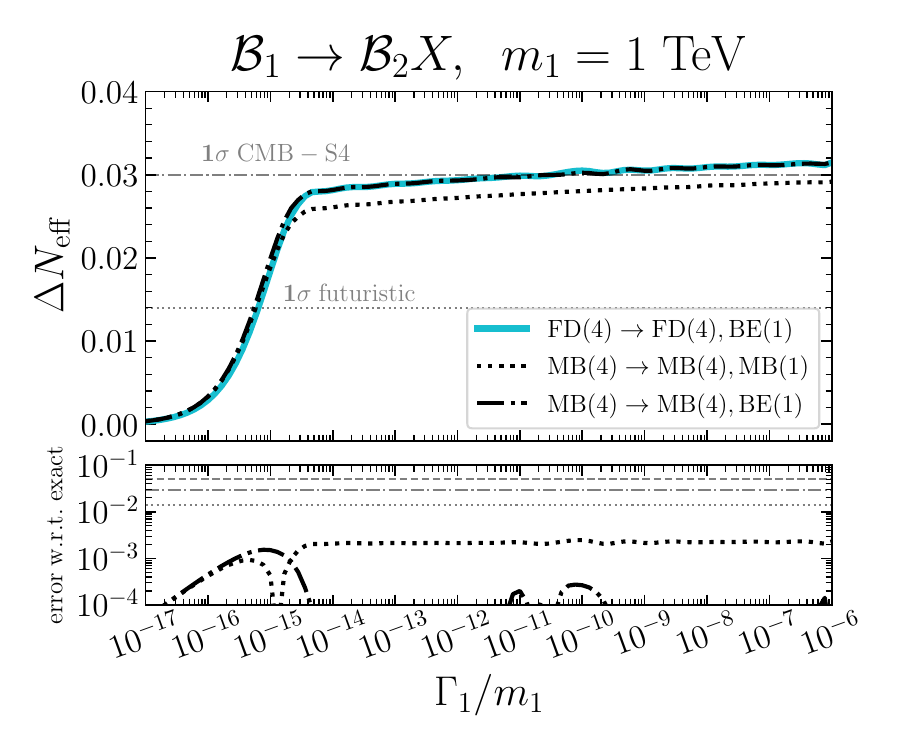}	\includegraphics[width=.49\textwidth]{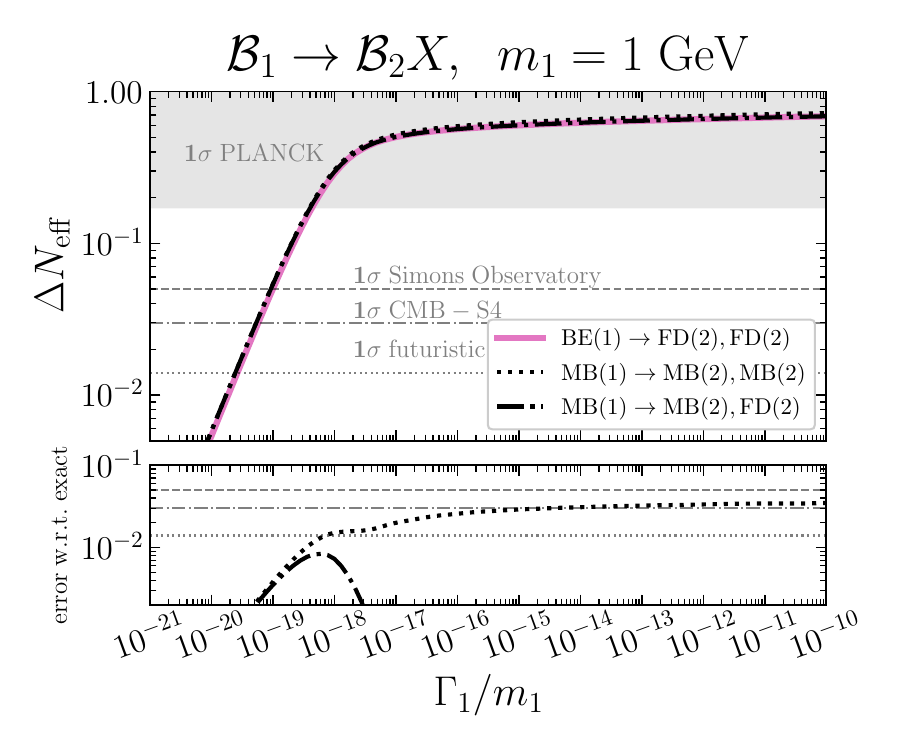}
	\includegraphics[width=.49\textwidth]{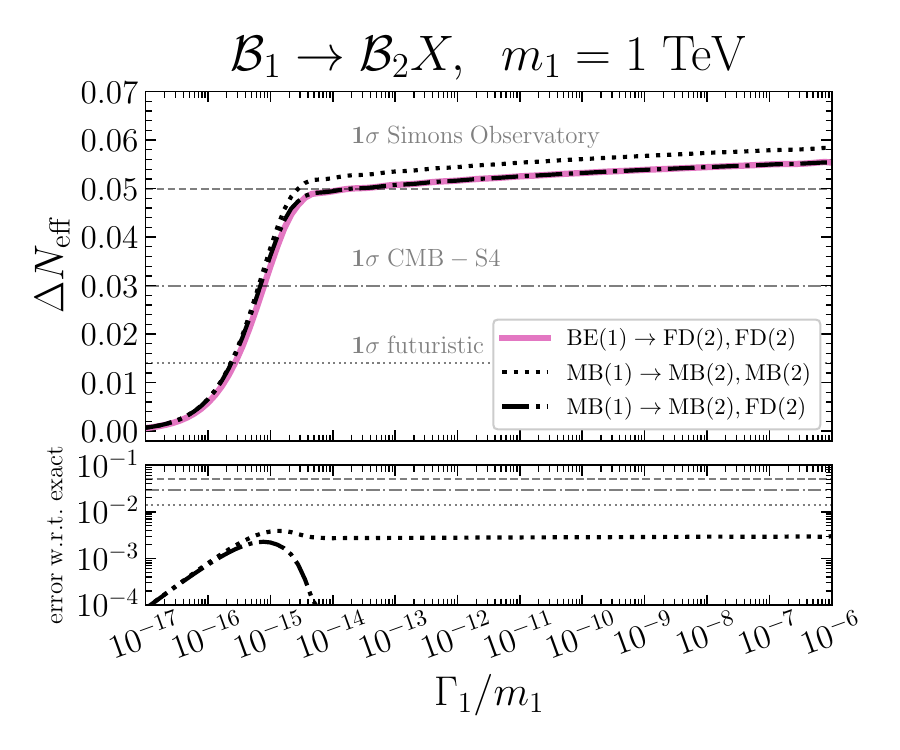}
	\caption{Predicted $\dneff$ for dark radiation produced via two-body decays as a function of the coupling strength $\Gamma_1/m_1$ for $ m_1 = 1 \, {\rm GeV}$ (left) and $ m_1 = 1 \, {\rm TeV}$ (right). We consider bosonic (upper panels) and fermionic (lower panels) dark radiation. Solid lines show the exact results where we employ quantum statistical distributions for all particles, and we also report results from approximate analysis: all quantum statistical effects neglected (thick dotted black lines), and only $\dr$'s quantum statistics accounted for (thick dot-dashed black lines). In the lower part of each panel we quantify the errors made by the approximate results.}
	 \label{fig:quantum_stats_decays}
\end{figure}

We relax the assumption of classical statistics and we treat all particles involved in the production process, bath degrees of freedom $\bath_i$ and dark radiation $\dr$, quantum mechanically. The first thing to decide is the spin of the dark radiation particle $\dr$, and we consider two cases: spin-0 (bosonic) dark radiation produced via the decay of a Dirac fermion; spin-1/2 (fermionic) dark radiation described by a Weyl field and produced via the decay of a spin-0 bath particle. The spin of the final state bath particle is determined accordingly. We label the first case FD(4)$\to$FD(4)+BE(1), where the number in brackets denotes the number of internal degrees of freedom (i.e., $g_1=g_2=4$, $g_X=1$), and the second case BE(1)$\to$FD(2)+FD(2) (i.e., $g_1=1,\ g_2=2$, $g_X=2$). As done before, we consider the decaying particle to be the only massive particle. The collision term for this case is provided by Eq.~\eqref{eq:C12final}, and the explicit expressions for the function $D(k,T)$ for the different statistics are shown in Tab.~\ref{tab:stats}. Fig.~\ref{fig:quantum_stats_decays} shows the prediction for $\dneff$ for the two different mass values $m_1 = 1 \, {\rm GeV}$ (left) and $m_1 = 1 \, {\rm TeV}$ (right). We compare the full numerical solutions obtained by employing the proper quantum statistical distributions (solid lines) with the approximate analysis done in the previous section where all particles are treated with MB statistics (dotted lines), and with a different approximate analysis where we adopt the MB distribution for all bath particles but we treat $\dr$ particles with the correct quantum statistics (solid dotted lines). For each plot, we show also the error of each approximate analysis with respect to the exact solution. We notice how treating all particles with MB statistics works up to a 10\% error over the whole range of coupling strengths $g_1 \Gamma_1/m_1$ with the absolute error becoming detectable by futuristic $\dneff$ probes $\sim 0.01$. Instead, the analysis where we treat only the dark radiation quantum mechanically suffers from the same 10\% error at small couplings, but agrees very well with the exact result at larger values ($g_1 \Gamma_1/m_1>10^{-18}$ for $m_1 = 1 \, {\rm GeV}$, and $g_1 \Gamma_1/m_1>10^{-15}$ for $m_1 = 1 \, {\rm TeV}$). Thus neglecting the bath particle statistics in the case of decays does not impact the calculation of $\dneff$ as long as the coupling is sufficiently strong. Nevertheless, the error made using this approximation for smaller couplings is not detectable by future $\dneff$ probes. The underlying physics responsible for these results is illustrated in Fig.~\ref{fig:C_over_H} where we can observe how different statistics change the temperature at which the interaction starts to be relevant (i.e., the transition from ${\cal C}/H< 1$ to ${\cal C}/H> 1$), but have no impact on the decoupling epoch (i.e., the transition from ${\cal C}/H> 1$ to ${\cal C}/H< 1$). Thus decoupling for a given momentum $k$ happens at the same temperature regardless of the statistics (see the discussion in App.~\ref{app:thermalization} for further details, and in particular Fig.~\ref{fig:Tdec}).

\begin{figure}
	\centering
	\includegraphics[width=.49\textwidth]{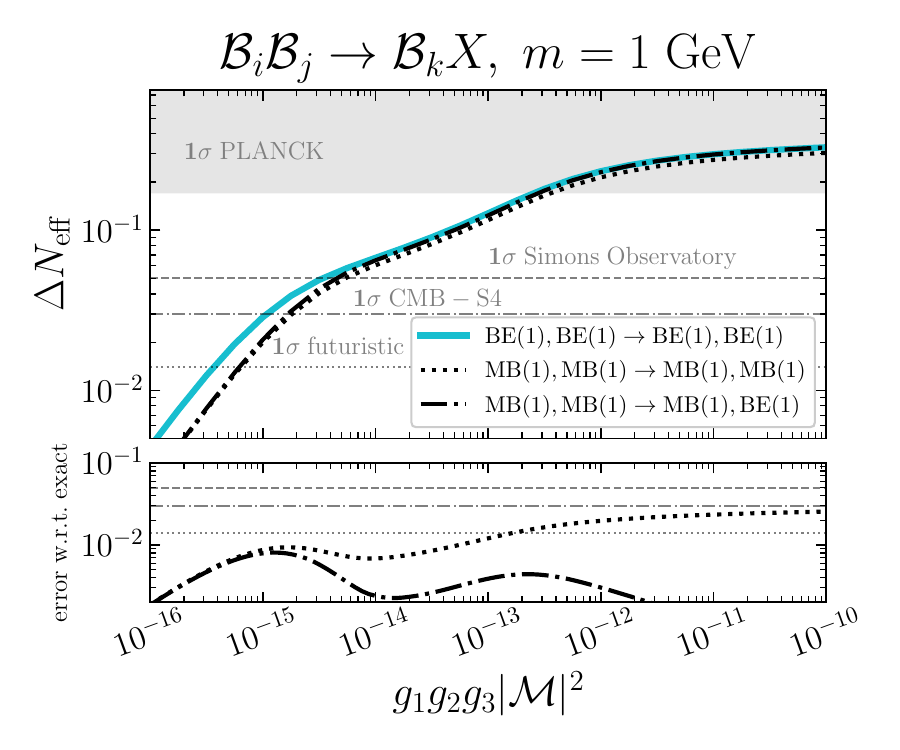}
	\includegraphics[width=.49\textwidth]{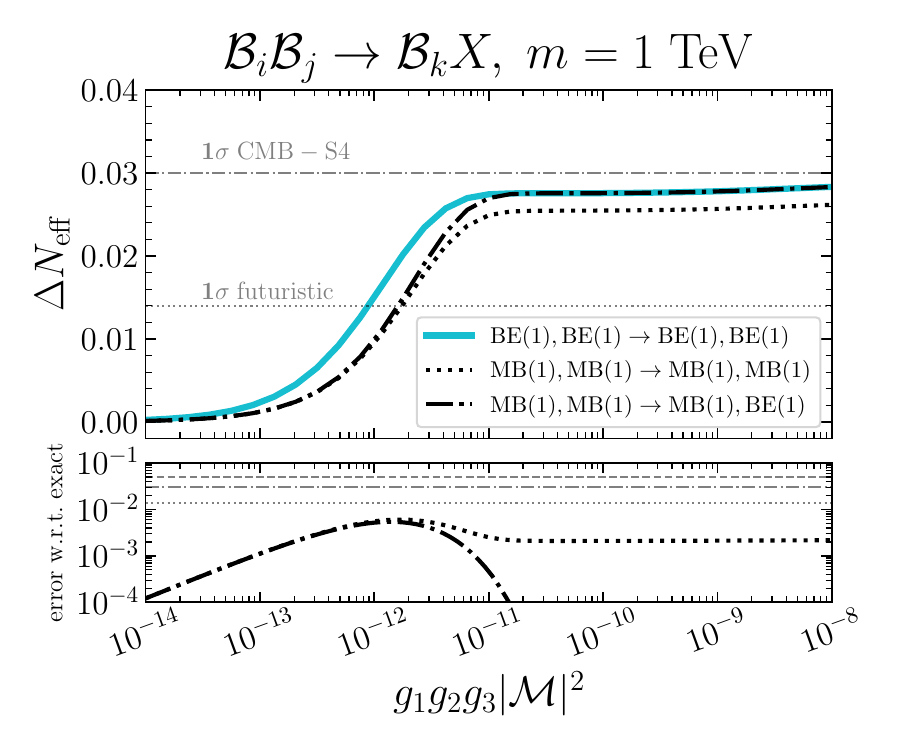}
	\includegraphics[width=.49\textwidth]{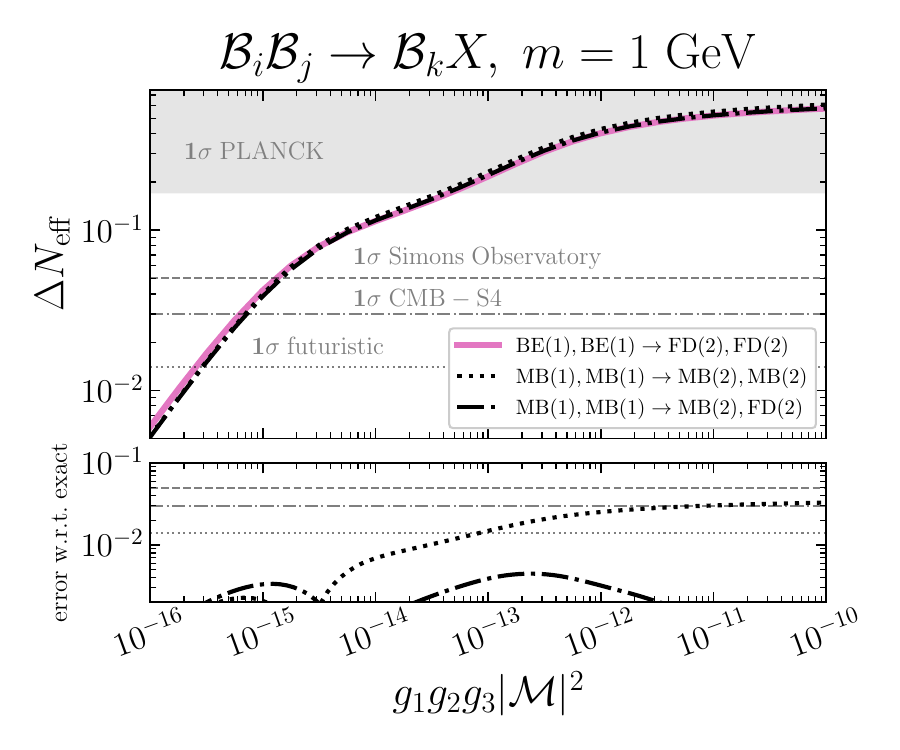}
	\includegraphics[width=.49\textwidth]{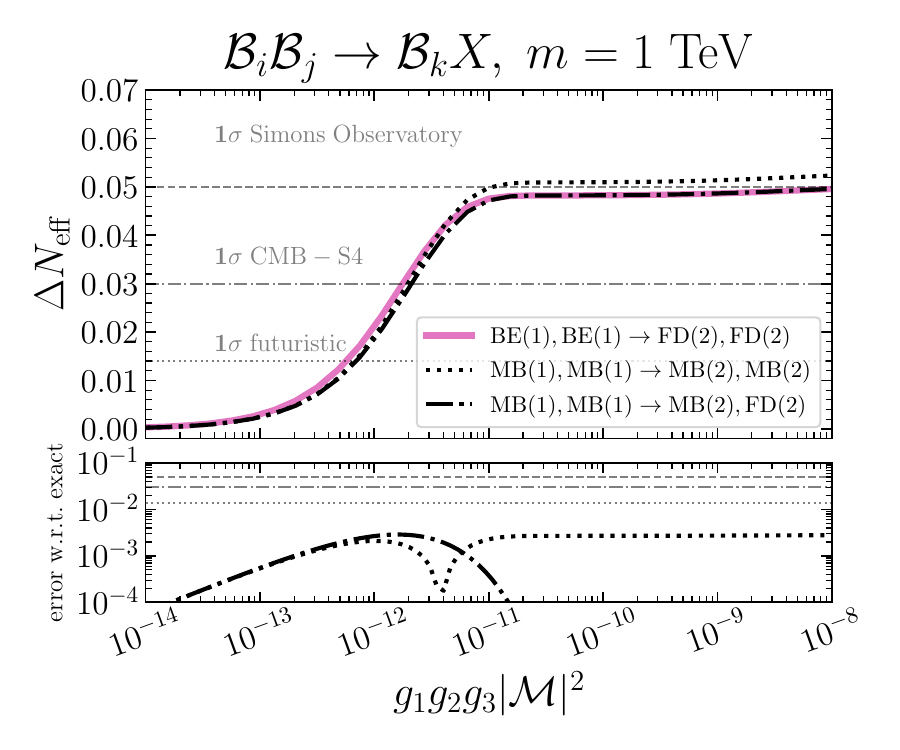}
	\caption{Predicted $\dneff$ for production via scatterings as a function of the interaction strength $g_1g_2g_3 |{\cal M}|^2$ for $ m = 1 \, {\rm GeV}$ (left) and $ m = 1 \, {\rm TeV}$ (right). Notation as in Fig.~\ref{fig:quantum_stats_decays}.}
	\label{fig:quantum_stats_scatterings}
\end{figure}

We consider now bosonic and fermionic dark radiation produced via collisions. The former is assumed to be a spin-less particle produced via collisions of other spin-0 degrees of freedom, and the latter a is Weyl fermion produced via collisions of spin-0 bosons and a Weyl fermion. For both cases, only the bosons ${\cal B}_1$ and ${\cal B}_2$ are massive and have equal mass $m$. Fig.~\ref{fig:quantum_stats_scatterings} compares the results obtained taking into account the quantum statistics of all the particles involved (solid colored lines) with the approximation of treating all particles with the MB statistics, and the improved method with the correct statistics just for the $\dr$ particle. The error at small couplings ($g_1g_2g_3 |{\cal M}|^2<10^{-15}$ for $m = 1 \, {\rm GeV}$ and $<10^{-12}$ for $m = 1 \, {\rm TeV}$) is at least 10\% up to even 100\% for all cases. For larger couplings, the improved method is overall better than treating all particles with the MB statistics. However, differently from the decay case, the scattering rate is the sum of three different processes and the total scattering rate is much more sensitive to the statistics with respect to decay. This can be understood referring to Fig.~\ref{fig:C_over_H_sca}, and in more detail in Fig.~\ref{fig:Tdec} of App.~\ref{app:thermalization}: considering different statistics and same momentum, we see that the decoupling temperature (i.e., ${\cal C}_{2 \rightarrow 2}/H$ drops below one after being larger than one) depends on the statistics. Thus assuming MB statistics in the collision term ${\cal C}_{2 \rightarrow 2}$ is not an approximation as good as in the decay case even at large couplings.
 
\section{Comparison with Approximate Methods}
\label{sec:comparison}

We take one step back and consider approximate methods to estimate $\dneff$ that rely upon ordinary differential equations. In this section, we always adopt MB statistics for both bath particles and dark radiation. The main goal here is to compare the errors that we make when we adopt these approximate methods, and choosing statistical distributions that offer analytical collision rates helps our case. Furthermore, we have seen how the full quantum interaction rates are well approximated by the MB statistics. A careful review of the approximate methods is provided in App.~\ref{app:approx}. Before comparing with the full results obtained via a momentum space analysis, we summarize them briefly here.
\begin{itemize}
\item \textbf{Instantaneous Decoupling.} The underlying assumption is that $\dr$ particles thermalize and eventually decouple. The departure from thermal equilibrium is a sudden phenomenon at a specific temperature $\TD$ that is determined via a comparison between the interaction and the expansion rates. At temperatures lower than $\TD$, dark radiation particles free stream, and one can evaluate their energy density at recombination by employing entropy conservation. This approach is outlined in App.~\ref{subsec:ID}. Once we determine $\TD$, the resulting contribution to $\dneff$ results in $\dneff \simeq 0.027 \, g_\dr \xi_{\rho_\dr} \, \left( 106.75 / g_{*s}^\bath(\TD) \right)^{4/3}$, where the dimensionless coefficient $\xi_{\rho_\dr}$ accounts for the statistical properties of $\dr$ and it is given explicitly in Eq.~\eqref{eq:rhodrapp} for the different cases. Within this framework, one can account for quantum statistical effects for both bath and dark radiation particles since the interaction rates are evaluated for equilibrium distribution. 
\item \textbf{Tracking the number density.} This method allows us to deal with the details of the decoupling epoch, and to account for situations where dark radiation never achieves thermalization. The mathematical tool is an ordinary differential equation for the $\dr$'s number density $n_\dr$ that is derived from the integro-differential Boltzmann equation in momentum space. We review carefully this derivation in App.~\ref{subsec:BEn}. Along the way, we need to assume that assume that $\phi$ particles are in kinetic equilibrium and described by MB statistics. No assumption is necessary about the statistics of bath particles. The Boltzmann equation takes the final form given in Eq.~\eqref{eq:BEforn}. Furthermore, we need to convert the solution for $n_\dr$ into a correspondent energy density $\rho_\dr$ in order to evaluate $\dneff$. A further assumption that is required for this last step is to assume a thermal shape in chemical equilibrium for the dark radiation's PSD. 
\item \textbf{Tracking the energy density.} The assumptions of kinetic equilibrium and MB statistics for the dark radiation are still unavoidable if one wants to deal with an ordinary differential equation. However, tracking $\rho_\dr$ would have the indisputable benefit that it does not require any conversion to evaluate $\dneff$. In particular, we do not need to assume a thermal shape in chemical equilibrium for the dark radiation PSD anymore. We review this method in App.~\ref{subsec:BErho} where we also explain how to account consistently for the presence of $\dr$'s in the energy and entropy densities as done in the main text. Methods analogous to this one were used by Refs.~\cite{Escudero:2018mvt,EscuderoAbenza:2020cmq,Luo:2020fdt,Adshead:2022ovo,Esseili:2023ldf}.

\end{itemize}

Among these three methods, the last one where we track the energy density has its obvious advantages but it also comes with its own limitations. In particular, it is not possible to go beyond $\dr$'s kinetic equilibrium and to include quantum statistical effects for $\dr$ without dealing with the problem in momentum space. 

Before presenting the explicit comparison, we find it useful to introduce another approximate procedure that still relies upon working in momentum space. We remind the reader that the general system to solve for an accurate momentum space analysis is the one given in Eq.~\eqref{eq:BoltzSystemGeneral}. This fourth approximate method provides a simplified version of it. 
\begin{itemize}
\item \textbf{Phase space without feedback on the bath.} The approximation for his case counts on the fact that the contribution of the dark radiation to the total energy density is negligible, and more precisely of a percent order with respect to the thermal bath for even the largest couplings we consider here. Thus it is reasonable to consider a standard cosmological history with the Hubble parameter given by $H = (\pi \sqrt{g^\bath_{\star\rho}(T)} \, T^2)/(3 \sqrt{10} M_{\rm Pl})$. Consequently and consistently with this approximation the evolution of the bath energy density is determined by the above relation. Thus the system collapses to a single equation for the dark radiation PSD, namely Eq.~\eqref{eq:Cfsingle1}, and we can employ entropy conservation to use the temperature as the evolution variable
\be 
\frac{df_X(q,T)}{d\log T} = -\frac{{\cal C}(q,T)}{H} \left( 1 + \frac{1}{3}\frac{d\log\gsts(T)}{d\log T} \right) \left( 1 - \frac{f_\dr(q, T)}{f_\dr^{\rm eq}(q, T)}  \right)\ .
\ee 
Here, we define the comoving momentum $q=(k/T)  (\gsts(T)/\gsts(m))^{1/3}$ as in Ref.~\cite{DEramo:2020gpr}.
\end{itemize}

\begin{figure}
	\centering
	\includegraphics[width=.49\textwidth]{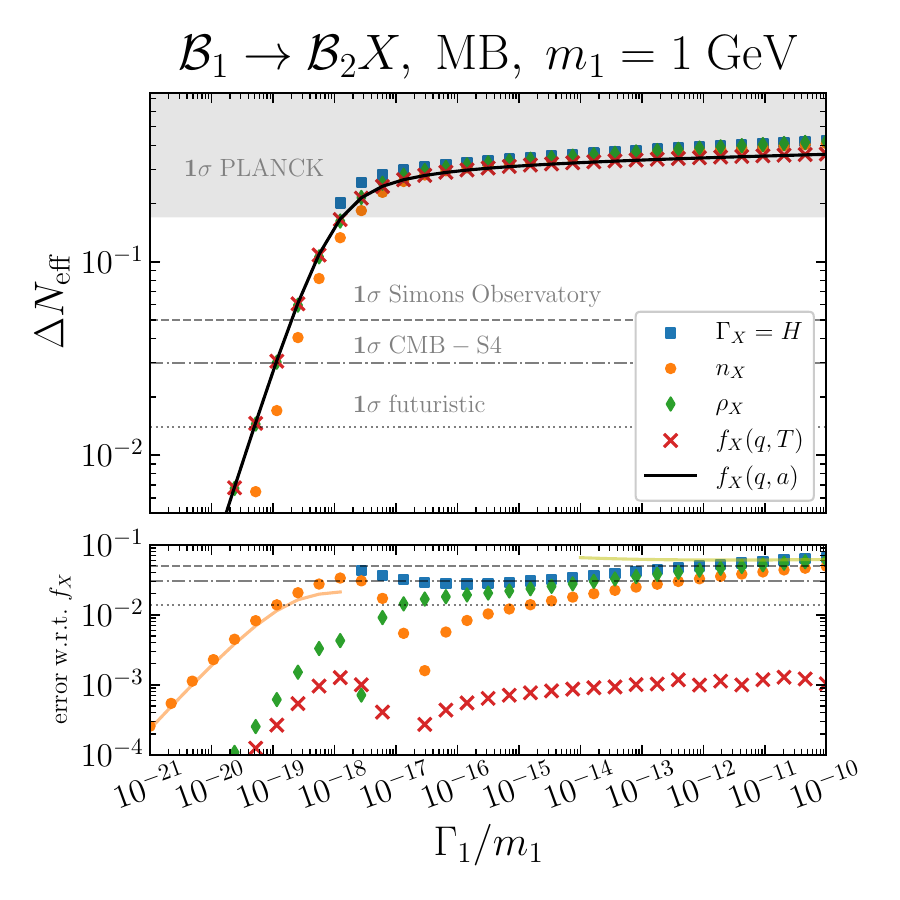}
	\includegraphics[width=.49\textwidth]{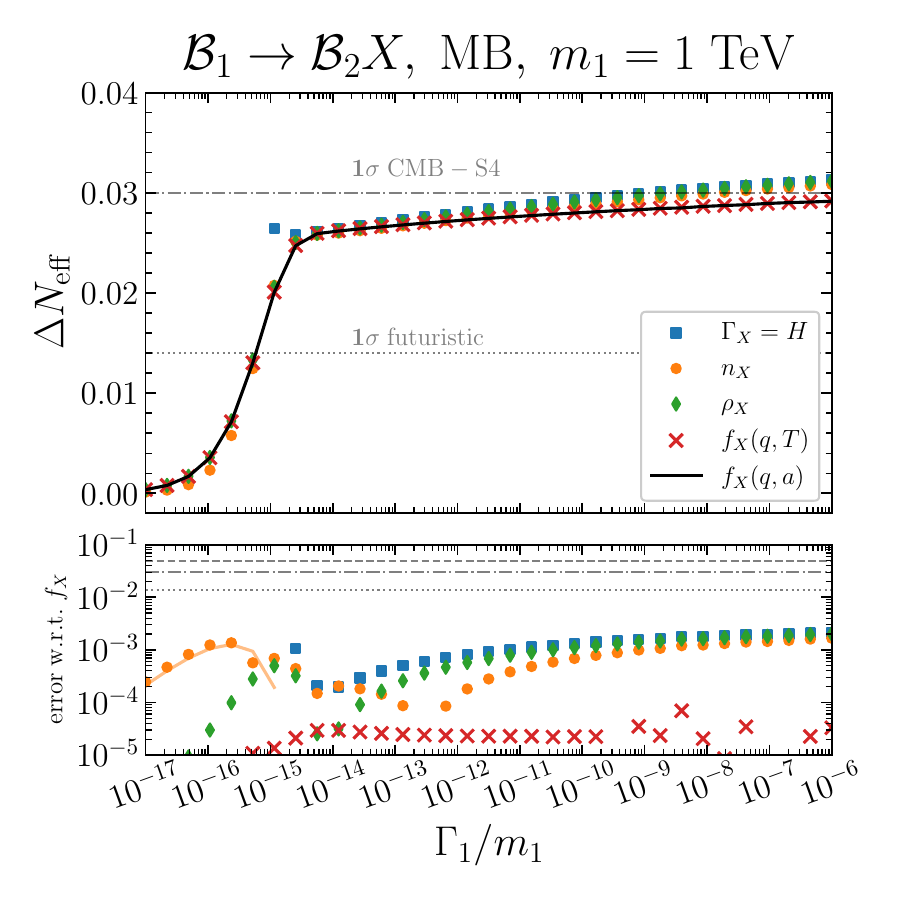}
		\caption{Comparison of $\Delta N_{\rm eff}$ predictions for production via two-body decays. All particles are treated with MB statistics, and we consider the two benchmarks values $ m_1 = 1$ GeV (left) and $ m_1 = 1$ TeV (right) for the decaying particle mass. We vary the value of $ \Gamma_1/m_1 $ over the horizontal axis, and we report the prediction for $\Delta N_{\rm eff}$ on the vertical one. The different horizontal lines identify current bounds as well as the sensitivities of future CMB probes. Exact results from a momentum space analysis are shown as black solid lines. The approximate methods we compare with are: instantaneous decoupling (blue squares), tracking the $\dr$ number density (orange up-sided triangles), and tracking the $\dr$ energy density (green down-sided triangles). We also show the results of a phase space solution without including the dark radiation feedback on the thermal bath (red crosses). The lower panels show the absolute errors on $\Delta N_{\rm eff}$ from the exact solution; the orange lines show the estimate of the error due to the conversion between $ n_\dr $ and $ \rho_\dr $.}
	\label{fig:dneff_MB}
\end{figure}

Fig. \ref{fig:dneff_MB} shows the comparison for the case of the two-body decay for the two usual benchmarks $ m_1 = 1$ GeV (left) and $ m_1 = 1$ TeV (right). The limitations of the instantaneous decoupling approximation are clear since different comoving momenta decouple at different temperatures (see Fig.~\ref{fig:Tdec} in App.~\ref{app:thermalization}). Moreover, if the coupling is not strong enough to achieve thermalization, the method is not applicable at all. 

Small couplings are troublesome also for the $n_\dr$ method. For the $m_1=1$ GeV case, the error can be sizable with the most striking differences appearing in the low coupling region where dark radiation never thermalizes. In general, we notice that to reproduce the exact prediction for $\dneff$ with the $n_\dr$ method we need higher couplings. It is interesting to investigate further the origin of this result. When we convert the final number density $n_\dr$ to an associate energy density and compute $ \dneff $, we have to assume an equilibrium distribution for the dark radiation. This assumption is clearly not completely justified at small couplings, and this leads to an underestimate of the dark radiation energy density. The details of this conversion are described in App.~\ref{subsec:BEn}, and the final relation that allows us to perform this operation can be found in Eq.~\eqref{eq:from_n_to_rho}. Thus the conversion is done using the temperature of the dark radiation $T^\rho_\dr$ found from the energy density, and this has to be contrasted with the $T_\dr^f$ defined from the PSD and defined in Eq.~\eqref{eq:TX}. Therefore the error between the $n_\dr$ method and the phase space solution at small couplings can be estimated by the ratio between these two temperatures, $ \dneff^{\rm err}(n_\dr,f_\dr) \sim (1-T_\dr^\rho/T_\dr^f) \times \dneff(f_\dr)$ (the difference between these temperatures is elaborated in App.~\ref{app:thermalization} and shown explicitly in Fig.~\ref{fig:f_sol_MB_T}). This is illustrated by the orange lines in the lower plots in Fig.~\ref{fig:dneff_MB}. Instead, the error at larger coupling is due to the fact that we are solving for an integrated quantity of $ f_\dr $, hence implicitly assuming kinetic equilibrium. Integrated methods lead to an underestimate of the decoupling temperature with respect to phase space solution in the case of decays (see Fig.~\ref{fig:Tdec}). Moreover, even if this effect is sub-leading, we are not accounting for the feedback of dark radiation into the SM plasma. We will further explain these two aspects below. 

The $ \dneff $ found via the $ \rho_\dr $ method agrees very well with the result from the PSD computation at low couplings, but it shows again a sizable (larger than the future experimental sensitivity) difference at couplings large enough that thermal equilibrium is reached during the evolution. Specifically, the $ \rho_\dr $ solution overestimates the amount of dark radiation. This again can be explained from the fact that kinetic equilibrium is not attained during the decoupling phase, in which the interaction rate drops below the Hubble parameter. Fig.~\ref{fig:f_sol_MB_kin} of App.~\ref{app:thermalization} shows the factor $ (f_\dr/f_\dr^{\rm eq})/(R_\dr/R_\dr^{\rm eq}) $ as a function of the momentum at different epochs. Well after decoupling, for $T \ll m_1$, this quantity depends on the momentum bin also for large couplings. From a superficial examination of this figure, one could conclude that since this ratio is much larger for small couplings the $ \dneff $ computed from $ \rho_\dr $ and $ f_\dr $ should roughly agree for large coupling and disagree strongly for small couplings. This would lead to the wrong conclusion that for couplings for which thermality is attained at least once in the thermal history of the dark radiation solving for the energy density will give the same $ \dneff $ as solving for the PSD, while the result will differ for smaller couplings. Fig.~\ref{fig:dneff_MB} shows exactly the opposite behavior. The reason for this resides in the different importance of the back-reaction terms (the one proportional to $ f_\dr/f_\dr^{\rm eq}  $ in the equations for the PSD and the one proportional to $ R_\dr/R_\dr^{\rm eq} $ in the equation for $ R_{\dr} $) at different values of the coupling $ \Gamma_1/m_1 $.  Indeed for $ \Gamma_1/m_1 \ll 10^{-16}$ we are in the ``freeze-in''-like case $(R_\dr \ll R_\dr^{\rm eq}  $ , $ f_\dr \ll f_\dr^{\rm eq})  $ for which we can safely neglect the back-reaction term in the Boltzmann equations (since $ \dr $ particles never reach equilibrium) which, respectively become
\be
\frac{dR_\dr}{d\log A} \approx \frac{A^4}{H} \frac{{\epsilon}_{\dr}}{T_I^4} \ , \qquad\qquad\qquad \qquad \frac{df_\dr}{d\log A} \approx \frac{{\cal C}_{1\to2}}{H}\ .
\ee
Instead, for $ \Gamma_1/m_1 > 10^{-16}$ the equations depend on back-reaction terms
\be 
\frac{dR_\dr}{d\log A} =  \frac{A^4}{H} \frac{{\epsilon}_{\dr}}{T_I^4}  \bigg[ 1- \left( \frac{R_\dr}{R_{\dr}^{\rm eq}}\right) \bigg]\ , \qquad \qquad 
\frac{df_\dr}{d\log A} = \frac{{\cal C}_{1\to 2}}{H}\left[1-\left( \frac{f_\dr}{f_{\dr}^{\rm eq}}\right) \right]\ .
\ee
Thus if dark radiation particles thermalize and kinetic equilibrium after decoupling is not attained (as shown by Fig.~\ref{fig:f_sol_MB_kin} at large momenta),  the backreaction in the energy density equation is stronger than the backreaction in each momentum bin and one explains the suppression in the energy density obtained from the integration of the PSD solution with respect to the solution for $ \rho_\dr $. This discussion shows that, while the solution of the Boltzmann equation is accurate as long as $ \dr $ particles are in thermal equilibrium, the relic energy density of dark radiation particles cannot be determined correctly by solving for the energy density because spectral distortions determine a difference in the right-hand sides of the Boltzmann equations, that albeit very small, are exponentially enhanced in the solution of the differential equation. Instead, if dark radiation is produced by a pure freeze-in, i.e. $ R_\dr \ll R_\dr^{\rm eq}  $ at all times, then these spectral distortions have no consequence on the right-hand sides of the Boltzmann equations, thus the solution for $ f_\dr $ and $ R_\dr $ coincide at the percent level. 

The above discussion advocates for a complete solution in phase space to avoid overestimating the amount of dark radiation, in particular in the strongly coupled regime. Inspecting closely Fig.~\ref{fig:dneff_MB}, we notice that the most interesting region is the transition between small-coupling and large coupling, around $ \Gamma_1/m_1 \sim 10^{-18}$ as the error between the approximate methods and the exact is large, experimentally detectable and the value of $ \dneff $ is not excluded by Planck \cite{Planck:2018vyg}, but will be probed by future searches.

\begin{figure}
	\centering
	\includegraphics[width=.49\textwidth]{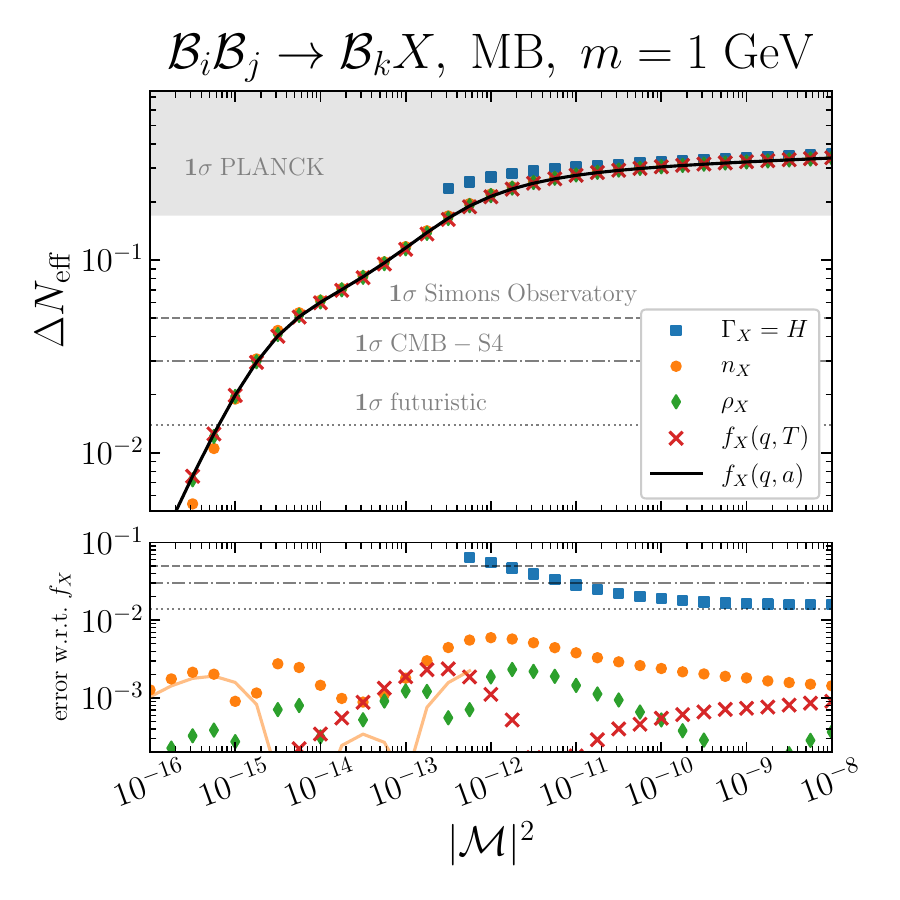}
	\includegraphics[width=.49\textwidth]{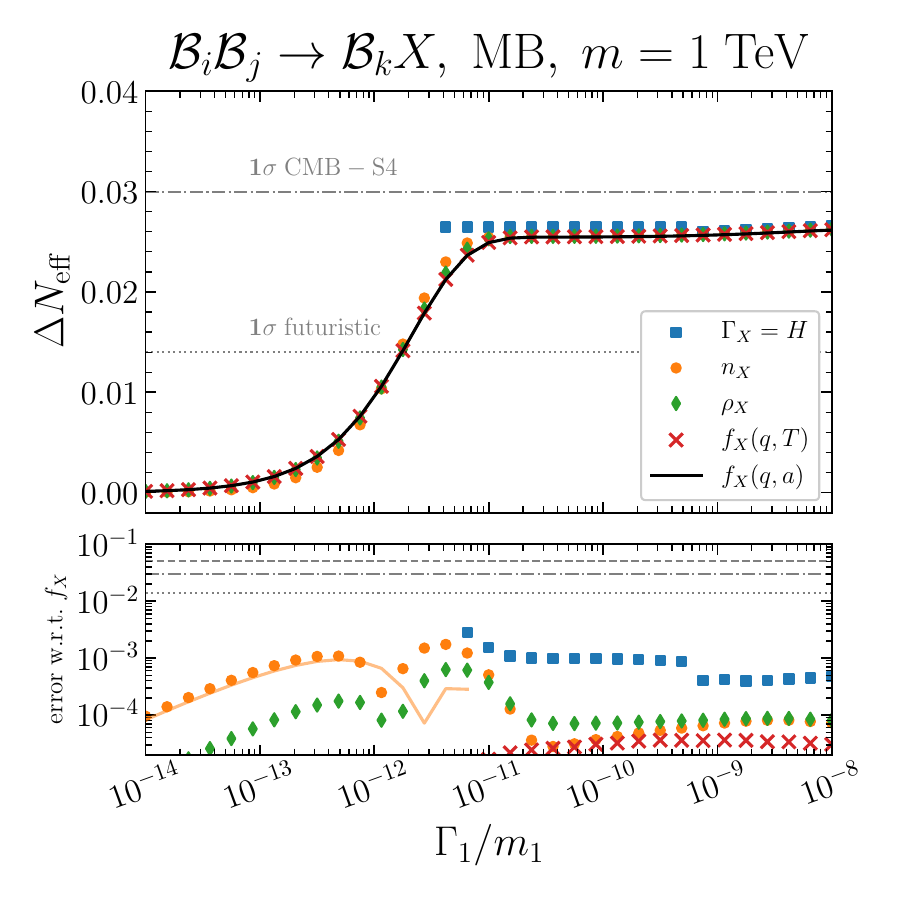}
		\caption{Comparison of $\Delta N_{\rm eff}$ predictions for binary scatterings in the MB approximation as a function of the squared matrix element $ |{\cal M}|^2 $. We pick the same mass spectrum adopted in the paper with a single overall mass scale $m$, and we present results for $ m = 1 \, {\rm GeV}$ (left panel) and $ m = 1 \, {\rm TeV}$ (right panel). The  notation is the same as in Fig.~\ref{fig:dneff_MB}. }
		\label{fig:dneff_MB_sca}
\end{figure}

The case of binary scatterings is overall analogous to the two-body decay case but with a few peculiarities visible in Fig.~\ref{fig:dneff_MB_sca}. Solving for integrated quantities, both $n_\dr$ and $\rho_\dr$ prove to be a much better approximation than in the case of decays. This is because in the scattering case, the kinetic equilibrium is more easily assured, as there are three contributing processes. Because of this, a momentum which decoupled in a process can still be coupled with another. This results in an overall flatter ${\cal C}/H$ when compared to decays, something we knew already from the comparison between Fig.~\ref{fig:C_over_H} and \ref{fig:C_over_H_sca}. Thus when momenta reach equilibrium they remain coupled for a longer time, ensuring kinetic equilibrium on a higher level. Furthermore, different comoving momenta decouple at roughly very similar temperatures and these temperatures are close to the decoupling temperature estimated through $\Gamma_\dr = H $, as clear from Fig.~\ref{fig:Tdec}, signaling that the relic distribution is closer to the equilibrium one. 

It is important to emphasize that for this study we have only considered theories where the scattering matrix element is constant and independent of the external momenta. This is not the case in several explicit models, and the momentum dependence in the matrix element will alter the result in Fig.~\ref{fig:C_over_H_sca}. In particular, the plot for ${\cal C}/H$ will exhibit less flat lines, and larger spectral distortions are likely. Thus it is not obvious that the remarkably good agreement with approximate methods would persist with non-constant matrix elements.

\section{Conclusions}
\label{sec:conclusion}

The vast amount of present and cosmological upcoming data offers a unique opportunity to probe beyond SM physics. Several motivated frameworks feature the presence of light and weakly-coupled particles. The strong motivation for their existence is behind the astonishing experimental effort hunting for them. Terrestrial searches produce solid bounds that do not rely upon any assumption about the cosmological history, but they also come with their own limitations. They require coupling strengths large enough that these particles are indeed produced, and they are able to constrain their interactions with light degrees of freedom. The QCD axion is an emblematic case since terrestrial searches are able to probe directly only axion couplings with photons, electrons, and nucleons. Cosmological surveys provide a complementary tool to test these frameworks because of the high temperatures and densities achieved through the expansion history. Weakly-coupled particles can still be produced copiously at early times. Moreover, the primordial thermal bath contains all SM (and possibly beyond the SM) degrees of freedom in thermal equilibrium, and this provides us with a way to test the interactions of these new feebly interacting states with heavier SM particles.

In this work, we focused on new hypothetical degrees of freedom that are coupled to SM particles and light enough that they provide an additional contribution to the radiation part of the energy budget at early times. We quantified their abundance in terms of an effective number of additional neutrino species, $\dneff$, as it is conventionally done in the literature. The fraction to the energy budget due to radiation can be measured at two key moments in cosmic history: at the time of BBN when the Universe is approximately one second old, and at the last scattering surface when the Universe is much older (380,000 years) but still relatively young compared to its present age. The remarkable agreement of SM predictions with the observations at both BBN and CMB was the first motivation for our work since $\dneff$ is a very powerful new physics probe. At the same time, the astonishing sensitivity of current observations and the even more impressive future projections motivated our refinement of the calculation for $\dneff$. With only a few exceptions, this quantity is obtained via the solution of ordinary differential equations that derive from integrations over the phase space of the Boltzmann equation in momentum space. Or, even less accurately, the abundance of new relativistic degrees of freedom is estimated by assuming an instantaneous decoupling. We took one step back to the momentum space Boltzmann equation, and we developed a general framework to compute $\dneff$ without any approximation. In particular, we never assumed chemical or kinetic equilibrium for dark radiation at any time, and we tracked the complete distribution in phase space at any moment of the cosmological history. This allowed us to deal also with small couplings such that dark radiation never thermalizes with the primordial bath, and to track the details of decoupling when it thermalizes instead. An additional benefit of our framework was that we never had to employ Maxwell-Boltzmann statistical distributions and we could fully incorporate quantum statistical effects. Finally, we coupled the Boltzmann equation in momentum space with another one describing the evolution of the bath energy density, and we accounted for the energy exchanged between the visible and invisible sectors.

For the sake of concreteness, we focused on situations where bath processes producing dark radiation are two-body decays or binary collisions, and we assumed that only one dark radiation particle is produced in the final state. The decay case allowed us to work in a model-independent fashion and solve the Boltzmann equation in momentum space for a wide range of masses and couplings. On the contrary, binary collisions require the knowledge of the scattering matrix elements and these are model-dependent. For the purpose of this study, we consider theories where matrix elements are independent of the momentum. For both cases, we produced numerical solutions for the phase space distribution of dark radiation and used them to investigate for what couplings the assumption of chemical or kinetic equilibrium is justified and to investigate the details of the decoupling process. And we used the output of the numerical integration of the Boltzmann system to evaluate $\dneff$. Crucially, we found spectral distortions in the final distribution that have an impact on the resulting $\dneff$ detectable in the future. We compared these exact results with approximate momentum-space analysis employing classical statistics, and also with the output of approximated methods. Our last appendix contains material where these methods are reviewed. Figs.~\ref{fig:quantum_stats_decays} and \ref{fig:quantum_stats_scatterings} show when quantum corrections impact the final result by an amount larger than future experimental sensitivity. Likewise, Figs.~\ref{fig:dneff_MB} and \ref{fig:dneff_MB_sca} illustrate the error made by approximate analysis. We observe that for decays this error could be larger than the sensitivity of future CMB surveys, and therefore our analysis in momentum space is necessary if one wants accurate predictions. From the same figures, one may conclude that approximate methods based on integrated Boltzmann equations work rather well for scattering. This is indeed what we find from our numerical solutions. However, once we apply our formalism to realistic models the approximation of constant scattering matrix elements is likely to be incorrect, and consequently the collision terms would have a stronger momentum dependence that can ultimately induce spectral distortions. This is something that we defer to future work. 

Our work can be extended along several other lines. For example, we could go beyond a single production and include an arbitrary number of dark radiation particles in the final state. Several models have double production as the leading process and single production is absent. As discussed in the paper, the resulting Boltzmann equation would be more complicated since the collision operator appearing on the right-hand side of the Boltzmann equation in momentum space would remain an operator and not a regular function as was the case for single production. We always considered in this study the production during a radiation-dominated epoch with the primordial bath dominating the energy budget. However, it is straightforward to extend our study to cases when the thermal bath constitutes a sub-dominant component as long as we eventually recover the RD era before the BBN epoch~\cite{Allahverdi:2020bys}; all it takes is using the appropriate expression for the Hubble expansion rate. 

Finally, we mention a natural application of our framework to a concrete particle physics framework: light and weakly-coupled axion-like particles (ALPs). A leading candidate within this class is the QCD axion given its strong theoretical motivation from the strong CP problem. Early studies estimated the abundance of relativistic axions via the instantaneous decoupling approximation (see, e.g., Ref.~\cite{Baumann:2016wac}). Predictions for $\dneff$ obtained by tracking the number density of the relativistic axion population and then converting it to a corresponding energy density (thus assuming chemical equilibrium) followed up by Refs.~\cite{Ferreira:2018vjj,DEramo:2018vss,Arias-Aragon:2020qtn,Arias-Aragon:2020shv,DEramo:2021lgb,DEramo:2021psx,Green:2021hjh,DEramo:2021usm}. Recently, axion production via pion scattering was analyzed in momentum space by Refs.~\cite{Notari:2022ffe,Bianchini:2023ubu} that reported spectral distortions in the final phase space distribution. This confirms what was written above about the agreement between our exact solutions and the ones derived via approximate methods since a non-trivial momentum dependence in the transition amplitudes is likely to alter how decoupling happens for different momenta. It is worth exploring spectral distortions in the axion distribution function for production via derivative couplings to SM fermions. Furthermore, we could observe the cosmological effects of a non-vanishing QCD axion mass~\cite{Melchiorri:2007cd,Hannestad:2007dd,Hannestad:2010yi,Archidiacono:2013cha,Archidiacono:2015mda,DiValentino:2015zta,DiValentino:2015wba}. With the phase space distribution in hands, it is possible to reconsider existing analyses~\cite{Giare:2020vzo,Ferreira:2020bpb,Caloni:2022uya,DEramo:2022nvb,DiValentino:2022edq} and update the axion mass bound for production channels different than pion scatterings. We leave these developments of our general formalism to future studies. 

\paragraph*{Acknowledgements}

The authors thank Aleksandr Chatrchyan, Miguel Escudero, Andrzej Hryczuk, Hyungjin Kim, Bardia Najjari, Philip S{\o}rensen, Margherita Putti and Edoardo Vitagliano for useful discussions. A.L thanks Guglielmo Frittoli for the essential support. F.H. acknowledges the support and hospitality of DESY theory workshop 2021 in Hamburg, Germany and Astroparticle Symposium 2021 at Pascal Institute in Saclay, Paris, France. This work is supported in part by the Italian MUR Departments of Excellence grant 2023-2027 ``Quantum Frontiers''. The work of F.D. is supported by the research grant “The Dark Universe: A Synergic Multi-messenger Approach” number 2017X7X85K under the program PRIN 2017 funded by the Ministero dell’Istruzione, Università e della Ricerca (MIUR) and by Istituto Nazionale di Fisica Nucleare (INFN) through the Theoretical Astroparticle Physics (TAsP) project. F.D. acknowledges support from the European Union’s Horizon 2020 research and innovation programme under the Marie Skłodowska-Curie grant agreement No 860881-HIDDeN. The work of A.L. was supported by the Deutsche Forschungsgemeinschaft under Germany’s Excellence Strategy - EXC 2121 Quantum Universe - 390833306.
  
\appendix

\section{Phase space integrals}
\label{app:PSint}

We provide here the computational details for the evaluation of the collision terms appearing in the Boltzmann equations solved in this work. As done throughout the paper, we divide the discussion into two cases: two-body decays and binary scatterings. The formalism developed in this appendix generalizes the work done in Ref.~\cite{DEramo:2020gpr} by including the inverse processes in the collision terms and quantum statistical effects. We keep the discussion as general as possible and the results derived here are valid also for scenarios beyond the ones studied in this paper. The phase space integrals performed below can be employed also if one considers particle production via decays or scattering of degrees of freedom not in thermal equilibrium. Although our study focuses on the production of massless particles, we keep in this appendix finite $\dr$ mass effects since they do not bring any additional complications. In what follows, the $X$ particle has four-momentum $K=(\omega, \vec{k})$ which is never integrated over, and the on-shell condition $\omega^2 = k^2 + m_\dr^2$ always holds. We provide general results for finite $\dr$ mass, and we also report the limiting expressions for $m_\dr = 0$ that are used in the main text of this paper. 

\subsection{Two-body decays} 
\label{app:PSintA}

The collision term for the two-body decay is given explicitly in Eq.~\eqref{eq:C_decays}. The squared matrix element is constant, and as discussed in the main part of the paper we can trade it with the partial decay width. As a result, we can write the collision term as follows
\be
\mathcal{C}_{1 \rightarrow 2}(k, t) = \frac{16 \pi m_1}{g_2 g_X \mathcal{Y}_2} \, \Gamma_1 \, \times \, \mathcal{I}_{1\rightarrow 2}[f_{\bath_{1}} (1\pm f_{\bath_{2}})]  \ ,
\label{eq:C12app}
\ee
where we introduce the following functional
\be
\mathcal{I}_{ 1\rightarrow  2  }[{\cal F}] \equiv \frac{1}{2 \omega} \int d\Pi_1 d\Pi_2 \; (2\pi)^4 \delta^{(4)} (P_1 - P_2 - K) \mathcal{F} \ .
\label{eq:I2decayDEF} 
\ee
As anticipated, we keep the discussion general without specifying the explicit form for $\mathcal{F}$. Only at the end of the manipulations performed below we plug $\mathcal{F} = f_{\bath_{1}} (1\pm f_{\bath_{2}})$, and consistently with our notation the distributions $f_{\bath_{1}}$ and $f_{\bath_{2}}$ are the equilibrium ones. Thus our formalism is valid also if the particles participating in the two-body decays are not in equilibrium.

The four-momentum $K$ is fixed and we do not integrate over it. The phase space factor for the decaying particle $\bath_1$ can be written in the manifestly Lorentz invariant form
\be
d\Pi _1 = g_1 \frac{d^3 p_1}{(2 \pi)^3 2 E_1} = g_1 \frac{d^4 P_1}{(2\pi)^3}  \delta(P_1^2 -m_1^2) \Theta(P_1^0) \ ,
\ee
where $\Theta(x)$ is the Heaviside step function. We integrate over $d^4 P_1$ by using the delta function that imposes $P_1 = P_2 + K$. In particular, $P_1^0 = E_2 + \omega > 0$ and the Heaviside step function is always equal to one. We employ polar coordinates for the integration over $d^3 p_2$, and we choose the polar axis to be the direction of the fixed spatial momentum $\vec{k}$ of the $\dr$ particle. We define the polar angle $\theta$ via $\vec{p}_2 \cdot \vec{k} = p_2 k \cos\theta$, and we find the expression
\be
\mathcal{I}_{ 1\rightarrow  2 }[{\cal F}] = \frac{g_1 g_2}{8 \pi \omega} \int \frac{p_2^2 dp_2}{E_2} \, d\cos\theta \, \frac{d \phi}{2\pi}  \, \delta((P_2+K)^2-m_1^2) \, {\cal F}\ .
\ee
The integration over the azimuthal angle $\phi$ is straightforward since $\mathcal{F}$ cannot depend on it, whereas the integration over the polar angle $\theta$ can be performed by imposing the remaining Dirac delta function whose argument explicitly reads
\be
h_d(\cos\theta) \equiv (P_2+K)^2-m_1^2 = - \Delta^2 + 2 E_2 \omega - 2 p_2 k \cos\theta \ . 
\ee
In the above equation, we have introduced the quantity $\Delta^2 \equiv m_1^2 - m_2^2 - m_\dr^2 \geq 2 m_2 m_X \geq 0$ if the decay is kinematically allowed. It is useful to identify the angle $\theta_\star$ where the argument of the delta function vanishes, $h_d(\cos\theta_\star) = 0$, and it explicitly reads
\be
\cos\theta_\star = \frac{2 E_2 \omega - \Delta^2}{2 p_2 k} \ .
\ee
Finally, we use the well-known relation
\be
\delta(h_d(\cos\theta)) = \left| \frac{d h_d}{d \cos\theta} \right|^{-1}_{\cos\theta = \cos\theta_\star} \delta( \cos\theta - \cos\theta_\star) = \frac{1}{2 p_2 k} \delta( \cos\theta - \cos\theta_\star) \ ,
\label{eq:diracoff}
\ee
and after using the relation $p_2 dp_2 = E_2 dE_2$ we find
\be
\mathcal{I}_{ 1\rightarrow 2  }[{\cal F}] = \frac{g_1 g_2}{16 \pi \omega k} \int d E_2 \, d\cos\theta \delta(\cos\theta - \cos\theta_\star) {\cal F} = 
\frac{g_1 g_2}{16 \pi \omega k} \int_{E_2^-}^{E_2^+} d E_2 \, {\cal F}\ .
\ee
The last integration over $d \cos\theta$ gives a non-vanishing result as long as $-1\leq \cos\theta_\star \leq 1$, and this imposes a condition on the domain for the $d E_2$ integral  
\begin{align}
E_2^- = & \, \frac{\omega \Delta^2 - k \sqrt{\Delta^4 - 4 m_2^2 m_\dr^2}}{2 m_\dr^2} \; \xrightarrow{m_\dr \rightarrow 0} \; \frac{\Delta^2}{4 k} + \frac{m_2^2 k}{\Delta^2}  \ , \\
E_2^+ = & \, \frac{\omega \Delta^2 + k \sqrt{\Delta^4 - 4 m_2^2 m_\dr^2}}{2 m_\dr^2} \; \xrightarrow{m_\dr \rightarrow 0} \; + \infty \ ,
\end{align}
where we report also the limiting expressions for the massless $\dr$ case (for which $\omega = k$).  

We conclude with the application of this general result to the scenario of thermal production via two-body decays introduced in Sec.~\ref{sec:2bodydec}. In the main text of our paper, we write the collision term as in Eq.~\eqref{eq:C12final}. The identification of the function $D(k,T)$ only requires a direct comparison with Eq.~\eqref{eq:C12app}, and we find
\be
D(k,T) = \frac{16 \pi k^2}{T} \, \times \, \frac{\mathcal{I}_{1\rightarrow 2 }[f_{\bath_{1}} (1\pm f_{\bath_{2}})]}{g_1 g_2} \ .
\label{eq:Ddef}
\ee

\subsection{Binary scatterings}
\label{app:PSintB}

The phase space integrals for scatterings pose new challenges since the matrix elements do not need to be constant. This is in contrast with the two-body decays just discussed. Nevertheless, it is possible to perform an analogous derivation that does not require information about the underlying process until the last step. We start by defining the functional
\begin{equation}
{\cal I}_{2\rightarrow 2}[{\cal F}] \equiv \frac{1}{2 \omega} \int d\Pi_1 d\Pi_2 d\Pi_3 \; (2\pi)^4 \delta^{(4)} (P_1 + P_2 -P_3-K) {\cal F}  \ . 
\label{eq:I2scattDEF} 
\end{equation}
Upon choosing $\mathcal{F} = |{{\cal M}_{\bath_1\bath_2 \to \bath_3 \dr}}|^2 f_{\bath _1}f_{\bath_2}(1\pm f_{\bath_3})$, this corresponds to Eq.~\eqref{eq:colltermscat} for the specific process $\bath_1 \bath_2 \rightarrow \bath_3 \dr$ (i.e., the first term on the right-hand side of Eq.~\eqref{eq:collratescattering}). The other contributions to the total collision rate can be obtained by permutations. As discussed before, our analysis is valid  for a generic $\cal F$ which is a function of scalars products between momenta and moduli of spatial momenta $p_i,k$. In particular, we do not make any assumption in this appendix about the functional dependence of the squared matrix element.

\begin{figure}
	\centering
	\includegraphics[width=0.75\textwidth]{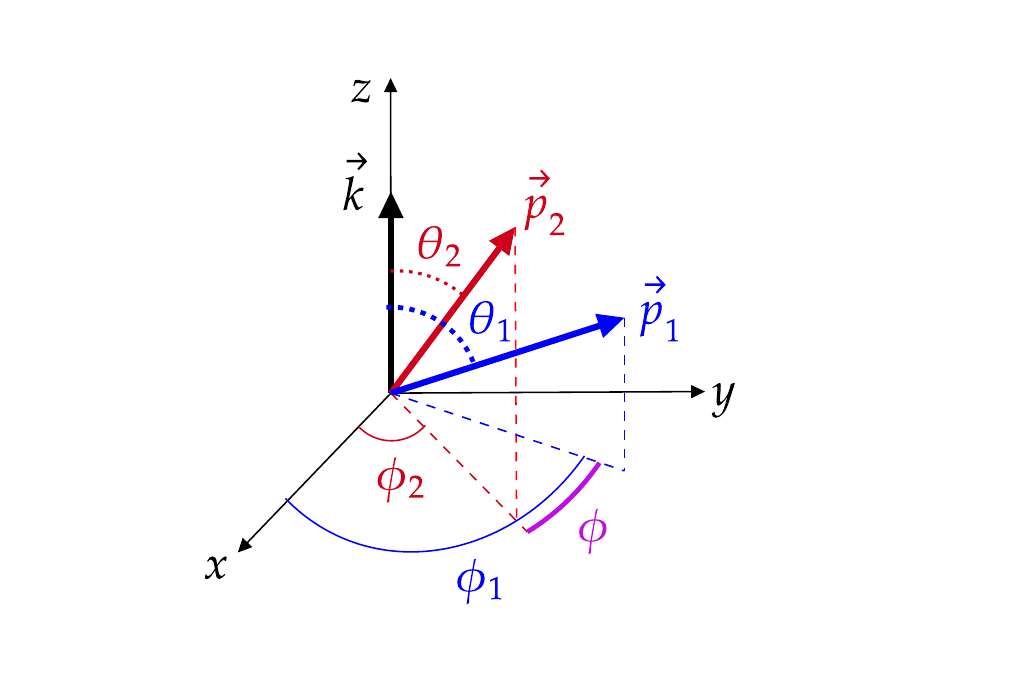}
	\caption{Polar coordinate system chosen for the phase space integration.}
	\label{fig:frame}
\end{figure}

We start again from the manifestly Lorentz invariant phase space measure
\begin{equation}
d\Pi _3 = \frac{g_3}{(2\pi)^3} d^4 P_3 \delta(P_3^2 -m_3^2) \Theta(P_3^0)\ ,
\end{equation}
and we integrate over $d \Pi_3$ with the delta function that fixes $P_3 = P_1 + P_2 - K$. We introduce the polar coordinate system illustrated in Fig.~\ref{fig:frame} for the other integrals. In such a frame, the vector $\vec{k}$ is fixed and we measure polar angles with respect to its direction 
\be
\vec{p}_i = p_i \left( \sin\theta_i \cos\phi_i, \sin\theta_i \sin\phi_i, \cos\theta_i \right) \qquad \qquad (i = 1,2) \ .
\ee
The quantity $\mathcal{F}$ can only depend on the following scalar products
\be
\frac{\vec{p}_1 \cdot \vec{k}}{p_1 \, k} =  \cos\theta_1 \equiv  c_1 \ , \qquad \frac{\vec{p}_2 \cdot \vec{k}}{p_2 \, k} =  \cos\theta_2 \equiv c_2  \ , \qquad \frac{\vec{p}_1 \cdot \vec{p}_2}{p_1 \, p_2} = c_{1} c_{2} + s_{1} s_{2} \cos(\phi_1 - \phi_2) \ .
\label{eq:SP} 
\ee
As it is manifest from this expression, and obvious from symmetry considerations, only the difference $\phi_1 - \phi_2$ between the two azimuthal angles is physical. For this reason, we change the azimuthal angles variables and introduce $\phi \equiv \phi_1 - \phi_2$ and $\Phi = (\phi_1 + \phi_2) / 2$. Thus the remaining phase space measure to be integrated over results in
\be
d^3 p_1 d^3 p_2 = p_1^2 dp_1 \; p_2^2 dp_2 \; dc_{1} \, dc_{2} \; d\phi \, d\Phi \ .
\ee
The integral in Eq.~\eqref{eq:I2scattDEF} now reads
\begin{equation}
{\cal I}_{2\rightarrow 2}[{\cal F}]= \frac{1}{2 \omega}\frac{g_1g_2g_3}{(2\pi)^4} \int  \, \frac{p_1 ^2dp_1}{2 E_1}\frac{p_2 ^2dp_2}{2 E_2}\;dc_{2} \,dc_{1} \; d\phi \, \frac{d\Phi}{2 \pi} \,\delta(P_3^2-m_3^2) \Theta(E_3) {\cal F} \ .
\label{eq:I2scattTEMP}
\end{equation}
The integral over $d \Phi$ is straightforward, and the delta function allows us to integrate over $d \phi$. The argument of this delta function explicitly reads
\be
h_s(\phi) \equiv (P_1+P_2-K)^2 - m_3^2 = \Delta^2 + 2 \epsilon + 2 k (p_1 c_1 + p_2 c_2) - 2p_1 p_2 (c_1 c_2 + s_1 s_2 \cos\phi )  \ ,
\ee
where we define $\Delta^2 \equiv m_1^2 + m_2^2 + m_\dr^2 - m_3^2$ and $\epsilon \equiv E_1 E_2 - (E_1 + E_2) \omega$. The angles $\phi_{\star r}$ (with $r$ labeling different solutions) for which the argument of the delta function vanishes satisfy
\begin{equation}
\cos(\phi_{\star r}) = \frac{\Delta^2/2 + \epsilon -  p_1 p_2 c_1 c_ 2 +  k ( p_1 c_1 + p_2 c_2 )}{ p_1 p_2  s_1 s_2} \ .
\label{eq:cosphistar}
\end{equation}
The above condition identifies the value of the cosine of the angles $\phi_{\star r}$ and therefore there are two possible values within the integration domain $0 \leq \phi \leq 2 \pi$, one in the interval $[0,\pi]$ and the other in $[\pi,2\pi]$. We replace the delta function with the expression
\be
\delta(h_s(\phi)) = \sum_{r=1}^2  \left| \frac{d h_s}{d \phi} \right|^{-1}_{\phi = \phi_{\star r}} \delta(\phi - \phi_{\star r}) \ ,
\ee
The angular dependence of the integrand in Eq.~\eqref{eq:I2scattTEMP} (and in particular of the quantity ${\cal F}$) can only depend on the scalar products of spatial momenta given in Eq.~\eqref{eq:SP} and therefore only on the cosine of the relative azimuthal angle $\phi$. For this reason, the two different roots $\phi_{\star 1}$ and $\phi_{\star 2}$ will give an indentical contribution. We consider one of them and we multiply the full expression by an overall factor of $2$. The delta function becomes
\be
\delta(h_s(\phi)) = 2 \times \frac{1}{2 p_1 p_2 s_1 s_2 \sqrt{1 - \cos^2\phi_{\star 1}}}\delta(\phi - \phi_{\star 1}) = \frac{\Theta(\mathcal{S}(p_1, p_2, k, c_1, c_2))}{\sqrt{\mathcal{S}(p_1, p_2, k, c_1, c_2)}} \delta(\phi - \phi_{\star 1}) \ ,
\ee
where we introduce the following function
\be
\mathcal{S}(p_1, p_2, k, c_1, c_2) = 4 p_1^2 p_2^2 s_1^2 s_2^2-\left[2 k (p_1 c_1 + p_2 c_2) - 2 p_1 p_2 c_1 c_2 + \Delta^2 + 2 \epsilon \right]^2  \ . 
\ee
The Heaviside theta function ensures that the argument of the square root is positive, and this condition is equivalent to imposing that the zeroes of the argument of the delta function gives cosines in Eq.~\eqref{eq:cosphistar} that have absolute values equal or less than one. 

The final expression for the phase space integral results in
\begin{equation}
{\cal I}_{2\rightarrow 2}[{\cal F}]= \frac{1}{2 \omega}\frac{g_1g_2g_3}{(2\pi)^4} \int  \, \frac{p_1 ^2dp_1}{2 E_1}\frac{p_2 ^2dp_2}{2 E_2}\;dc_{2} \,dc_{1} \; \Theta(E_1 + E_2 - \omega) 
\frac{\Theta(\mathcal{S}(p_1, p_2, k, c_1, c_2))}{\sqrt{\mathcal{S}(p_1, p_2, k, c_1, c_2)}} {\cal F} \ .
\label{eq:I2scattTEMP2}
\end{equation}
The remaining four-dimensional integral can be evaluated through Monte Carlo techniques. We reduce the integration domain and make it finite via the change of variables $x_i = \exp(-p_i/\Lambda_i)$ with $dp_i = - \Lambda_i/x_i dx_i$. We choose $\Lambda_i \sim 5 k$ to achieve good convergence, and this leads to an acceptance rate $> 50 \%$. If $ \cal F $ contains thermal distribution functions with temperature $ T $ an excellent choice is $ \Lambda \sim 10 T $.  The result is given by
\begin{equation}
{\cal I}_{2\rightarrow 2}[{\cal F}] \approx \frac{V}{\omega} \frac{g_1g_2g_3}{(2\pi)^4} \mean{ \frac{\Lambda_1 \Lambda_2}{x_1 x_2}\frac{p_1 ^2 }{2 E_1}\frac{p_2 ^2}{2 E_2}  \Theta(E_1+E_2-\omega)\frac{\Theta(\mathcal{S}(p_1, p_2, k, c_1, c_2)}{\sqrt{\mathcal{S}(p_1, p_2, k, c_1, c_2)}} {\cal F} }  \ .
\end{equation}
where $V=1^2\times 2^2=4$ is the integration variables volume and $\mean{ \cdot }$ is the average over $N$ samples of $(x_1,x_2,c_1,c_2)$. We checked that our numerical results agree with the analytical ones for the explicit cases considered in Ref.~\cite{DEramo:2020gpr} at the 0.1 \% level with $10^7$ samples.

The general expression in Eq.~\eqref{eq:I2scattTEMP2} is quite useful for a Monte Carlo evaluation of the collision rate. Its validity is rather general since there is no assumption about the explicit for of the function $\cal F$ besides the obvious fact that it is only dependent on scalar products between the four-momenta involved and the sizes of the spatial momenta. For binary collisions, and if we want to account for all the quantum effects, the specific expression is $\mathcal{F} = |{{\cal M}_{\bath_1\bath_2 \to \bath_3 \dr}}|^2 f_{\bath _1}f_{\bath_2}(1\pm f_{\bath_3})$. We conclude this Appendix with an analytical derivation of the functional in Eq.~\eqref{eq:I2scattTEMP2} when we employ MB statistics for bath particles. This approximation turns out to be often useful. Thus we evaluate now the analytical expression for the functional ${\cal I}_{2\rightarrow 2}[{\cal F}]$ when ${\cal F} = {\cal F}_{\rm MB}  =|{\cal M}(s,t)|^2 \exp[ - (E_1 + E_2) / T]$. Here, we express the squared matrix element in terms of the Mandelstam variables
\be
s = (P_1+P_2)^2 = (P_3 +K)^2,\qquad\qquad t=(P_1-P_3)^2 = (P_2-K)^2 \ .
\ee 
It is convenient to start from the definition in Eq.~\eqref{eq:I2scattDEF}, and rewrite the phase space measure as follows (see, e.g., Ref.~\cite{DEramo:2020gpr})
\be
d\Pi_1 d\Pi _2 d\Pi_3 (2\pi)^4 \delta^{(4)}(P_1+P_2-P_3-K) = \frac{g_1 g_2 g_3}{128\pi^3}\frac{dsdtdE_3}{k\sqrt{\lambda(s,m_3,m_X)}} \ ,
\ee 
where we define $\lambda(a,b,c)\equiv[a-(b+c)^2][a-(b-c)^2]$. Thus we rewrite the integral 
\be 
\begin{split}
{\cal I}_{2\rightarrow 2}[{\cal F}_{\rm MB}] =  \frac{g_1g_2g_3}{256\pi^3\omega} & \int_{s_-}^\infty \frac{ds}{\sqrt{\lambda(s,m_3,m_X)}} \int^{E_3^+(s)}_{E_3^-(s)}dE_3 \; \times \\ &  \int_{t_-(s)}^{t_+(s)}dt\ |{\cal M}(s,t)|^2  \exp[ - (E_1 + E_2) / T]  \ .
\end{split}
\label{eq:I22simple}
\ee 
Here $s_- = \max[(m_1+m_2)^2,(m_3+m_X)^2]$. The limits of integration on $E_3$ come from the values of $s$ in the frame in which $X$ has four-momentum $ (\omega,\vec{k}) $, and they are found by solving the equation
\be 
s = m_3^2 +m_X^2+2(E_3^\pm \omega \mp k\sqrt{{E_3^\pm}^2-m_3^2}   )
\ee
The limits on the Mandelstam variable $t$ are obtained by evaluating $ t $ in the center-of-mass frame, where the momenta of particles are known functions of $s$ and the mass spectrum
\be
\hat{p}_{1,2}=\frac{\sqrt{\lambda(s,m_1,m_2)}}{2\sqrt{s}},\qquad \qquad \hat{p}_{3,X}=\frac{\sqrt{\lambda(s,m_3,m_X)}}{2\sqrt{s}} \ ,
\ee 
so that
\be 
t_\pm = m_2^2 +m_X^2 -2\left(\sqrt{m_2^2+\hat{p}_{1,2}^2}\sqrt{m_X^2+\hat{p}_{3,X}^2}\mp \hat{p}_{1,2}\hat{p}_{3,X}\right)
\ee
The integration over $ E_3 $ is trivial because $\exp[-(E_1+E_2)/T]=\exp[-(\omega+E_3)/T]$ after using energy conservation. Finally, if the squared matrix element is constant (as we assume in the main text), the integration over $ t $ is also trivial
\be \label{eq:I22MBAmplConst}
\left.{\cal I}_{2\rightarrow 2}[{\cal F}_{\rm MB}]\right|_{{\cal M}^2}  = \frac{g_1g_2g_3}{256\pi^3} \frac{|{\cal M}|^2 T e^{-\omega/T}}{\omega k}\int_{s_-}^\infty \frac{ds}{s} \sqrt{\lambda(s,m_1,m_2)} \bigg(e^{-E^-_3/T}-e^{-E^+_3/T} \bigg)\ .
\ee 
Specializing this formula to the massless $X$ case and to the mass spectra of the three processes studied in the main text, we finally reproduce Eq.~\eqref{eq:CscFinal}.
 
\section{Spectral Distortions and Dark Radiation Temperature}
\label{app:thermalization}

In this Appendix we scrutinize how the different bins of momenta get populated. We explain the origin of the spectral distortions, we check whether kinetic equilibrium is a reasonable approximation for the dark radiation PSD, and we quantify the dark radiation temperature.
\begin{itemize}
\item \textbf{Momentum Bin Decoupling.} As already discussed in the main text, Figs.~\ref{fig:C_over_H} and \ref{fig:C_over_H_sca} make it manifest how different momenta decouple at different times. We define the decoupling temperature $\TD$ from the condition $ {\cal C}(k, \TD) \simeq H(\TD) $ (with the $g_i$ factors in the collision rate as plotted in Figs.~\ref{fig:C_over_H} and \ref{fig:C_over_H_sca}). We focus on the cases when the overall mass scale of the production process is the GeV. If dark radiation is produced via scattering or decays of bath particles at the TeV scale, the resulting spectral distortions would be much smaller since there is no significant change in the number of bath effective degrees of freedom at such high temperatures. Things are rather different once we get closer to the confinement scale. As done in the main text, we normalize the physical momentum $k$ in units of the bath temperature $T$. Fig.~\ref{fig:Tdec} shows how the decoupling temperature $\TD$ depends on the momentum bin for two-body decays (left panels) and scatterings (right panels). Top panels have the interaction strength fixed, and feature results for different choices of the statistical distributions. For the decay case, the choice of the statistical distribution has no impact on the decoupling temperature meaning that attaining equilibrium will lead to the same outcome for all the statistics and differences can just arise at smaller couplings. Differences arise instead for binary scatterings. The results in the lower panels are produced for MB statistics, and compare the decoupling temperature for three different interaction strengths (solid lines) with the result obtained via the instantaneous decoupling approximation (dot-dashed lines) reviewed in App.~\ref{subsec:ID}. We notice how the mismatch is rather severe for decays where the instantaneous decoupling approximation underestimate typical decoupling temperatures by a factor of $2$. This in turn implies that the approximate method overestimates the energy density of dark radiation (i.e., $ \dneff $). This difference is milder for scatterings. Furthermore, the functional dependence of the decoupling temperature is also different for decays and scattering. While in the former case the decoupling temperature decreases steeply for $ k<T $, and flattens for large momenta, in the scattering case it is only weakly dependent on $ k/T $, slightly increasing over the relevant range.
\begin{figure}
	\centering
	\includegraphics[width=.47\textwidth]{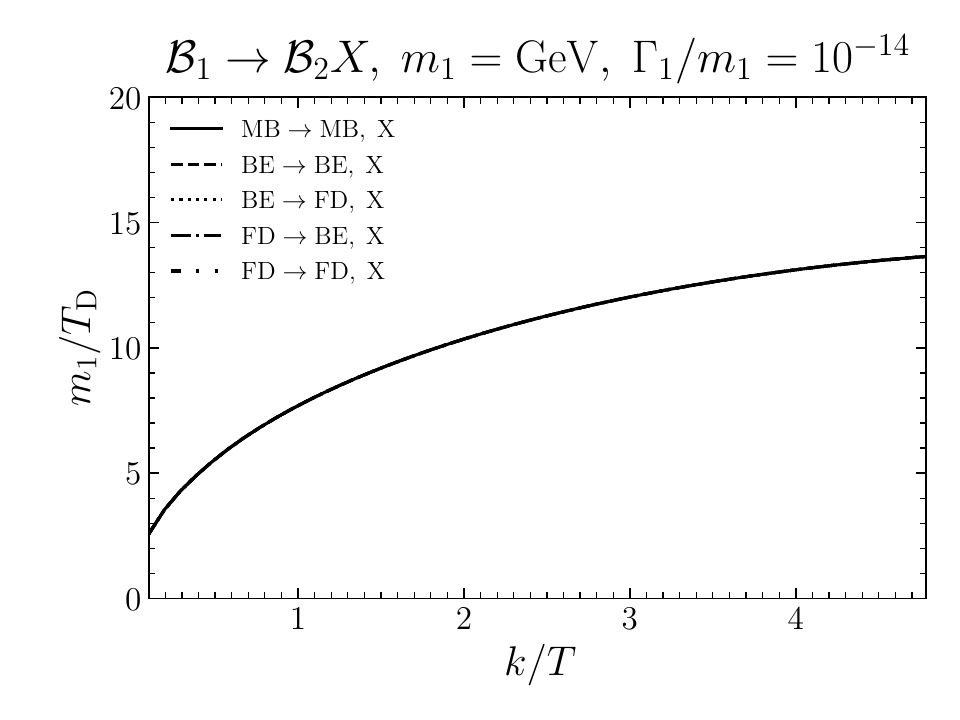}
	\includegraphics[width=.47\textwidth]{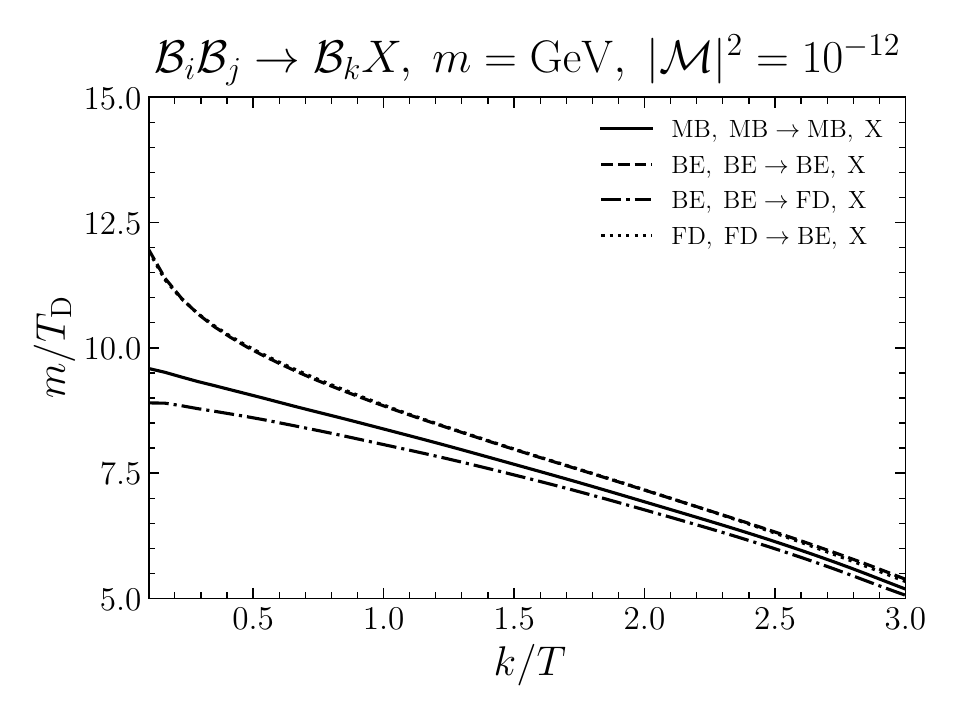} 
	\includegraphics[width=.47\textwidth]{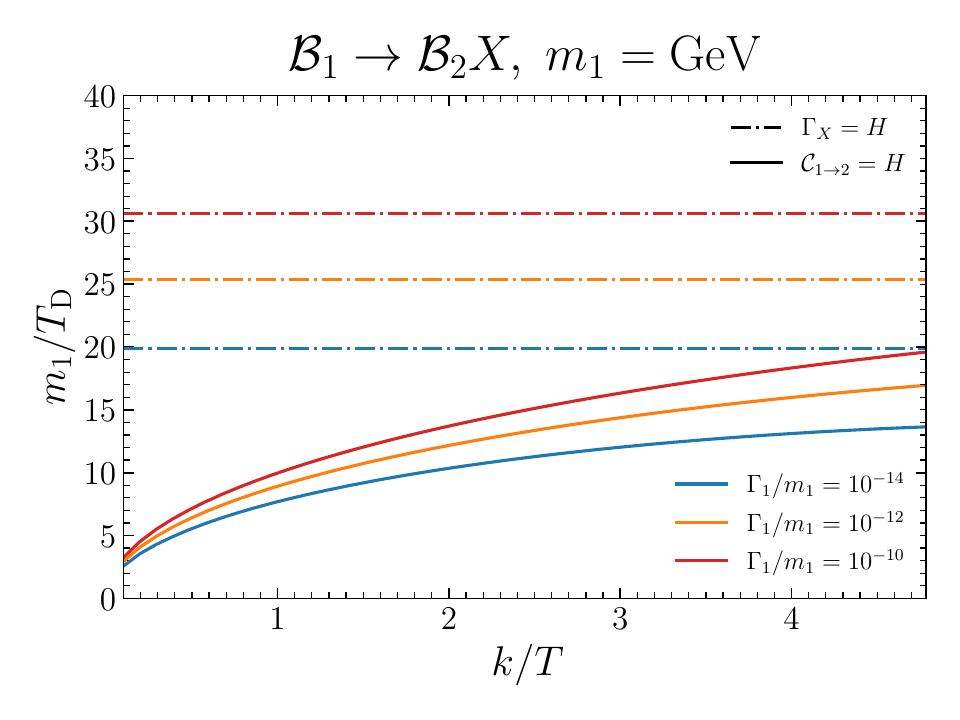}
	\includegraphics[width=.47\textwidth]{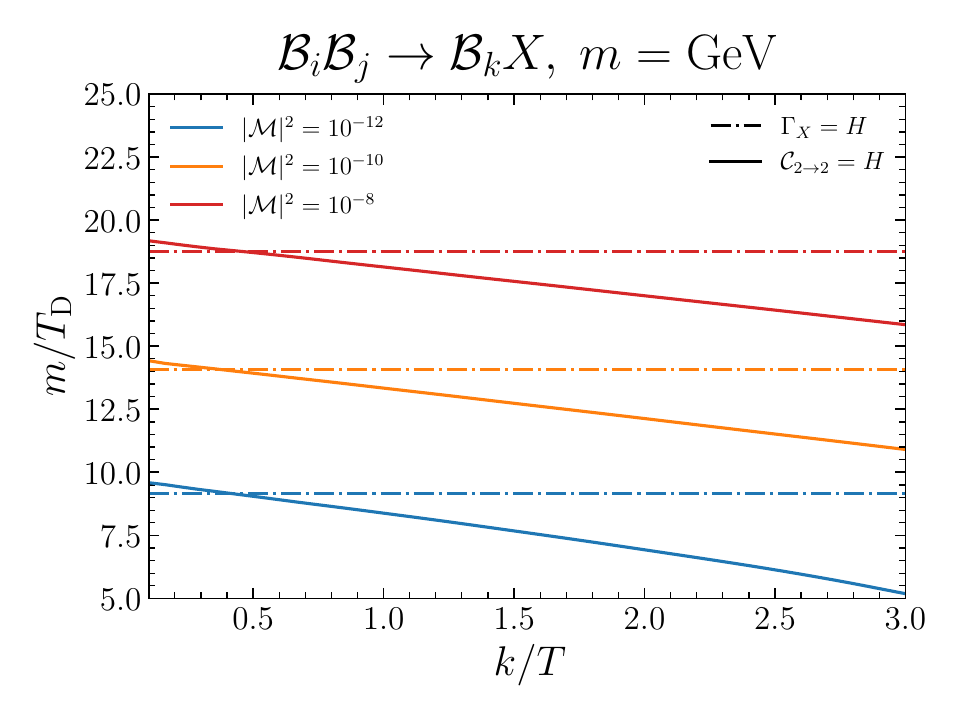} 
	\caption{The decoupling temperature for each momentum bin for decays (left panels) and scatterings (right panels). Top panels consider different statistics for the parameter space points in Fig.~\ref{fig:C_over_H} (left) and Fig.~\ref{fig:C_over_H_sca} (left). Bottom panels focus on the MB statistics but with different couplings. We show the decoupling temperature (solid lines) and compare it with the one obtained via the instantaneous decoupling approximation (dot-dashed lines).}
	\label{fig:Tdec}
\end{figure}
\begin{figure}
	\centering
	\includegraphics[width=.49\textwidth]{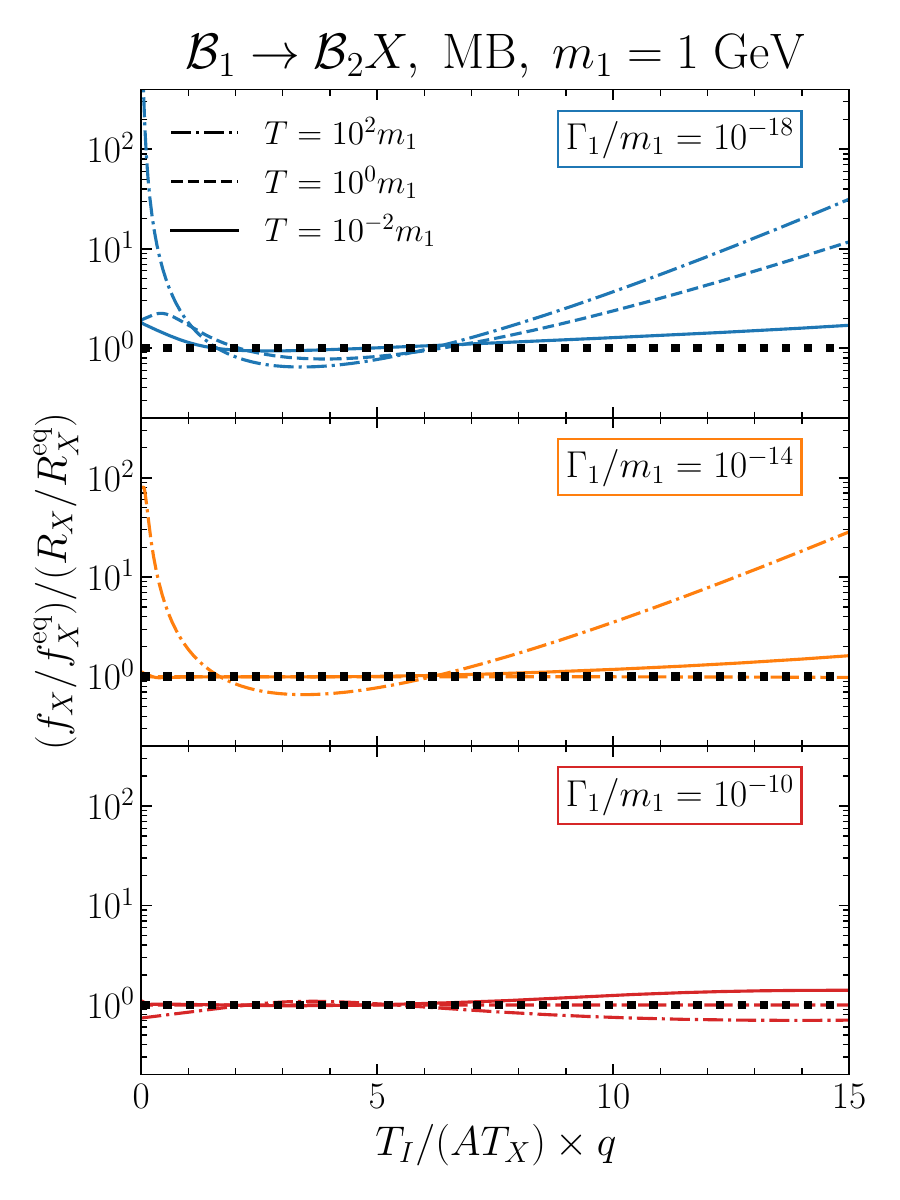}
	\includegraphics[width=.49\textwidth]{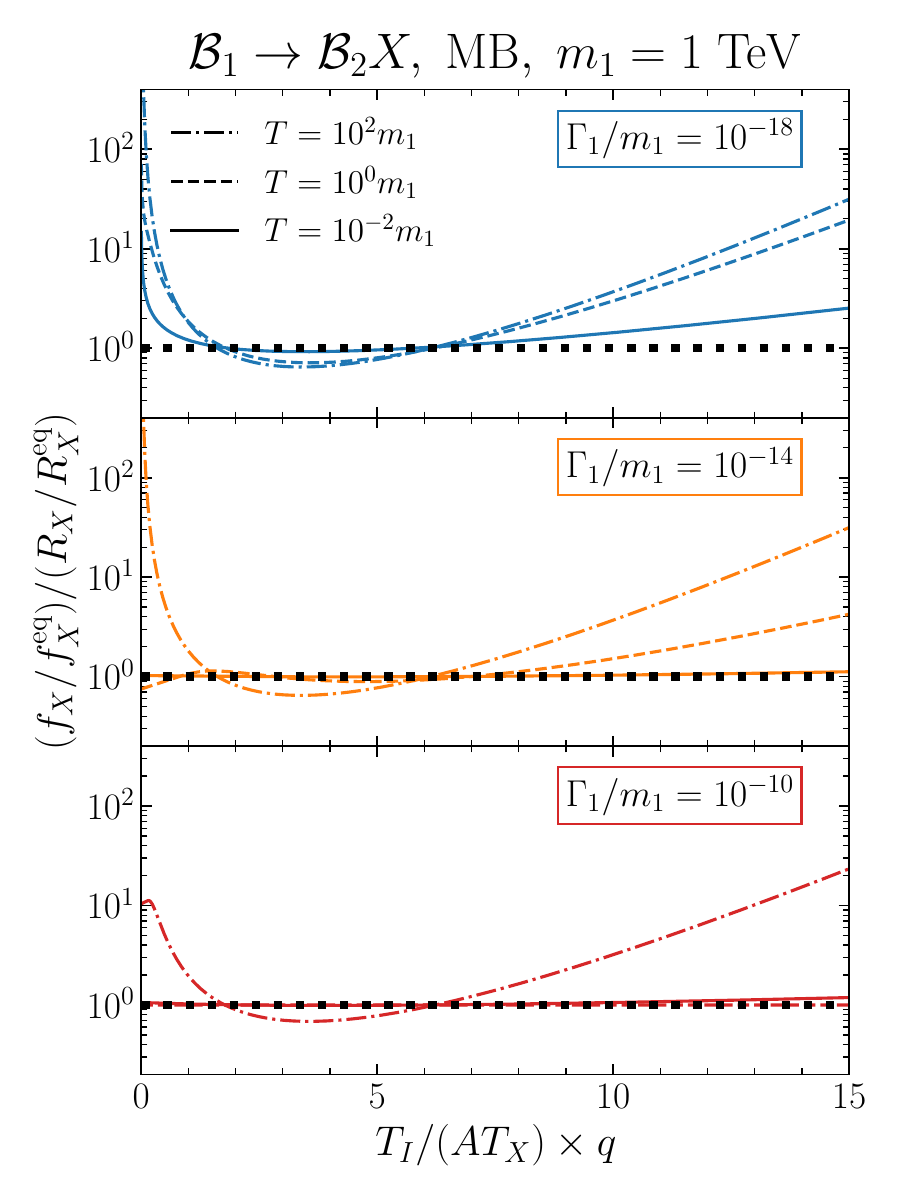}
	\includegraphics[width=.49\textwidth]{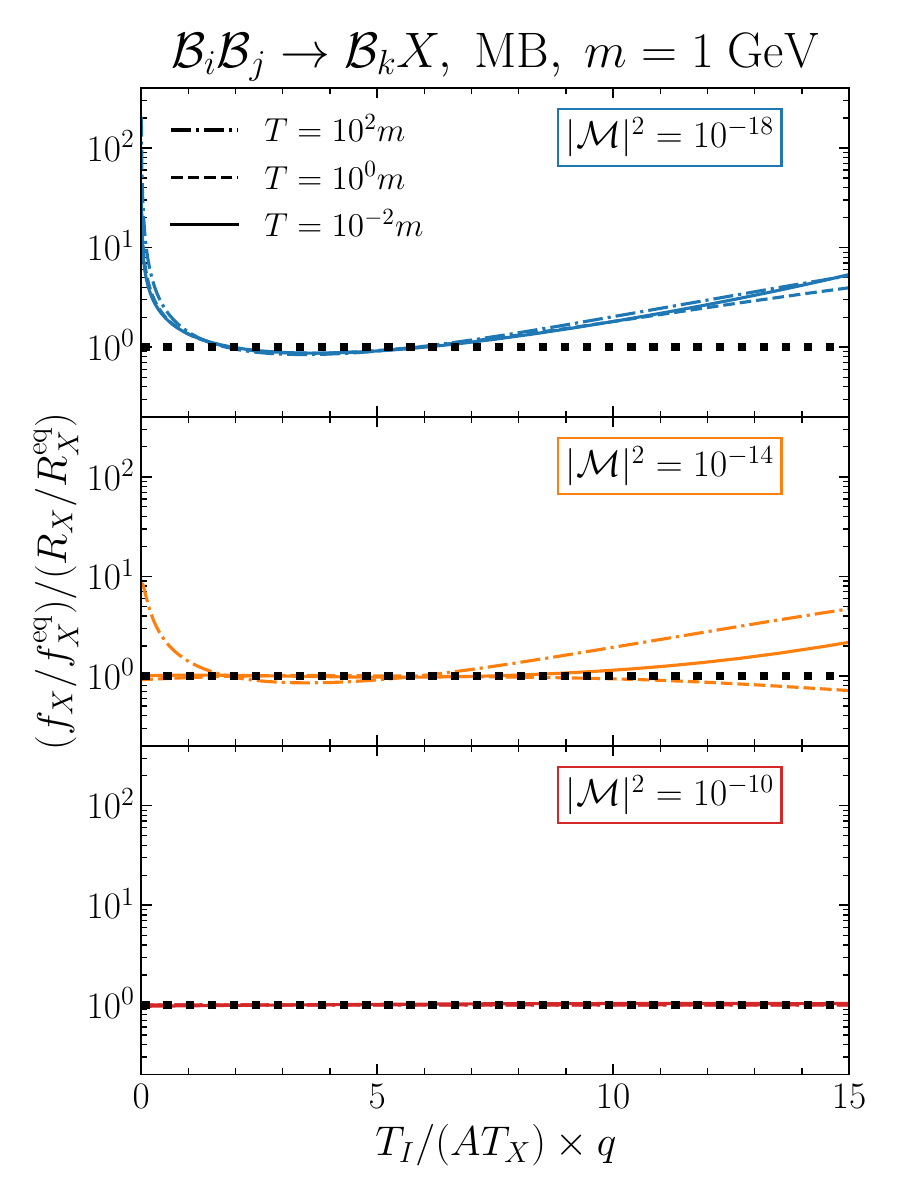}
	\includegraphics[width=.49\textwidth]{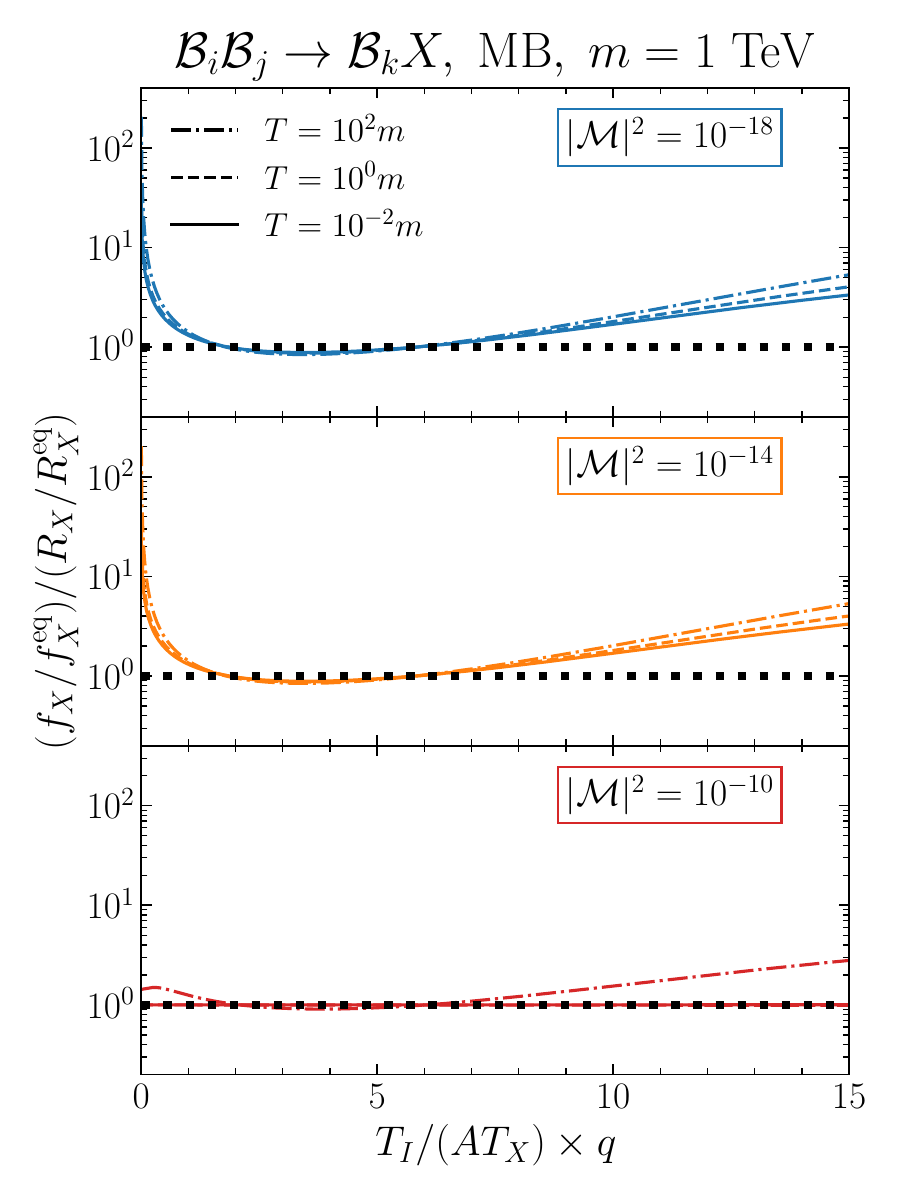}
	\caption{Numerical solutions for the ratio $ (f_X/f_X^{\rm eq})/(R_X/R_X^{\rm eq}) $ for two-body decays (top panels) and scatterings (bottom panels). We present the results at  the same temperatures chosen in Fig.~\ref{fig:f_sol_MB} and Fig.~\ref{fig:f_sol_MB_sca}. The horizontal lines mark the constant value of $1$ that is expected if thermal equilibrium is achieved. Spectral distortions are clearly visible.}
	\label{fig:f_sol_MB_kin}
\end{figure}
\item \textbf{Spectral Distortions and Kinetic Equilibrium.} The different decoupling temperature for the various momentum bins is at the hearth of the PSD spectral distortions. An effective way to visualize them is to plot $ (f_\dr/f_\dr^\eq)/(R_\dr/R_\dr^\eq) $ as a function of normalized comoving momenta $ T_I/(AT_X) \times q  $ at different temperatures. This is also useful to check whether or not kinetic equilibrium is a good assumption. If the  ratio $ (f_\dr/f_\dr^\eq)/(R_\dr/R_\dr^\eq) $ is constant and therefore independent on the momentum, then kinetic equilibrium is assured. If the ratio is also one, it means that we also have chemical equilibrium. We show this quantity in Fig.~\ref{fig:f_sol_MB_kin} for two-body decays (upper panels) and scatterings (lower panels) for the usual benchmark mass values adopted in this work. The different sub-plots for each panel are produced for different coupling values. For all cases, we notice non negligible deviations from kinetic equilibrium at large momenta. For small couplings, dark radiation do not ever thermalize and the deviation from kinetic equilibrium are rather large. Instead, for large interaction strengths, the fact that different momenta decouple at different times generates small spectral distortions. Comparing upper and lower panels, differences between the decay and scattering cases are clear. Spectral distortions for scatterings at strong coupling are milder and concentrated at very high comoving momenta, but three processes contribute to the production so it is easier for $\dr$ particles to retain kinetic equilibrium if they thermalize. This can be observed directly in Fig.~\ref{fig:Tdec} which shows that the dependence of decoupling temperature on comoving momentum is much weaker than in the decay case.
\item \textbf{Dark Radiation Temperature.} Another important quantity to track as the Universe expands is the dark radiation temperature. In particular, we want to compare it with the bath temperature $T$ until times well after the era of electron-positron annihilations. We defined in the main text the dark radiation temperature as the width of the PSD with the appropriate normalization. An alternative definition would be to identify the temperature as the fourth root of the energy density, a procedure that clearly works well only if we are close to thermal equilibrium. We present results here for both. We define $T_\dr^f$ as the dark radiation temperature obtained via Eq.~\eqref{eq:TX}. And we also define the temperature $T_\dr^\rho$ satisfying the relation $\rho_\dr \propto (T_\dr^\rho)^4$ with the correct numerical factors given in Eq.~\eqref{eq:rhodrapp}. In general, the former definition is the trustworthy one, especially in the case of small couplings where equilibrium is never achieved. We show in  Fig.~\ref{fig:f_sol_MB_T} the ratio $ T^\alpha_X/T $ with $ \alpha = f$ (solid lines) and $ \alpha = \rho$ (dashed lines). Notice that initially the temperature ratio is zero for $ T_\dr^\rho $ since $ f_\dr(A=1)=0 $, while it is finite for $ T_\dr^f $. Notice also how this ratio evolves accordingly to the number of degrees of freedom in the thermal bath after dark radiation production stops being efficient. If thermalization is reached (as in the case for red and orange lines), the difference between the final values of $T_\dr/T$ arises from non-instantaneous decoupling of the different momenta. This is the reason why these differences are milder in the scattering cases.
\begin{figure}
	\centering
	\includegraphics[width=.49\textwidth]{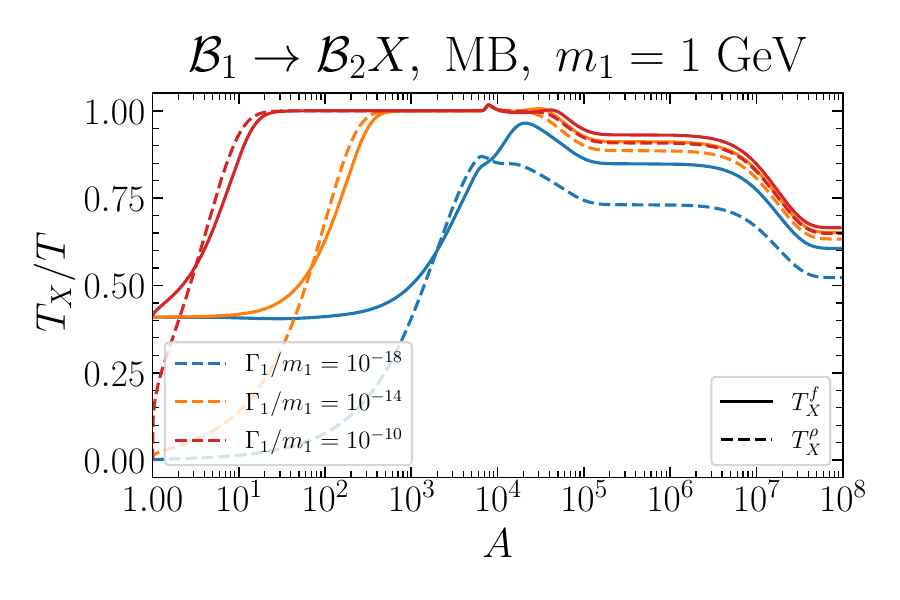}
	\includegraphics[width=.49\textwidth]{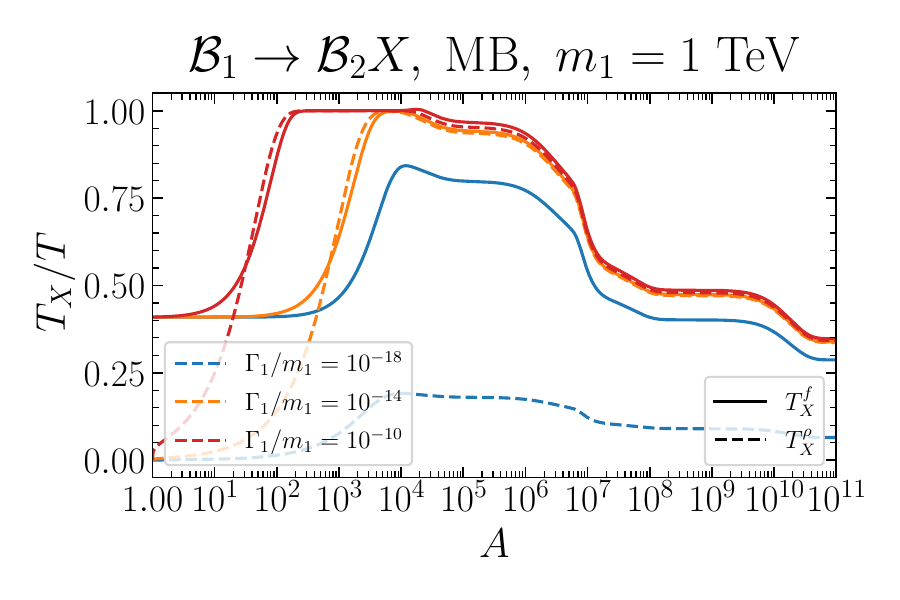}
	\includegraphics[width=.49\textwidth]{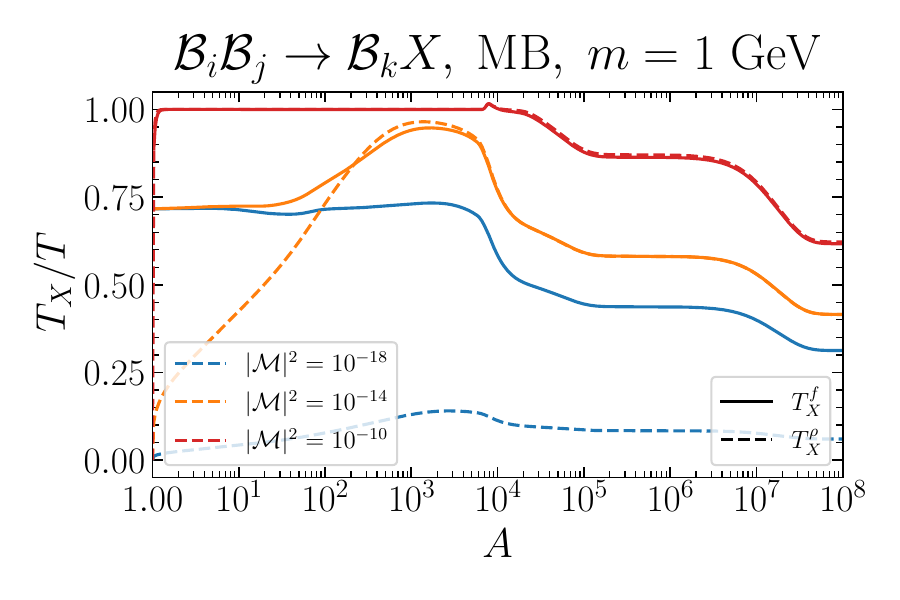}
	\includegraphics[width=.49\textwidth]{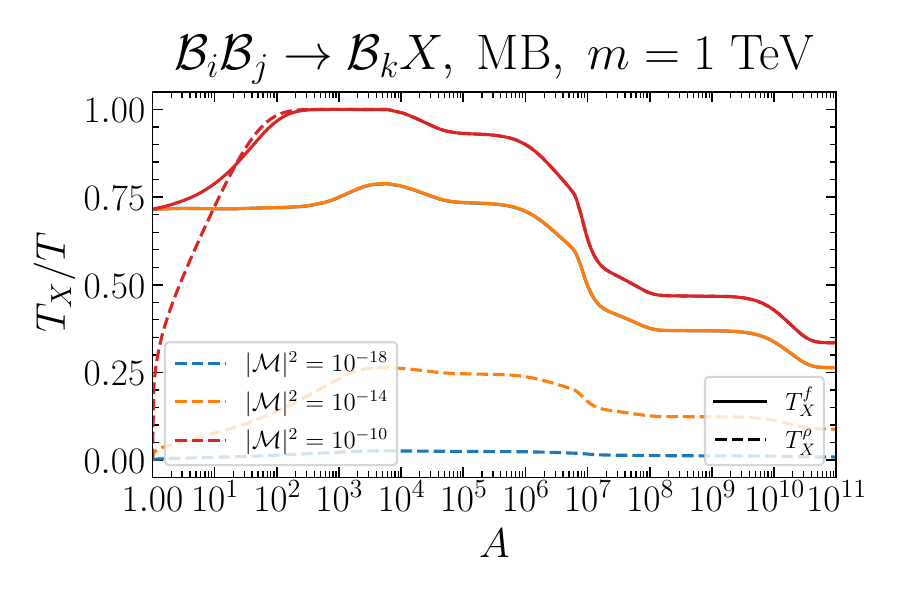}
	\caption{Ratio of dark radiation and thermal bath temperatures vs scale factor for decays (top panels) and scatterings (bottom panels). We consider $T_\dr$ defined both from the width of the PSD (solid lines) and the fourth root of the energy density (dashed lines).}
	\label{fig:f_sol_MB_T}
\end{figure}
\end{itemize}

\section{Approximate Methods}
\label{app:approx}

In this last appendix, we collect and review three approximate procedures that are alternative methods to estimate $\dneff$. First, we present an approach that relies upon the instantaneous decoupling approximation. The second method tracks the evolution of the dark radiation number density $n_\dr$ and it converts its freeze-out value to the corresponding energy density. This last step, which is needed to identify $\dneff$, requires the further assumption of an exact thermal distribution. We address this limitation thanks to a third improved approximate procedure that instead keeps track of the dark radiation energy density $\rho_\dr$. For this last method, we also account for the energy exchanged between the bath and the dark radiation which is not always accounted for in this kind of analysis. 

Before presenting the methods, we find it useful to collect some well-known results on equilibrium thermodynamics for the case when the dark radiation PSD has a thermal shape with a temperature $T_X$. When the dark radiation is in \textit{chemical equilibrium} we have 
\be
f_\dr^{\rm eq}(k,t) = 
\left\{ \begin{array}{ccl} \left( \exp[k / T_\dr] - 1 \right)^{-1} & $\qquad$  & \text{BE} \\ 
	\left( \exp[k / T_\dr] + 1 \right)^{-1} & $\qquad$ & \text{FD} \\ 
	\exp[- k / T_\dr] & $\qquad$ & \text{MB} \end{array}\right. \ ,
\label{eq:fXeq}
\ee
where we write the distribution in terms of the dark radiation temperature $T_\dr$ that is a function of the cosmic time $t$. We neglect the mass of $\dr$ and therefore the dark radiation has the dispersion relation $\omega = k$. In this appendix, we employ for practicality the symbol $T_\dr$ for the $\dr$'s temperature appearing in the equilibrium PSD. This is generally different from the bath temperature $T$, and it is also different from $T_\dr$ defined by Eq.~\eqref{eq:TX} and adopted in the main text. It is straightforward to evaluate the number density
\be 
n^{\rm eq}_\dr = g_\dr \int \frac{d^3 k}{(2\pi)^3} f^{\rm eq}_\dr(k, t) = g_\dr \, \xi_n \frac{\zeta(3)}{\pi^2} T_X^3 \ , \qquad \qquad  
\xi_n  = \left\{ \begin{array}{ccl} 1 & $\qquad$  & \text{BE} \\ 
	3/4 & $\qquad$ & \text{FD} \\ 1/\zeta(3) & $\qquad$ & \text{MB} \end{array}\right.    \ .
\label{eq:ndr}
\ee
Likewise, the energy density results in
\be
\rho^\eq_\dr  = g_\dr \int \frac{d^3 k}{(2\pi)^3} \, k \, f^{\rm eq}_\dr(k, t) = g_\dr \, \xi_\rho \frac{\pi^2}{30} T_X^4 \ ,  \qquad \qquad  
\xi_\rho  = \left\{ \begin{array}{ccl} 1 & $\qquad$  & \text{BE} \\ 7/8 & $\qquad$ & \text{FD} \\ 90/\pi^4 & $\qquad$ & \text{MB} \end{array}\right. \ .
\label{eq:rhodrapp}
\ee

Throught this review of approximate methods, we also discuss the \textit{kinetic equilibrium} regime. If $\dr$ particles thermalize in the early Universe, kinetic equilibrium is well justified. However, this results breaks down in the weak coupling regime. Nevertheless, the production source of dark radiation is from decays and scatterings of particles in thermal equilibrium, and the final PSD of $\dr$ is aware of the underlying bath temperature. Things would be drastically different from other production mechanisms such as dark radiation from inflaton decays where final state particles are produced monochromatic (unless elastic processes can efficiently redistribute particle momenta and induce a thermal profile for the PSD). The dark radiation PSD in the kinetic equilibrium regime takes the form   
\be
f_\dr^{\rm k.eq}(k, t) = 
\left\{ \begin{array}{ccl} \left( \exp[(k - \mu_\dr) / T_\dr] - 1 \right)^{-1} & $\qquad$  & \text{BE} \\ 
	\left( \exp[(k - \mu_\dr) / T_\dr] + 1 \right)^{-1} & $\qquad$ & \text{FD} \\ 
	\exp[- (k - \mu_\dr) / T_\dr] & $\qquad$ & \text{MB} \end{array}\right. \ ,
\label{eq:feqkineq}
\ee
where $\mu_X$ is the $\dr$'s chemical potential. Finally, it is useful for the subsequent discussion to find the number densities in the kinetic equilibrium regime
\be
n^{\rm k.eq}_\dr = n_\dr^{\rm eq} \times
\left\{ \begin{array}{ccl}  {\rm Li}_{3}(e^{\mu_X/T_\dr})/\zeta(3) & $\qquad$  & \text{BE} \\ 
	(-4/3){\rm Li}_{3}(-e^{\mu_X/T_\dr})/\zeta(3) & $\qquad$ & \text{FD} \\ 
	e^{\mu_X/T_\dr} & $\qquad$ & \text{MB} \end{array}\right. \ ,
\label{eq:n_neq}
\ee
where in the quantum statistical cases we have the polylogarithm function ${\rm Li}_3(z)$. Similar relations hold for the energy densities
\be
\rho^{\rm k.eq}_\dr  = \rho_\dr^{\rm eq} \times  
\left\{ \begin{array}{ccl}  90{\rm Li}_{4}(e^{\mu_\dr/T_\dr})/\pi^4 & $\qquad$  & \text{BE} \\ 
	(-720/7){\rm Li}_{4}(-e^{\mu_\dr/T_\dr})/\pi^4 & $\qquad$ & \text{FD} \\ 
	e^{\mu_\dr/T_\dr} & $\qquad$ & \text{MB} \end{array}\right. \ .
\label{eq:rho_rhoeq}
\ee

\subsection{Instantaneous decoupling}
\label{subsec:ID}

This first approach assumes that $\dr$ particles thermalize with the primordial bath at early times, and subsequently, decouple instantaneously. The temperature when such a decoupling happens, which is dubbed the \textit{freeze-out} temperature $\TD$, can be estimated through a comparison between two competing effects. On the one hand, collisions favor thermalization. On the other hand, the Hubble expansion dilutes and cools the universe down making it hard for particles to stay in thermal contact. We identify the freeze-out temperature by imposing that at the decoupling moment the collision rate is equal to the Hubble expansion rate
\be
\Gamma_{\dr}(\TD) = H(\TD) \,.
\ee
The dark radiation energy density at the decoupling time is still the equilibrium one, whereas $\dr$ particles free-stream afterward and their energy density decreases with the scale factor as $\rho_\dr \propto a^{-4}$. We define the interaction rate as follows\footnote{This results from the integration of the collision operator in Eq.~\eqref{eq:Cfgeneralfinal} with the appropriate normalization.}
\be
\begin{split}
	\Gamma_{\dr} \equiv \frac{\ell}{n_{\dr}^{\rm eq}} & \int \prod_{i = 1}^{n} d\Pi_i  \prod_{j = n+1}^{n+m}  d\Pi_j \prod_{r = 1}^{\ell}  d\mathcal{K}_r \; (2\pi)^4\delta^4 
	\left(P_{\rm in} - P_{\rm fin}  \right) |\mathcal{M}|^2  \\ &  \times 
	\prod_{i = 1}^n f_{\bath_{i}}\prod_{j=n+1}^{n+m} (1\pm f_{\bath_{j}})   \prod_{r = 1}^{\ell} \left(1 \pm f^{\rm eq}_{\dr}(k_r) \right) \ .
\end{split}
\label{eq:rateGAMMAH}
\ee
The above expression is general and valid for any choice of the statistical distributions. The bath particles PSDs are the equilibrium ones in Eq.~\eqref{eq:feq}, and consistently with our hypothesis we set the PSD for the $\dr$ to the equilibrium expression in Eq.~\eqref{eq:fXeq}. We normalize $\Gamma_{\dr}$ with a factor proportional to the inverse equilibrium number density in Eq.~\eqref{eq:ndr}. The Hubble rate is given by the Friedmann equation, $H = \sqrt{\rho_\bath + \rho_\dr} / (\sqrt{3} \, M_{\rm Pl})$. The visible contribution to the energy density reads $\rho_\bath = (\pi^2/30) g^\bath_{\star\rho}(T) T^4$ whereas the invisible one before decoupling is given by the expression in Eq.~\eqref{eq:rhodrapp}. 

Before the decoupling epoch, all the thermodynamic quantities for the dark radiation are the equilibrium ones with $T_X = T$. After decoupling, the use of comoving quantities ease the analysis. The entropy density, defined as $s(T) = (2 \pi^2 / 45) g_{\star s}(T) T^3$, turns out to be quite useful. The effective number of entropic degrees of freedom is the sum of two contributions, $g_{\star s}(T) = g^\bath_{\star s}(T) +  g^\dr_{\star s}(T)$. The former is a property of the thermal bath, the latter follows from the fact that $f_X$ remains thermal after decoupling but with $T_\dr \neq T$ because bath particles become non-relativistic through the expansion, and it explicitly reads
\be
g^\dr_{\star s}(T) = g_\dr \xi_\rho \left( \frac{T_\dr}{T} \right)^3 \ , \qquad \qquad 
\frac{T_X}{T} = \left\{ \begin{array}{ccccccl}
	1 & & & & & $\quad$  & T \geq \TD \\
	\left(g^\bath_{\star s}(T) / g^\bath_{\star s}(\TD)\right)^{1/3} & & & & & $\quad$  & T < \TD 
\end{array} \right. \ .
\label{eq:gstarsX}
\ee
We define the dimensionless ratio ${\cal R}_{\dr}(T) \equiv \rho_\dr(T) / s(T)^{4/3}$. Entropy conservation ($s \propto a^{-3}$) imposes ${\cal R}_\dr(T < \TD) = {\cal R}_\dr(\TD)$, and in particular we have ${\cal R}_\dr(T_{\rm CMB}) = {\cal R}_\dr(\TD)$. 

Finally, we evaluate $\Delta N_{\rm eff}$ via Eq.~\eqref{eq:dneff} and we find
\be
\Delta N_{\rm eff} = \frac{8}{7} \left( \frac{11}{4} \right)^{4/3} \frac{{\cal R}_\dr(T_{\rm CMB}) \, s(T_{\rm CMB})^{4/3}}{\rho_\gamma(T_{\rm CMB})} = 
g_\dr \xi_\rho \; \frac{4}{7} \left( \frac{11}{4} \right)^{4/3} \, \left( \frac{g_{\star s}(T_{\rm CMB}) }{g_{\star s}(\TD) } \right)^{4/3}\ . 
\ee
The result is proportional to the ratio, with the exponent of $4/3$, of the \textit{total number} (i.e., contributions from both the bath and $\dr$) of entropic degrees of freedom at the recombination and decoupling epochs. Once we plug the explicit temperature dependence of the $\dr$'s contribution, as given in Eq.~\eqref{eq:gstarsX}, we find
\be
\Delta N_{\rm eff} = g_\dr \xi_\rho \; \frac{4}{7} \left( \frac{11}{4} \right)^{4/3} \, \left( \frac{g^\bath_{\star s}(T_{\rm CMB}) }{g^\bath_{\star s}(\TD) } \right)^{4/3}\ . 
\label{eq:dneff_istant}
\ee
In this last expression there is only the bath contribution. If we have a primordial thermal bath made of only SM particles ($g^\bath_{\star s}(T) = g^{\rm SM}_{\star s}$), the contribution at the recombination epoch results in $g^{\rm SM}_{*s}(T_{\rm CMB}) = 2 + N_{\rm eff}^{\rm SM} \times (7/11) \simeq 3.94$.

\subsection{Tracking the number density}\label{subsec:BEn}

Another approximate method commonly employed in the literature is to track the dark radiation number density defined from the PSD as follows
\be
n_\dr = g_\dr \int \frac{d^3 k}{(2\pi)^3}f_\dr(k, t)    \ .
\ee
We derive the Boltzmann equation describing the time evolution of this quantity. Our starting point is the Boltzmann equation in momentum space in Eq.~\eqref{eq:BEforF}, and we multiply both sides by $(g_X / \omega_1)$. For the collision operator, we take the general expression in Eq.~\eqref{eq:Cfgeneralfinal}. After we integrate both sides over the three components of the spatial momentum $\vec{k}_1$, we find 
\be
\begin{split}
	\frac{dn_{\dr} }{dt} + & \; 3 H n_{\dr}  =  \ell \times \int \prod_{i = 1}^{n} d\Pi_i  \prod_{j = n+1}^{n+m}  d\Pi_j \prod_{r = 1}^{\ell}  d\mathcal{K}_r \; (2\pi)^4\delta^4 
	\left(P_{\rm in} - P_{\rm fin}  \right) |\mathcal{M}|^2  \\ &  \times 
	\prod_{i = 1}^n f_{\bath_{i}}\prod_{j=n+1}^{n+m} (1\pm f_{\bath_{j}})   \prod_{r = 1}^{\ell} \left(1 \pm f_{\dr}(k_r) \right) 
	\left[ 1 - \prod_{r = 1}^{\ell} e^{\omega_r / T} \frac{f_{\dr}(k_r)}{1 \pm f_{\dr}(k_r)}  \right]   \ .
\end{split}
\label{eq:ngeneral}
\ee
The above result is still not an ordinary differential equation, and we need some further assumptions to achieve a simpler equation to handle. Given the thermal origin of the produced $\dr$ particles, it is reasonable to assume kinetic equilibrium and take the PSD given in Eq.~\eqref{eq:feqkineq}. If we have $f_\dr = f_\dr^{\rm k.eq}$ then the following relation holds
\be
\prod_{r = 1}^{\ell} e^{\omega_r / T_\dr} \frac{f_{\dr}(k_r)}{1 \pm f_{\dr}(k_r)} = \left(e^{\mu_\dr / T_\dr}\right)^\ell \ .
\ee
This result is valid for any statistical distribution of the dark radiation degree of freedom, and it allows us to take the term inside the square brackets in Eq.~\eqref{eq:ngeneral} outside of the integral since it is independent on the momenta. We find the following Boltzmann equation 
\be
\begin{split}
	\frac{dn_{\dr} }{dt} +3 H n_{\dr}  = & \; \ell \, \left[ 1 - \left(e^{\mu_\dr / T_\dr}\right)^\ell   \right]  \times \int \prod_{i = 1}^{n} d\Pi_i  \prod_{j = n+1}^{n+m}  d\Pi_j \prod_{r = 1}^{\ell}  d\mathcal{K}_r \\ & \; (2\pi)^4\delta^4 \left(P_{\rm in} - P_{\rm fin}  \right) |\mathcal{M}|^2 \prod_{i = 1}^n f_{\bath_{i}}\prod_{j=n+1}^{n+m} (1\pm f_{\bath_{j}})   \prod_{r = 1}^{\ell} \left(1 \pm f_{\dr}(k_r) \right) \ .
\end{split}
\label{eq:ngeneral2}
\ee
The term inside the square bracket can be related to the $\dr$'s number density. Consistently with our assumption of kinetic equilibrium, namely that the PSD distribution has the functional form given in Eq.~\eqref{eq:feqkineq}, the explicit expression for $n_\dr = n^{\rm k.eq}$ is provided by Eq.~\eqref{eq:n_neq}. However, even after we assume kinetic equilibrium the resulting Eq.~\eqref{eq:ngeneral2} is still not an ordinary differential equation for $ n_\dr $ because of the terms containing $ f_\dr$ on the right-hand side. Our only option is to approximate $1 \pm f_\dr \approx 1$, and this is equivalent to neglect quantum degeneracy effects for $\dr$. From now on, we employ MB statistics for the dark radiation.

To summarize, once we assume kinetic equilibrium as well as MB  for the $\dr$'s, we obtain the Boltzmann equation describing the number density evolution
\be
\frac{dn_{\dr} }{dt} + 3 H n_{\dr} = \gamma_\dr \left[1 - \left(\frac{n_\dr}{n_\dr^{\rm eq}}\right)^\ell \right]   \ . 
\label{eq:BEforn}
\ee
Here, the temperature-dependent function $\gamma_\dr$ accounts for the averaged number of $\dr$ particles produced per unit of time and volume, and it explicitly reads
\be
\gamma_\dr  \equiv \ell \, \int \prod_{i = 1}^{n} d\Pi_i  \prod_{j = n+1}^{n+m}  d\Pi_j \prod_{r = 1}^{\ell}  d\mathcal{K}_r  (2\pi)^4\delta^4 
\left(P_{\rm in} - P_{\rm fin}  \right) |\mathcal{M}|^2 \prod_{i = 1}^n f_{\bath_{i}}\prod_{j=n+1}^{n+m} (1\pm f_{\bath_{j}})   \ .
\label{eq:gamman}
\ee
This general expression for the production rate accounts for the quantum degeneracy effects of bath particles. This has to be compared with the expression in Eq.~\eqref{eq:rateGAMMAH} that accounts also for the quantum degeneracy effects of the dark radiation particle. This is possible because within the instantaneous decoupling approximation we assume that $\dr$ thermalizes, and we can compute the phase space integral by using its equilibrium distribution. 

Our next step is to solve Eq.~\eqref{eq:BEforn} and evaluate the $\dr$'s abundance at the recombination time. As it is often done in the literature, it is convenient to switch to the comoving number density $Y_\dr = n_\dr/s$ to factor out the cosmological dilution due to the Hubble expansion, and to employ the thermal bath temperature as the evolution variable. Upon imposing entropy conservation (i.e., $d(s a^3)/ dt = 0$), we find the following form of the Boltzmann equation for the comoving number density
\begin{equation}
\frac{dY_\dr}{d \log T} = - \frac{\gamma_\dr}{s\, H} \bigg(1+\frac{1}{3}\frac{d\log g_{\star s}}{d\log T}\bigg)  \left[1 - \left(\frac{Y_\dr}{Y_\dr^{\rm eq}}\right)^\ell \right] \ .
\end{equation}
Here, $Y_\dr^\eq =n_\dr^\eq /s$ is the comoving equilibrium number density. It is enough to solve the Boltzmann equation up to the point when production stops being effective (e.g., couplings get too small and/or bath particles participating in the production become non-relativistic and disappear from the bath) and $Y_\dr$  reaches a constant asymptotic value. 

We still do not know how to account for the contribution of $\dr$ particles to the energy and entropy densities. The former affects the Boltzmann equation once we express the Hubble parameter $H$ via the Friedmann equation whereas the latter appear in the Boltzmann equation both explicitly and implicitly via the functions $s$ and $Y_\dr^{\rm eq}$. Typically, this extra contribution is neglected or it is considered to be maximal
\be
g_{* \alpha} (T)  \simeq  \left\{ \begin{array}{ccl} g^\bath_{*\alpha}(T) & $\qquad$  & \text{Neglecting $\dr$} \\ g^\bath_{*\alpha}(T) + g_\dr \xi_{\rho_\dr}  & $\qquad$ & \text{Thermalized $\dr$} \end{array}\right.  \qquad \qquad \qquad (\alpha = \rho, s) \ .
\label{eq:gstarsn}
\ee
Under this highly common assumption, where one adopts one of the two extreme choices, the equation for $ Y_\dr $ can be integrated and we can find its asymptotic value. 

The last step is converting the asymptotic comoving number density into an asymptotic comoving energy density. A standard assumption is that the $\dr$'s PSD has the thermal shape in Eq.~\eqref{eq:fXeq} with a generic temperature $T_\dr$ not necessarily equal to the one for the bath. We emphasize how this is a further assumption with respect to the ones already made in this sub-section. The Boltzmann equation in Eq.~\eqref{eq:BEforn} has been derived under the assumption of \textit{kinetic equilibrium} (as well as MB) for the dark radiation, whereas the distributions in Eq.~\eqref{eq:fXeq} are valid if \textit{chemical equilibrium} holds. Assuming kinetic equilibrium only would not be sufficient to make this conversion because of the unknown chemical potential $\mu_\dr$ entering the expressions in Eqs.~\eqref{eq:n_neq} and \eqref{eq:rho_rhoeq}. Thus the only way to proceed is to make the further assumption of chemical equilibrium for this conversion, with number and energy densities given in Eqs.~\eqref{eq:ndr} and \eqref{eq:rhodrapp}, and this allows us to find the relation 
\be \label{eq:from_n_to_rho}
\rho_\dr = g_\dr \xi_{\rho_\dr} \frac{\pi^2}{30}  \left(\frac{\pi^2}{\zeta(3)} \frac{n_\dr}{g_\dr \xi_{n_\dr}} \right)^{4/3}  \ .
\ee
While it is true that we assumed MB statistics for $\dr$ to derive the Boltzmann equation, this relation can be specialized to different cases for the dark radiation statistical properties upon choosing the appropriate $\xi$ factors. We find that using the appropriate statistics instead of the MB improves the error with respect to the full solution. 

Finally, we determine the contribution to $\dneff$ as defined in Eq.~\eqref{eq:dneff}
\be
\dneff = g_\dr \xi_{\rho_\dr} \frac{4}{7} \left( \frac{11}{4} \right) ^{4/3}\left( {\frac{2\pi^4}{45 \zeta(3)}} \frac{g^\bath_{*s}(T_{\rm CMB})}{g_\dr\xi_{n_\dr}} \right)^{4/3} Y_\dr(T_{\rm CMB})^{4/3} \ .
\label{eq:dneff_Y_std}
\ee
The final result for $\dneff$ is expressed in terms of the asymptotic solution of the Boltzmann equation $Y_\dr(T_{\rm CMB})$. We include only the entropic degrees of freedom from the bath $g^\bath_{*s}(T_{\rm CMB})$. The analogous contribution from dark radiation $g^\dr_{*s}(T_{\rm CMB})$ is between the two extremes in Eq.~\eqref{eq:gstarsn}, and it is likely to be suppressed with respect to the factor of $g_\dr \xi_{\rho_\dr}$ because threshold effects in the thermal bath between decoupling and recombination that lead to $T_\dr < T$ (see Eq.~\eqref{eq:gstarsX}). Thus taking the ``Thermalized $\dr$'' case is valid to solve the Boltzmann equation but it would not be appropriate at recombination.
 
We conclude this sub-section with an improvement that includes also the $\dr$'s contribution to the entropy and energy densities quantified as follows (see Eq.~\eqref{eq:gstarsX}) 
\begin{align}
g_{*s}^\dr(T) = & \, g_{*s}(T) - g_{*s}^\bath(T) = g_\dr\xi_{\rho_\dr}\left(\frac{T_\dr}{T}\right)^{3} = 
\xi_{\rho_\dr}\frac{2\pi^4}{45\zeta(3)} \frac{g_{*s}(T)}{\xi_{n_\dr}}  Y_\dr (T) \ , \\
g_{*\rho}^\dr(T) = & \, g_{*\rho}(T) - g_{*\rho}^\bath(T) = g_\dr\xi_{\rho_\dr}\left(\frac{T_\dr}{T}\right)^4 = 
g_\dr \xi_{\rho_\dr} \left( \frac{2\pi^4}{45 \zeta(3)} \frac{g_{\star s}(T)  }{g_\dr \xi_{n_\dr}} \right)^{4/3} Y_\dr(T)^{4/3} \ .
\end{align}
Here, we still assume that the $\dr$'s PSD is a chemical equilibrium one with an arbitrary temperature $T_\dr$. We can invert the first relation to find
\be
g_{*s} (T) =  \frac{g^\bath_{*s}(T)}{1-\frac{2\pi^4}{45 \zeta(3)}\frac{\xi_{\rho_\dr}}{\xi_{n_\dr}}Y_\dr(T)} \ . 
\ee
Once we know how the full contribution $g_{*s} (T)$ depends on the temperature, we can plug it into the expression for $g_{*\rho}^\dr(T)$ above and account also for the dark radiation contribution to the energy density. We use these improved estimates for the total number of effective degrees of freedom into the Boltzmann equation, and we can find a new differential equation for $Y_\dr(T)$. Finally, after we solve it and find the asymptotic value of $Y_\dr$, we can compute
\be
\Delta N_{\rm eff} = g_\dr \xi_{\rho_\dr} \frac{4}{7} \left( \frac{11}{4} \right) ^{4/3}\left( \frac{\frac{2\pi^4}{45 \zeta(3)g_\dr\xi_{n_\dr}}Y_\dr(T_{\rm CMB})}{1-\frac{2\pi^4}{45 \zeta(3)}\frac{\xi_{\rho_\dr}}{\xi_{n_\dr}}Y_\dr(T_{\rm CMB})}g^\bath_{*s}(T_{\rm CMB}) \right)^{4/3} \ .
\label{eq:DeltaNeffY}
\ee

\subsection{Tracking the energy density}\label{subsec:BErho}

As we have just seen the number density is not the most suitable quantity to compute $\Delta N_ {\rm eff}$. Here, we derive a Boltzmann equation describing the evolution of the energy density $\rho_\dr$. We proceed very similarly as in the previous case, up to Eq.~\eqref{eq:ngeneral2}, which in this case reads
\be
\begin{split}
	\frac{d\rho_{\dr} }{dt} +4 H \rho_{\dr}  = & \; \ell \, \left[ 1 - \left(e^{\mu_\dr / T}\right)^\ell \right] \int \prod_{i = 1}^{n} d\Pi_i  \prod_{j = n+1}^{n+m}  d\Pi_j \prod_{r = 1}^{\ell}  d\mathcal{K}_r \, (2\pi)^4\delta^4 \left(P_{\rm in} - P_{\rm fin}  \right) \\ &  \; k_1 \, |\mathcal{M}|^2 \prod_{i = 1}^n f_{\bath_{i}}\prod_{j=n+1}^{n+m} (1\pm f_{\bath_{j}})   \prod_{r = 1}^{\ell} \left(1 \pm f_{\dr}(k_r) \right) \ .
\end{split}
\label{eq:rhogeneral2}
\ee
We remind the reader that the expression above is obtained under the assumption of kinetic equilibrium. With the further assumption of MB statistics for the dark radiation we find
\be
\frac{d\rho_{\dr} }{dt} + 4 H \rho_{\dr} = \epsilon_\dr \left[1 - \left(\frac{\rho_\dr}{\rho_\dr^{\rm eq}}\right)^\ell \right]   \ . 
\ee
Here, the temperature dependent function $\epsilon_\dr$ accounts for the averaged energy of $\dr$'s transferred to dark radiation per unit time and volume. It explicitly reads
\be
\epsilon_\dr  \equiv \ell \, \int \prod_{i = 1}^{n} d\Pi_i  \prod_{j = n+1}^{n+m}  d\Pi_j \prod_{r = 1}^{\ell}  d\mathcal{K}_r  (2\pi)^4\delta^4 
\left(P_{\rm in} - P_{\rm fin}  \right) k_1 |\mathcal{M}|^2 \prod_{i = 1}^n f_{\bath_{i}}\prod_{j=n+1}^{n+m} (1\pm f_{\bath_{j}})   \ .
\label{eq:epsilon}
\ee

A possible strategy to proceed is mirroring the discussion that we have just presented for the number density. However, we improve here the solution technique by accounting for the feedback of the dark radiation on the thermal bath. This is something we account for also in the main text when we work in momentum space. The discussion here is analogous to the one presented in Sec.~\ref{sec:formalism}. We consider another Boltzmann equation describing the evolution of the bath energy density, and we couple them with the Friedmann equation. This procedure leads to the system
\be \left\{ 
\begin{split}
	&\frac{d\rho_{\dr} }{dt} +4 H \rho_{\dr} = \epsilon_\dr \left[1 - \left(\frac{\rho_\dr}{\rho_\dr^{\rm eq}}\right)^\ell \right] \\  
	&\frac{d\rho_\bath}{dt} +3(1+w_\bath) H \rho_\bath = -\epsilon_\dr \left[1 - \left(\frac{\rho_\dr}{\rho_\dr^{\rm eq}}\right)^\ell \right] \ , \\
	&H = \frac{\sqrt{\rho_\dr + \rho_\bath}}{\sqrt{3}\Mpl} \   
\end{split} \right.
\ee 

A convenient numerical strategy for the solution of this system is highlighted in Sec.~\ref{sec:results}. We employ the scale factor $a(t)$ as the evolution variable, and we start the evolution at the scale factor value $a_I$ defined from the condition that the bath temperature is equal to some reference value $T_I$. As done in the main text, we introduce $A = a / a_I$ and we work in terms of the dimensionless densities defined in Eq.~\eqref{eq:Rvar}. The system becomes 
\be \left\{ 
\begin{split}\label{eq:Xdr_XSM}
	&\frac{dR_\dr}{d\log A}   = \frac{A^4}{H} \frac{\epsilon_\dr}{T_I^4}   \left[1 - \left(\frac{R_\dr}{R_\dr^{\rm eq}}\right)^\ell \right]\\  
	&\frac{dR_\bath}{d\log A} + (3 w_\bath - 1) R_\bath = - \frac{A^4}{H} \frac{\epsilon_\dr}{T_I^4} \left[1 - \left(\frac{R_\dr}{R_\dr^{\rm eq}}\right)^\ell \right]\ ,\\
	&H = \frac{\sqrt{R_{\cal B} + R_\dr}}{\sqrt{3} M_{\rm Pl}} \frac{T_I^2}{A^2}
\end{split}\right. \ .
\ee
This solution method never assumes entropy conservation, and the bath time-temperature relation is the output of the system. The dependence of $\epsilon_\dr$ and $R_{\dr}^\eq$ on the scale factor are known once $T(A)$ is determined from
\begin{equation}
R_\bath(A) = \frac{\pi^2}{30} \gst^\bath(T) A^4 \left(\frac{T}{T_I} \right)^4  \ .
\end{equation}
We set initial conditions as in the main text, $R_\dr (A = 1) =0$ and $R_\bath(A =1) = (\pi^2/30) \gst^\bath(T_I)$. The final result for $ \dneff $ is given in Eq.~\eqref{eq:DNeffX}. 


\bibliographystyle{JHEP}
\bibliography{DarkRadiation}

\end{document}